\documentclass[aps,prx,reprint,amsmath,amssymb,superscriptaddress]{revtex4-2}
\usepackage{bm,graphicx,xcolor,hyperref,tabulary,titlesec,multirow,commath,enumitem,orcidlink,algorithmic}
\hypersetup{colorlinks, citecolor=blue, linkcolor=red, urlcolor=teal}
\setcitestyle{numbers,square,sort&compress}
\usepackage[utf8]{inputenc}
\usepackage[T1]{fontenc}
\begin{document}

\title{Glassy phase transition in immiscible steady-state two-phase flow in porous media}

\author{Santanu Sinha}
\email{santanu.sinha@ntnu.no}
\affiliation{PoreLab, Department of Physics, Norwegian University of Science and Technology, 7491 Trondheim, Norway}

\author{Humberto Carmona}
\email{carmona@fisica.ufc.br}
\affiliation{Departamento de Física, Universidade Federal do Ceará, Fortaleza, Brazil}

\author{Jos{\'e} S. Andrade Jr.}
\email{soares@fisica.ufc.br}
\affiliation{Departamento de Física, Universidade Federal do Ceará, Fortaleza, Brazil}

\author{Alex Hansen}
\email{alex.hansen@ntnu.no}
\affiliation{PoreLab, Department of Physics, Norwegian University of Science and Technology, 7491 Trondheim, Norway}

\date{\today}

\begin{abstract}
Immiscible two-phase flow in porous media is a widespread process that is controlled by the complex interplay between viscous and capillary forces in a disordered porous matrix. This results in a range of morphological fluid patterns with distinct pore-scale characteristics, while at the Darcy scale it exhibits crossovers between regimes with different macroscopic flow properties. Yet, it remains a scientific challenge to derive the steady-state Darcy-scale flow properties from the pore scale. Here we show that the characteristic Darcy-scale features can be predicted by mapping the pore-scale distribution of fluid droplets in two-phase flow onto the distribution of spins in a spin-glass model. We apply the Jaynes maximum entropy principle to the spin configurations, obtaining a configurational probability distribution with a spin-glass Hamiltonian. We then use Boltzmann machine learning trained on the pore-scale flow configurations from a dynamic pore-network model and construct a phase diagram for the two-phase flow system based on the order parameters and susceptibilities. We find that the critical line separating the paramagnetic phase from a spin-glass phase coincides with the transition where the dependence of the Darcy-scale flow rate on the imposed pressure gradient changes from linear to non-linear for intermediate saturations. The glassy phase of the spin model coincides with a flow regime characterized by hysteresis, frustration and strong fluctuations over a wide range of time scales. It is tempting to identify this flow regime as a dynamic glass state. The success of this approach is surprising, since two-phase flow is a non-equilibrium phenomenon, whereas the properties of the spin glass are obtained using equilibrium statistical mechanics. We offer an explanation for this.
\end{abstract}


\maketitle

\section{Introduction}
In the 1980s, interest in immiscible two-phase flow in porous media surged, coinciding with the fractal revolution \cite{nittman1985fractal, mandelbrot1977fractal, feder1988fractals}. One of the most important outcomes of this period was the regime map constructed by Lenormand et al. \cite{lenormand1988numerical} for drainage, in which a non-wetting fluid invades a porous medium saturated by a wetting fluid. The map was spanned by the capillary number, a dimensionless measure measuring the ratio of viscous to capillary forces, and the ratio of the invading to the defending fluid viscosities. It revealed three types of invasion patterns: capillary fingering at low capillary numbers where capillary forces dominate, viscous fingering at higher capillary numbers where viscous and capillary forces compete and the defending fluid is the more viscous, and stable displacement at high capillary numbers where the invading fluid is the more viscous.

Attractive as these fractal patterns are, relating them to the large-scale behavior of immiscible two-phase flow, the scale relevant to practical applications in hydrology, soil science and reservoir engineering, has remained a challenge. The difficulty arises from the disparity in the length scales involved, since the patterns observed in the 1980s are pore-scale features, whereas practical applications arise on scales at which the porous medium appears continuous. Bridging these two descriptions is therefore not straightforward.

Four scales may be distinguished in porous media \cite{feder2022physics}: (1) the sub-pore scale, where the hydrodynamics and thermodynamics of confined spaces matter; (2) the pore scale, where growth phenomena such as invasion percolation are relevant \cite{barabasi1995fractal}; (3) the Darcy scale, where the porous medium is treated as a continuum with effective properties; and (4) the reservoir scale, where geological heterogeneities must be resolved explicitly.

Upscaling consists in connecting descriptions of the problem on adjacent scales. Two upscaling problems are central. The first is connecting the pore-scale description of immiscible two-phase flow to the Darcy scale, which amounts to constructing effective Darcy-scale flow equations from pore-scale physics. The second concerns the disorder introduced by geological structure when moving from the Darcy scale to the reservoir scale where one must represent geological structures that influence flow at the meter scale and above within grid blocks hundreds of meters across.

It is the first upscaling problem that will be relevant here. Several approaches to it, getting from a pore scale description to Darcy scale, are reviewed in Ref. \cite{berg2026from}. Most of these approaches are based on homogenization techniques, i.e. averaging over pore scale variables combined with a closure assumption. A different approach builds on the Jaynes generalization of {\it equilibrium\/} statistical mechanics via the observation that configurational entropy based on the fluid distribution and velocity field on scales between the pore scale and the Darcy scale remains constant under steady-state flow \cite{hansen2023statistical, hansen2026immiscible}. The work we present here is based on this approach.

The simplest two-phase flow state is steady-state flow.  At the Darcy scale, this means that the flow variables have well-defined, stationary time averages, that the saturation field has no gradients, and that the time-averaged pressure gradient driving the flow is spatially uniform. In the laboratory, steady state is established by simultaneously injecting the immiscible fluids into the sample \cite{tallakstad2009steady, tallakstad2009steadyb}.

The Payatakes group conducted extensive studies of pore-scale steady-state flow in the 1990s and 2000s, see Avraam and Payatakes \cite{avraam1995flow}. They classified the {\it pore-scale} description of steady-state flow into four regimes. At very high capillary numbers, the steady state is {\it drop traffic flow}. The non-wetting fluid (the less wetting of the two; the other we call the wetting fluid) forms sub-pore-scale droplets that are transported through the porous medium, as traffic moves in a city, hence, the name. At lower capillary numbers, one encounters the {\it small ganglion regime}. Ganglions are fluid clusters spanning several pores. They are small in the sense that the internal pressure drop across a ganglion is too small to affect its motion, so they are carried along by the surrounding fluid. At further lower capillary numbers, we enter the {\it large ganglion regime}, where the ganglions are large enough that the internal pressure drop across them becomes significant for their motion. Finally, at sufficiently low capillary numbers, we find the {\it continuous pathway regime} when the pressure gradients are so small that ganglions no longer contribute significantly to flow, which is instead carried by connected pathways through the porous medium.

In 2009, Tallakstad et al.\ \cite{tallakstad2009steady, tallakstad2009steadyb} observed in a two-dimensional porous medium that the average steady-state flow velocity depends on the average pressure gradient as a power,
\begin{equation}
    \label{eqAH1}
    v = - m |\nabla P|^\alpha\;,
\end{equation}
where $m$ is the conductivity and $\alpha=1.85\pm0.27$. Several groups have since confirmed $\alpha>1$ in two- and three-dimensional media, with exponents that vary between studies \cite{rassi2011nuclear, sinha2012effective, yiotis2013blob, aursjo2014film, sinha2017effective, gao2017x, yiotis2019nonlinear, roy2019effective, gao2020pore, zbg21, fyhn2021rheology, zhang2022nonlinear, sales2022bubble, fsh23}. Yiotis et al.\ \cite{yiotis2013blob} showed that the power-law constitutive equation holds within a window of capillary numbers; outside this window the constitutive equation is linear. The power-law regime arises because an increasing number of interfaces mobilize as the pressure gradient increases. A quantitative explanation of this mechanism, predicting $\alpha=2$ is given in \cite{tallakstad2009steadyb}, and its dependence on the pore-size distribution is studied in \cite{rsh21, roy2019effective}.  

Berg et al.\ \cite{berg2026from} recently proposed a {\it Darcy-scale} classification of steady-state flow regimes. It is based on the constitutive equation relating flow velocity to pressure gradient, since pore-scale concepts have no meaning at the continuum Darcy scale. Berg et al.\ \cite{berg2026from} refer to the linear, power-law and linear constitutive-equation regimes as regimes I, II and III, respectively, and split regime I into Ia and Ib. In regime Ia the interfaces are immobilized, so the pressure drop across the sample does not fluctuate. At the pore scale, as the capillary number increases, interfaces begin to move and ganglions change shape, while transport through the system remains dominated by open channels \cite{armstrong2016role} and the associated time scales span a wide range \cite{mcclure2021capillary}. At the Darcy scale this manifests as the onset of fluctuations in pressure drop, accompanied by strong hysteresis. This defines regime Ib. Gao et al.\ \cite{gao2020pore} investigated the transition from regime Ib to regime II, demonstrating that the onset of the non-linearity is characterized by large pressure fluctuations, terming the two regimes the onset of dynamics and intermittent flow, respectively \cite{yiotis2019nonlinear,sbb19,zbg21}.

To our knowledge, regime II has not been systematically studied for global non-wetting saturations $S_n$ close to $0$ or $1$. As $S_n\to 0$ or 1, the exponent $\alpha$ in Eq. \ref{eqAH1} must approach unity. Whether this approach is a gradual crossover or an abrupt transition is not known.

We thus have three Darcy-scale transitions: Ia--Ib, Ib--II and II--III. We conjecture that the Ia--Ib transition is due to kinetic arrest \cite{foffi2000kinetic}, i.e.\ the time scales for interface motion exceed the time scale of the experiment as the capillary number is lowered. It is not a transition in the sense of a singularity in the Darcy-scale variables, and the same holds for the II–III transition. Roy et al.\ \cite{roy2024effective} studied the latter computationally and analytically using a capillary fiber-bundle model, finding that its location depends on system size: the larger the system, the more it shifts toward the Ib–II transition. It is not a phase transition in the ordinary sense.

It is the nature of the Ib--II transition which is the focal point of this paper. We will demonstrate that it is a spin-glass transition, and consequently {\it regime {\rm Ib} is a glassy flow phase} \cite{binder1986spin}. We approach the nature of the Ib--II transition and the Ib regime by using the {\it Boltzmann machine learning method} \cite{meshulam2025statistical}. Our starting point is a dynamic pore-network model \cite{sinha2021fluid} where a porous matrix, in this implementation, is represented by a bi-periodic network of links on a diamond lattice. The flow is simulated by time-stepping the interface positions, recalculating the network flow at each step. According to whether the instantaneous saturation in a link lies above or below the average saturation, we assign an up or down spin, yielding an Ising-type model on a bi-periodic lattice dual to that of the flow model. We then compute the average spin of each link and the pair correlations between all pairs of links. From these quantities, we obtain the configurational probability distribution using the Jaynes maximum entropy principle, which takes the Boltzmann form, the exponential of a quadratic spin Hamiltonian. We find that the spin model combined with the Metropolis algorithm reproduces the statistics of the configurations from the pore-network simulations, including the three-point spin correlation function.

How, then, can the flow in the network, which is a driven system, be mapped onto a spin model that is updated by an equilibrium algorithm? This is possible because the entropy describing the flow is not molecular entropy but the entropy associated with the fluid configurations. Molecular entropy is produced, for example, by viscous dissipation, whereas configurational entropy is not. This allows the flow to be treated as an {\it equilibrium process} at the level of pore-scale configurations. The idea forms the core of a generalized statistical mechanics which in turn leads to a generalized thermodynamics for immiscible two-phase flow in porous media \cite{hansen2023statistical,hansen2026immiscible}. That the equilibrium Metropolis Monte Carlo updates of the spin Hamiltonian reproduce the configurational statistics of the flow model is strong evidence for this picture.

We measure the magnetization, the magnetic susceptibility, the Edwards-Anderson order parameter and the spin-glass susceptibility \cite{binder1986spin} over a wide range of flow parameters, and identify a spin-glass critical line in the phase diagram of the spin model. When comparing it with the Ib--II transition, we find that the two coincide. We conclude that regime Ib is a glassy flow regime and that the transition to regime II, the power-law regime, is a spin-glass transition. Since regime Ib exhibits both hysteresis and fluctuations spanning a wide range of time scales, this interpretation is intuitively consistent.

It is worth asking what more is gained by identifying regime Ib as glassy and the Ib–II transition as a critical line, other than connecting Darcy scale transition to the pore-scale configurations. This identification makes the extensive literature on spin-glass theory relevant to immiscible two-phase flow in porous media, and it provides the spin-glass community with an experimentally realizable system to explore.

The paper is organized as follows. In Section \ref{sec_spin} we introduce the mapping between the flow problem and the Ising-type spin model. The dynamic pore-network model (DPN) generating the fluid configurations is presented in Section \ref{sec_model}. The implementation of the mapping between the two models, based on Boltzmann machine learning (BML), is presented in Section \ref{sec_verif}. Section \ref{sec_glass} discusses the consequences of the DPN--BML mapping, including the glass transition and its meaning for the flow problem. We summarize our work in the final section.

\section{Configurational probabilities}
\label{sec_spin}
In this section we will establish the mapping between the immiscible two-phase flow problem and an Ising-type spin model.  By running equilibrium Monte Carlo simulations \cite{binder2025guide} on this spin model, it will reproduce the saturation correlations found in the flow model.  This works as it is the fluid configurations at or above the pore scale that control the flow.  These configurations is described by a configurational entropy which remains constant in a steady state similar to an equilibrium state \cite{jaynes1957information, jaynes1957informationb}. 

For the present study, we restrict ourselves to (quasi-) two-dimensional (2D) porous media, though the approach is equally valid for three dimensions (3D). Common examples of such 2D systems used in the laboratory experiments are the Hele-Shaw cell filled with immovable glass beads \cite{tallakstad2009steady}, or the etched PDMS microfluidic chips \cite{khj14}, where snapshots of fluid configurations can be recorded using high resolution cameras. The most detailed configurations in these cases are governed by the resolution of the snapshots where a pixel can belong either to the solid matrix, or to one of the two fluids. The fluid configurations can then be specified by assigning spins variables with values based on the pixel types.  

In case of computational experiments, such as lattice-Boltzmann method (LBM) \cite{alk11} or dynamic pore network models \cite{jh12}, a configuration in the steady state can similarly be defined by associating spin variables to the fluid voxels or pores respectively. The spin values are then allocated according to the value of the corresponding voxels.  This is a data driven approach, which requires large scale configurational data for a wide range of parameters. We will here use a dynamic pore network (DPN) model \cite{sinha2021fluid}, which is relatively less computationally expensive compared to other methods such as LBM, and therefore is well suited here. We emphasize, however, that the whole approach can be implemented for other types of simulation methods or experiments.

Let us now consider a porous matrix consisting of $N$ pores. For the DPN model we will use here, a pore is represented by a link of a pore network. We associate a spin variable $\sigma_i$ to each link, where $i=1,2,\ldots,N$. Each link can contain bubbles of the wetting and non-wetting fluids. We define the state of a spin by,
\begin{equation}
    \displaystyle
    \label{eq_spindef}
    \sigma_i =
    \begin{cases}
        -1 & \text{if\;\;} s_i < \langle s\rangle \\
        +1 & \text{if\;\;} s_i \ge \langle s\rangle 
    \end{cases}
\end{equation}
where $s_i$ is the non-wetting saturation in the $i$th link, and $\langle s\rangle$ is the average pore-scale saturation of the non-wetting fluid. The set of the values of the spins, $\{\sigma_i\cdots\sigma_N\}$, defines a configuration $\{\sigma\}$.

Note that by assigning spins to the links in this way, it may seem that we have by construction made the overall magnetization, defined as the average spin over all links, always equal to zero. This however is not so. Suppose we have a network with $100$ links where $20$ of them are fully filled with non-wetting liquid and the rest $80$ are fully filled with the wetting liquid. We then have $\langle s\rangle = 0.2$, and $80$ of the links would be assigned spins pointing down and $20$ would be assigned spins pointing up, making the magnetization non-zero. 

We will now find a configurational probability $P(\{\sigma\})$ for a spin configuration $\{\sigma\}$ based on the Jaynes maximum entropy principle \cite{jaynes1957information, jaynes1957informationb}, such that two constraints, the average magnetization $\langle\sigma_i\rangle$ of each spin $i$ and the pairwise correlation $\langle \sigma_i\sigma_j \rangle$ between all spin pairs $i$ and $j$, can be reproduced by the probability distribution. The average $\langle\cdots\rangle$ here is over all the available configurations we can have from a simulation or experiment in the steady state. In terms of $P(\{\sigma\})$, these two constraints can be expressed as,
\begin{equation}
    \begin{aligned}
        \label{eq_constr}
        \displaystyle
        \langle \sigma_i \rangle         & = \sum_{\{\sigma\}} \sigma_iP(\{\sigma\}) \qquad\text{and}\\ 
        \langle \sigma_i\sigma_j \rangle & = \sum_{\{\sigma\}} \sigma_i\sigma_j P(\{\sigma\}) \; ,
    \end{aligned}
\end{equation}
where the sum is over all unique configurations. We also have in addition the normalization criterion,
\begin{equation}
    \label{eq_sump}
    \displaystyle
    \sum_{\{\sigma\}} P(\{\sigma\}) = 1 \;.
\end{equation}
With these constraints, we will maximize the Shannon information entropy,
\begin{equation}
    \label{eq_shannon}
    \displaystyle
    \Omega = -\sum_{\{\sigma\}}P(\{\sigma\})\ln P(\{\sigma\}) \;.
\end{equation}
This can be done by using the Lagrange multiplier method \cite{meshulam2025statistical}. We construct the Lagrangian with the constraints in Eqs. \ref{eq_constr}, \ref{eq_sump} and \ref{eq_shannon} as, 
\begin{equation}
    \begin{aligned}
    \label{eq_lagrange}
    \displaystyle
    \mathcal{L} = & - \sum_{\{\sigma\}} P(\{\sigma\}) \ln P(\{\sigma\})
                    + \theta \left[ \sum_{\{\sigma\}} P(\{\sigma\}) - 1\right]\\
                  & + \sum_i h_i \left[\sum_{\{\sigma\}} \sigma_i P(\{\sigma\}) - \langle \sigma_i \rangle \right]\\
                  & + \sum_i\sum_{j<i} J_{ij}\left[\sum_{\{\sigma\}}\sigma_i\sigma_j P(\{\sigma\}) - \langle \sigma_i\sigma_j \rangle \right]\;,
    \end{aligned}
\end{equation}
where $\theta$, $h_i$ and $J_{ij}$ are the Lagrange multipliers. We then set $\partial \mathcal{L}/\partial P(\{\sigma\}) = 0$ to find the maximum of $\mathcal{L}$ which provides,
\begin{equation}
    \label{eq_P}
    \displaystyle
    P(\{\sigma\}) = \frac{1}{Z}\exp\left[ \sum_i h_i\sigma_i + \sum_i\sum_{j<i} J_{ij}\sigma_i \sigma_j \right]\;.
\end{equation}
The normalization factor or the partition function $Z$ is given by,
\begin{equation}
    \begin{aligned}
        \label{eq_partition}
        \displaystyle
        Z & = \sum_{\{\sigma\}} P(\{\sigma\})\\
          & = \sum_{\{\sigma\}}\exp\left[\sum_i h_i\sigma_i + \sum_i\sum_{j<i} J_{ij}\sigma_i \sigma_j \right]\;.
    \end{aligned}
\end{equation}
The configurational probability can therefore be expressed in terms of an Hamiltonian $H$,
\begin{equation}
    \label{eq_PH}
    \displaystyle
    P(\{\sigma\}) \propto \exp\left[-H(\{\sigma\})\right] \;,
\end{equation}
where the ensuing spin model Hamiltonian is,
\begin{equation}
    \label{eq_H}
    \displaystyle
    H(\{\sigma\}) = -\sum_i h_i\sigma_i - \sum_i\sum_{j<i} J_{ij}\sigma_i \sigma_j\;.
\end{equation}
Here $h_i$ are field couplings and $J_{ij}$ are spin-spin couplings. 


The parameters $h_i$ and $J_{ij}$ respectively play the role of the field constants and the coupling constants determining the different phases of the system, set by the quenched disorder of the sample. Averages over this disorder are needed to be taken into account in addition to the thermal averages. The number of the couplings grow with the system size making it challenging for the theory. Edward and Anderson provided the first model in 1975 for spin glass \cite{ew75} which considers nearest neighbor interaction with $J_{ij}$s drawn from a binary or Gaussian distribution of random variables. Sherrington and Kirkpatrick (SK) introduced a solvable mean-field model \cite{sk75} where all the spin pairs have direct interactions, which was later solved by Parisi \cite{parisi2023nobel}. For a Gaussian distribution of the couplings in the SK model, their solution reveals a phase space with three phases --- ferromagnetic, paramagnetic and spin-glass phases, controlled by the standard deviation and mean of the distribution of the couplings \cite{binder1986spin}.

Given this similar form of the configurational probability, we will explore how we can characterize such phases for steady-state two-phase flow, and how the phases are related to different steady-state flow regimes. We use configurational data from the DPN model \cite{sinha2021fluid}, which we describe in the following.

\section{Dynamic pore network (DPN) model}
\label{sec_model}
To simulate immiscible two-phase flow in porous media using the dynamic pore network model, the pore space of a porous matrix is modeled by a network of links and nodes of certain geometry, through which the two fluids are transported \cite{jh12}. We are using a {\it dynamic\/} pore network model that takes the effect of both viscous and capillary forces into account to move the fluids. The fluid displacements are tracked by the position of all the fluid-fluid interfaces in the network. The links of the network consist of a narrow pore throat between two wider pore bodies, which are modeled by hour-glass shaped composite links. All the pore space of the medium is therefore accounted for by the links, and the nodes of the network denote only the positions of the link intersections. While we are only considering 2D networks here, the model can be used for any network topology.

\begin{figure}
    \centerline{\hfill\includegraphics[width=\columnwidth]{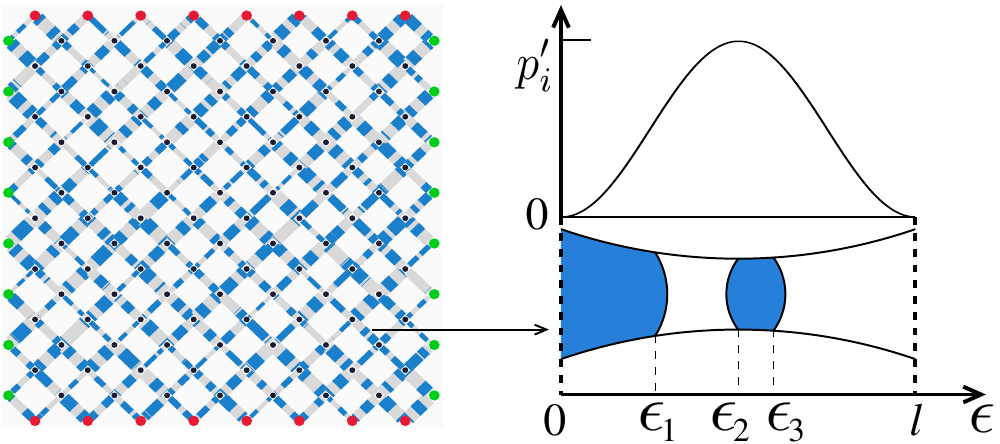}\hfill}
    \centerline{\hfill(a)\hfill\hfill(b)\hfill}
    \caption{\label{fig_model}(a) A pore network consisting $16\times 16$ links on a 2D diamond lattice. The bonds in the lattice represent composite pores and the black dots represent the position of nodes. The thickness of the bonds indicates their average diameter. The gray and blue parts of the links indicate wetting and non-wetting bubbles. There is periodic boundary condition in both the horizontal and vertical directions, which is indicated by the green nodes on the vertical boundaries being identical nodes. The same is true for the red nodes on the horizontal boundaries. The shape of one composite link in terms of interfacial pressure ($p_i'$) is shown in (b) bottom where $\epsilon_1$, $\epsilon_2$ and $\epsilon_3$ show the position of three interfaces. The variation of $p'_i$ at an interface along the length of the tube follows Eq. \ref{eq_pc}, which is indicated in (b) top.}
\end{figure}

We consider a network of $L\times L = N$ links embedded in a 2D diamond lattice. This is illustrated in Figure \ref{fig_model} (a) for a network of $16\times 16 = 256$ links. There are periodic boundary conditions in both directions as indicated by the blue and red nodes in the figure. This makes the system closed and the total volumetric saturation of the fluids therefore remains constant with time. All the links have the same length of $l=1\,{\rm mm}$, whereas the average radius $\rho_i$ of a link $i$ is drawn from a uniform distribution of random numbers in the range between $0.1l$ to $0.4l$. This is indicated by the variation in the widths of links in the figure. A typical distribution of two fluids inside the link is shown by the gray and blue colors indicating wetting and non-wetting bubbles. We do not take film or corner flow into account. For a fully developed laminar flow of incompressible fluids inside every link, the equation for the volumetric flow rate $\phi_i$ at any instance inside a link is given by \cite{w21,lenormand1988numerical},
\begin{equation}
    \displaystyle
    \label{eq_qrate}
    \phi_i = -\frac{\kappa_i}{l\mu_i}\left[\Delta p_i - \sum_k p'_i(\epsilon_k) \right] \;,
\end{equation}
where, $\Delta p_i$ is the pressure drop between the two nodes across the link $i$. Here $\kappa_i$ is the conductivity of the link, and we consider a circular cross section for which $\kappa_i=\pi \rho_i^4/8$ for the Hagen–Poiseuille flow. The term $\mu_i$ is the effective viscosity of the fluids present inside link $i$ at an instance. This is given by $\mu_i = s_i\mu_n + (1-s_i)\mu_w$, where $\mu_w$ and $\mu_n$ are the viscosities of the wetting and non-wetting fluids respectively, and $s_i$ is the volumetric saturation of the non-wetting fluid in that link, i.e., the proportion of the link volume that is occupied by the non-wetting fluid.

The second term ($p_i'$) inside the bracket in Eq. \ref{eq_qrate} corresponds to the interfacial pressure associated with the fluid-fluid interfaces. For a hour-glass shaped composite link, this pressure will vary along the link and will be a function of the interface position. We model this functional form with a modified Young–Laplace equation \cite{d92,shb13,amh98} given by,
\begin{equation}
    \displaystyle
    \label{eq_pc}
    \left|p_i'(\epsilon)\right| = \frac{2\gamma}{\rho_i}\left[1-\cos\left(\frac{2\pi\epsilon}{l}\right) \right] \;,
\end{equation}
where $\gamma$ is the surface tension between the two fluids and $\epsilon$ is the position of the interface. This is illustrated in Fig. \ref{fig_model} (b) which shows three interfaces at positions $\epsilon_1$, $\epsilon_2$ and $\epsilon_3$. The summation in Eq. \ref{eq_qrate} is over all the interfaces inside the link $i$ where proper signs for $p'_i$ need to be accounted depending on the direction of the capillary force.

Finally, we have from Kirchhoff law $\sum_{i\in n_i} \phi_i=0$ at every node for incompressible fluids. Here $n_j$ is the number of links connected to a node $j$. This leads to a set of linear equations which we solve by conjugate gradient method \cite{batrouni1988fourier} with necessary boundary conditions. The solutions provide local pressures at nodes, which are then used to calculate the link flow rates $\phi_i$. All the interfaces are displaced in the direction of the flow in respective links with a time step $\Delta t$, which is chosen in such a way that the maximum displacement of any interface should not exceed more than $10\%$ of the respective link. This is done by calculating $\Delta t = 0.1l/v_{\rm max}$ at every time step, where $v_{\rm max}$ is the fastest fluid velocity among all links. If there is a continuous flow channel across the network in the direction of global pressure drop, fluids in that channel should therefore need a minimum of $10L$ time steps to cross the whole network. As for this study we need the collection of configurations that are not correlated in time, we recorded configurations in steady state with an interval of every $2\cdot10L$ time steps.

To distribute the fluids from links to neighboring links through the nodes, we first calculate the total volume of the wetting and non-wetting fluids ($V_j^w$ and $V_j^n$ respectively, and $V_j^w+V_j^n=V_j$) arriving at every node $j$ from the connected links in which the flow is towards the node (incoming links). These volumes are then distributed as new bubbles in the connected links in which the flow is away from the node (outgoing links). The volume of these new wetting and non-wetting bubbles in an outgoing link $i$ are given by $V_i^w=\phi_i\Delta tV_j^w/V_j$ and $V_i^n=\phi_i\Delta tV_j^n/V_j$ respectively. This implies that the ratio of the injected fluid volumes ($V_i^w+V_i^n$) into different outgoing links is equal to the ratio of the flow rates $\phi_i$ in those links. In addition, the ratio between the $V_i^w$ and $V_i^n$ in one outgoing link is equal the ratio between the total incoming wetting and non-wetting volumes at the corresponding node.

Simulations are performed at constant capillary numbers $\rm Ca$ defined as,
\begin{equation}
    \label{eq_Ca}
    \displaystyle
    {\rm Ca}= \frac{Q\mu}{A\gamma}\;,
\end{equation}
where $Q$ is total flow rate of the fluids in the direction of global pressure drop $\Delta P$. Here $\mu$ is the total effective viscosity, and $A$ is the average cross-sectional area measured as $A=\sum_i\pi \rho_i^2/L$. Here we considered $\gamma = 0.03\;{\rm N/m}$, and equal viscosities for the two fluids with $\mu=0.1\;{\rm Pa.s}$. We then vary $\rm Ca$ by changing the total flow rate $Q$.

The model has been tested against numerous experimental \cite{sinha2017effective,roy2022co} and theoretical \cite{sinha2012effective,hsb18} results. It can simulate steady-state flow with different boundary conditions, as well as the drainage displacements to produce capillary and viscous fingers \cite{smf24}. The model is also applicable to 3D porous networks reconstructed from real samples \cite{sinha2017effective}.

\begin{figure*}
    \begin{minipage}{0.22\textwidth}
        \includegraphics[width=\textwidth]{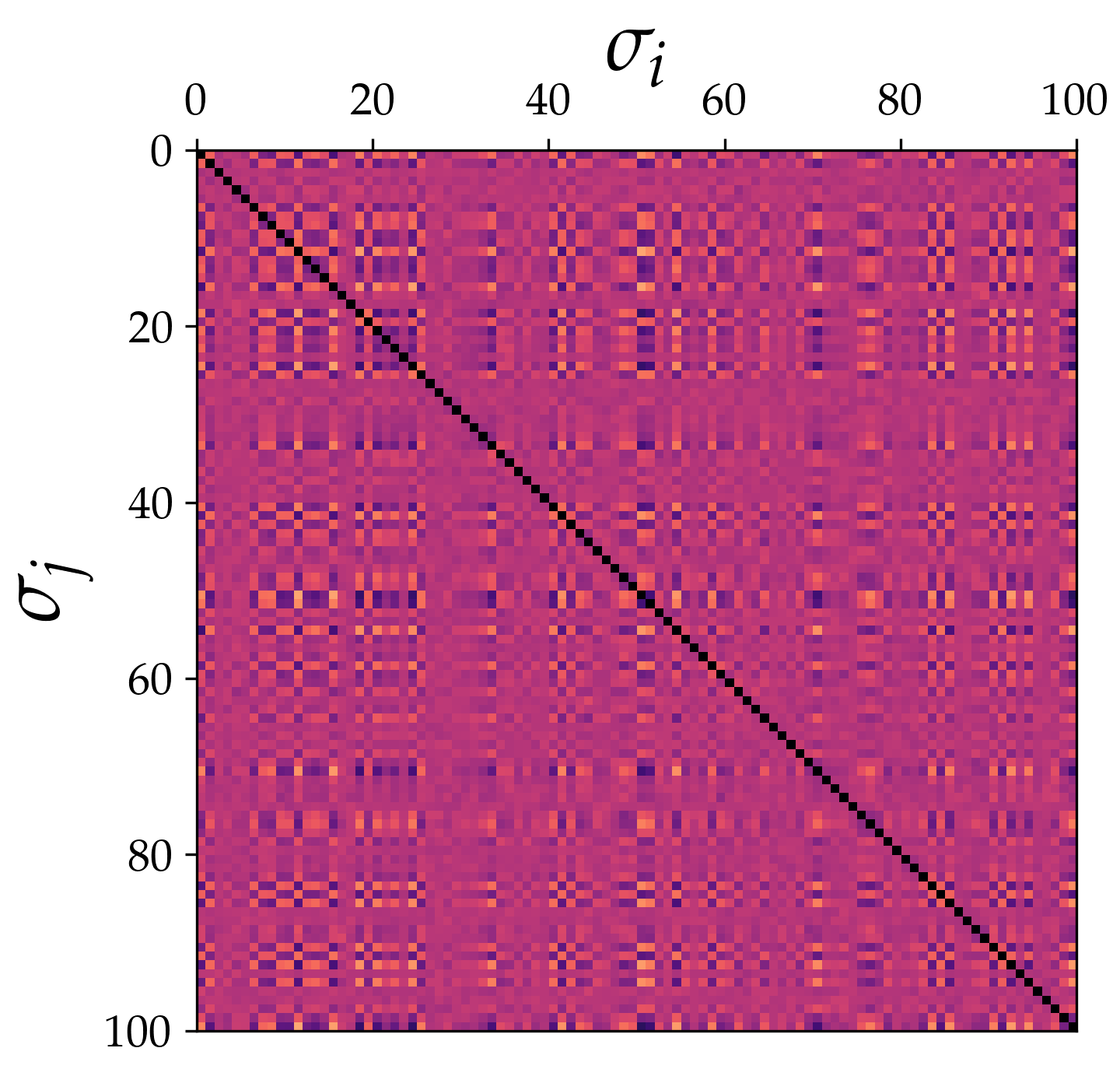}\\
        \includegraphics[width=\textwidth]{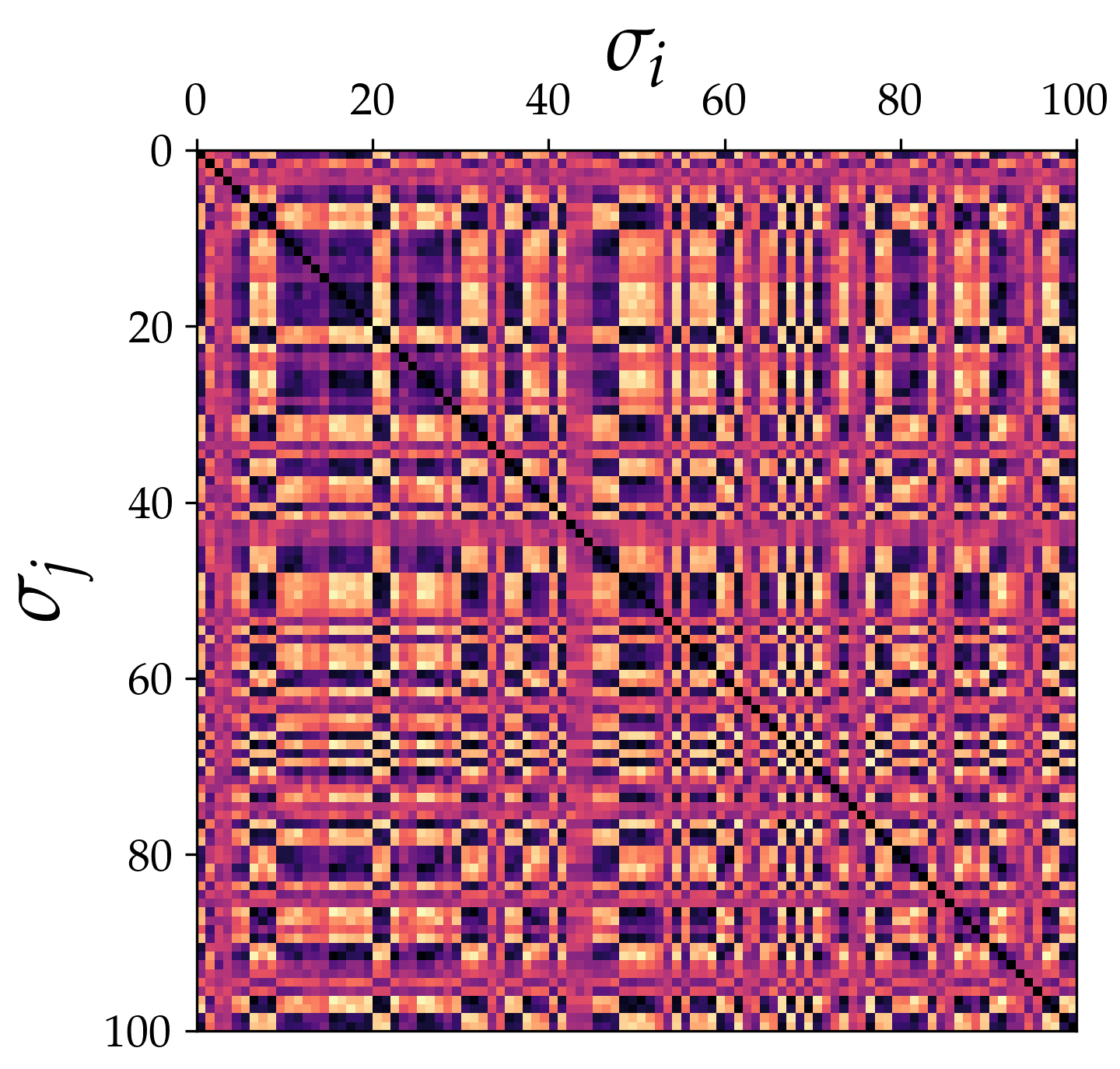}\\
        $\langle\sigma_i\sigma_j\rangle_{\rm Data}$
    \end{minipage}
    \begin{minipage}{0.22\textwidth}
        \includegraphics[width=\textwidth]{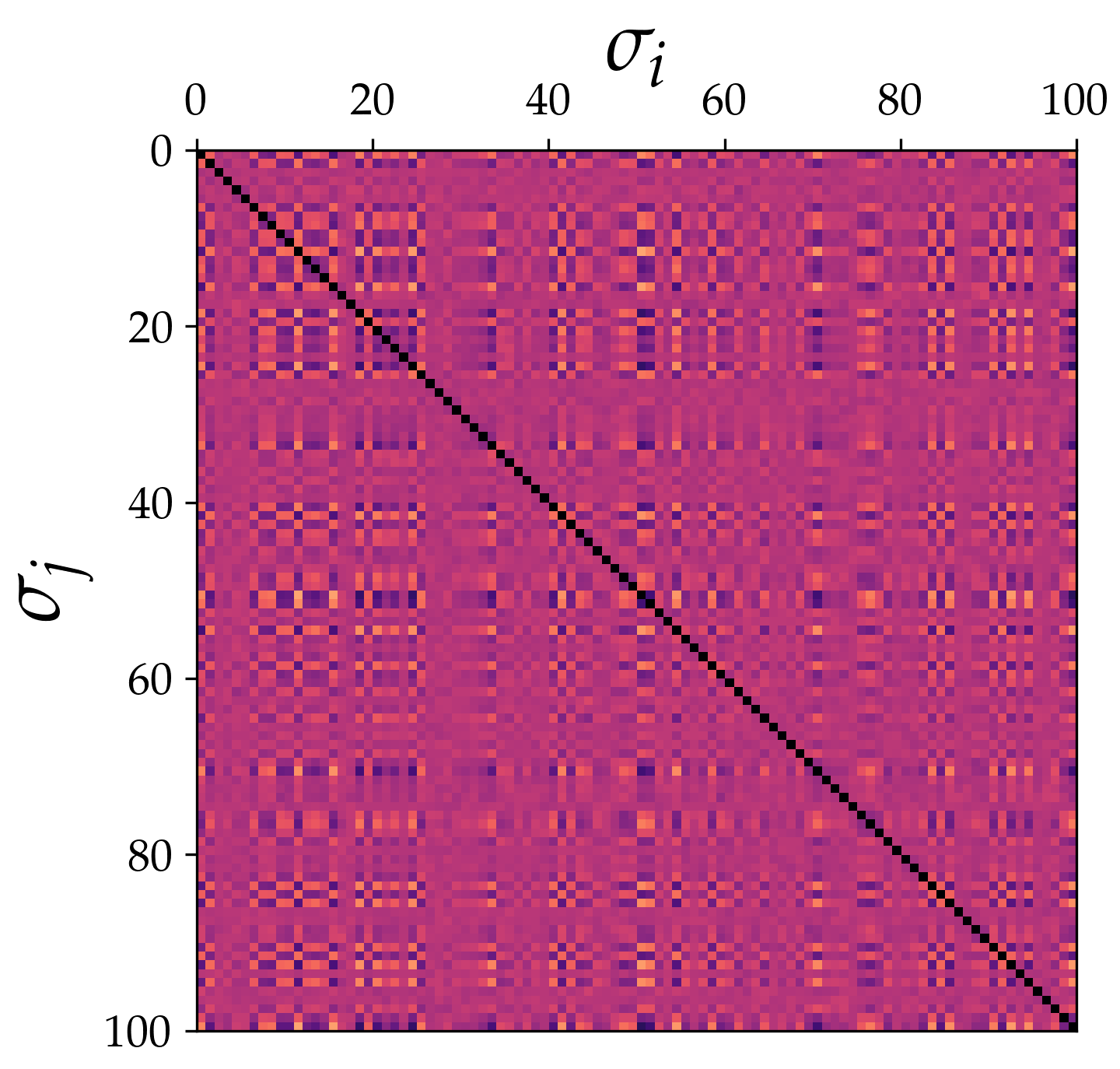}\\
        \includegraphics[width=\textwidth]{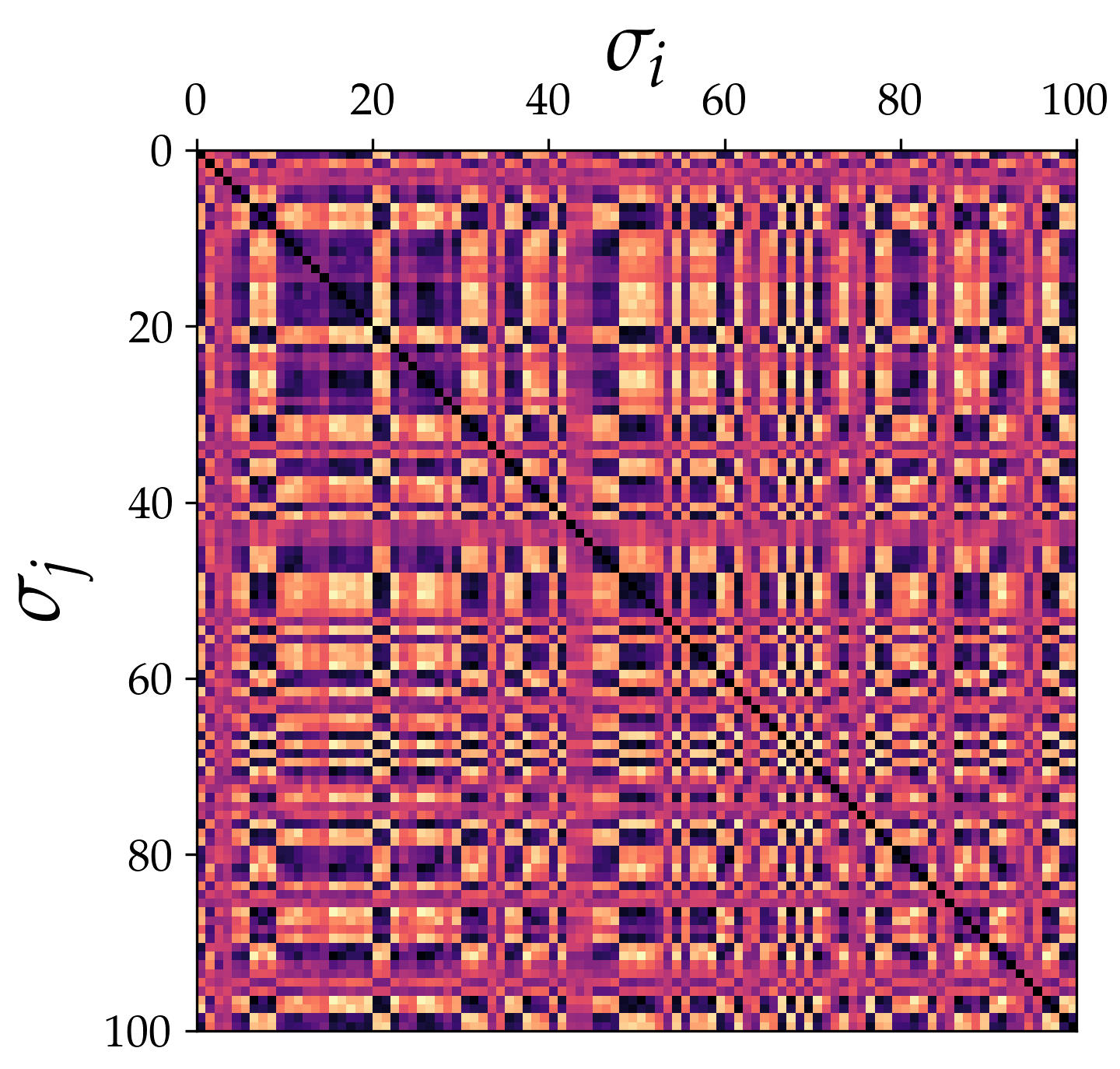}\\
        $\langle\sigma_i\sigma_j\rangle_{\rm MC}$
    \end{minipage}
    \begin{minipage}{0.25\textwidth}
        \includegraphics[width=\textwidth]{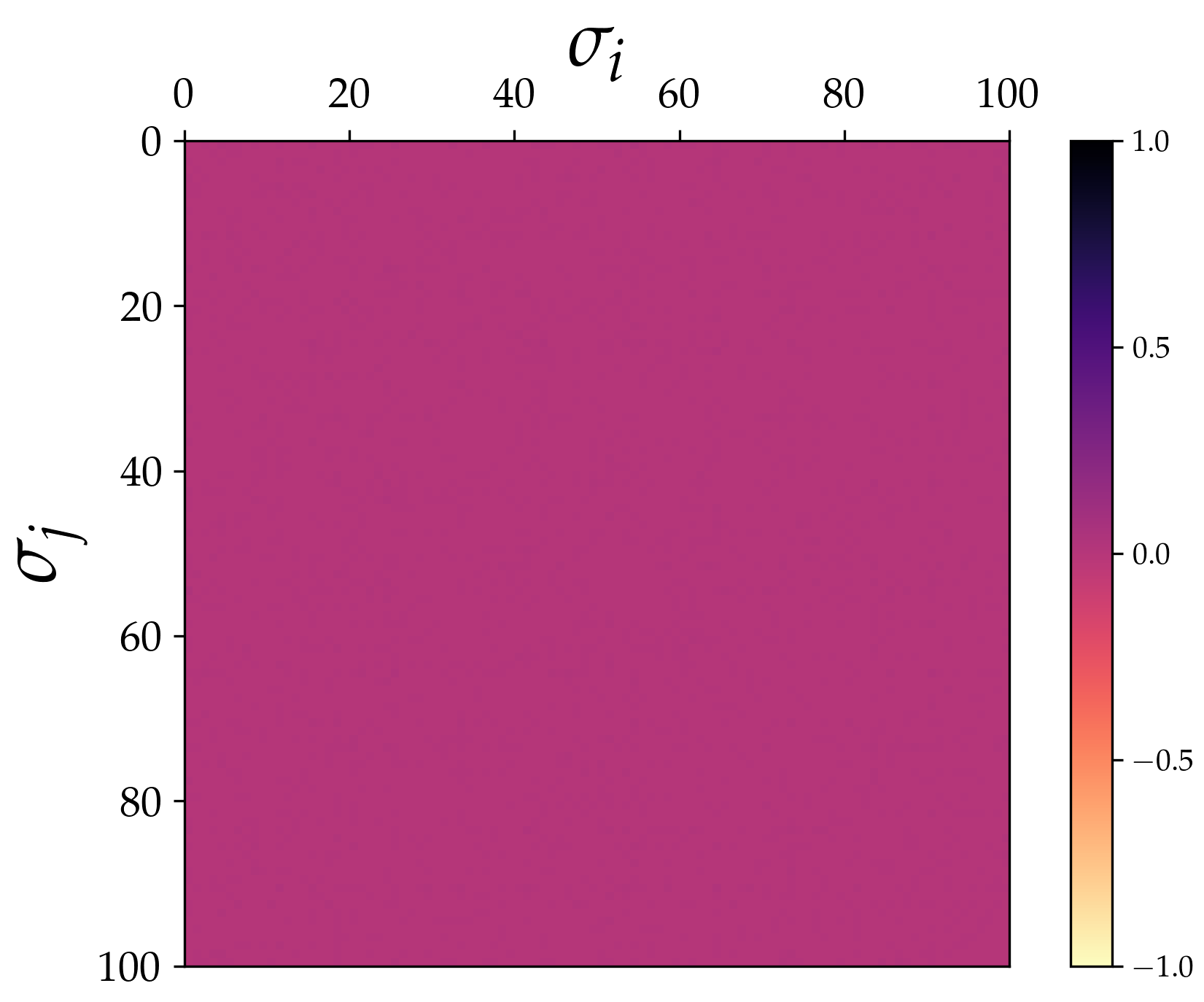}\\
        \includegraphics[width=\textwidth]{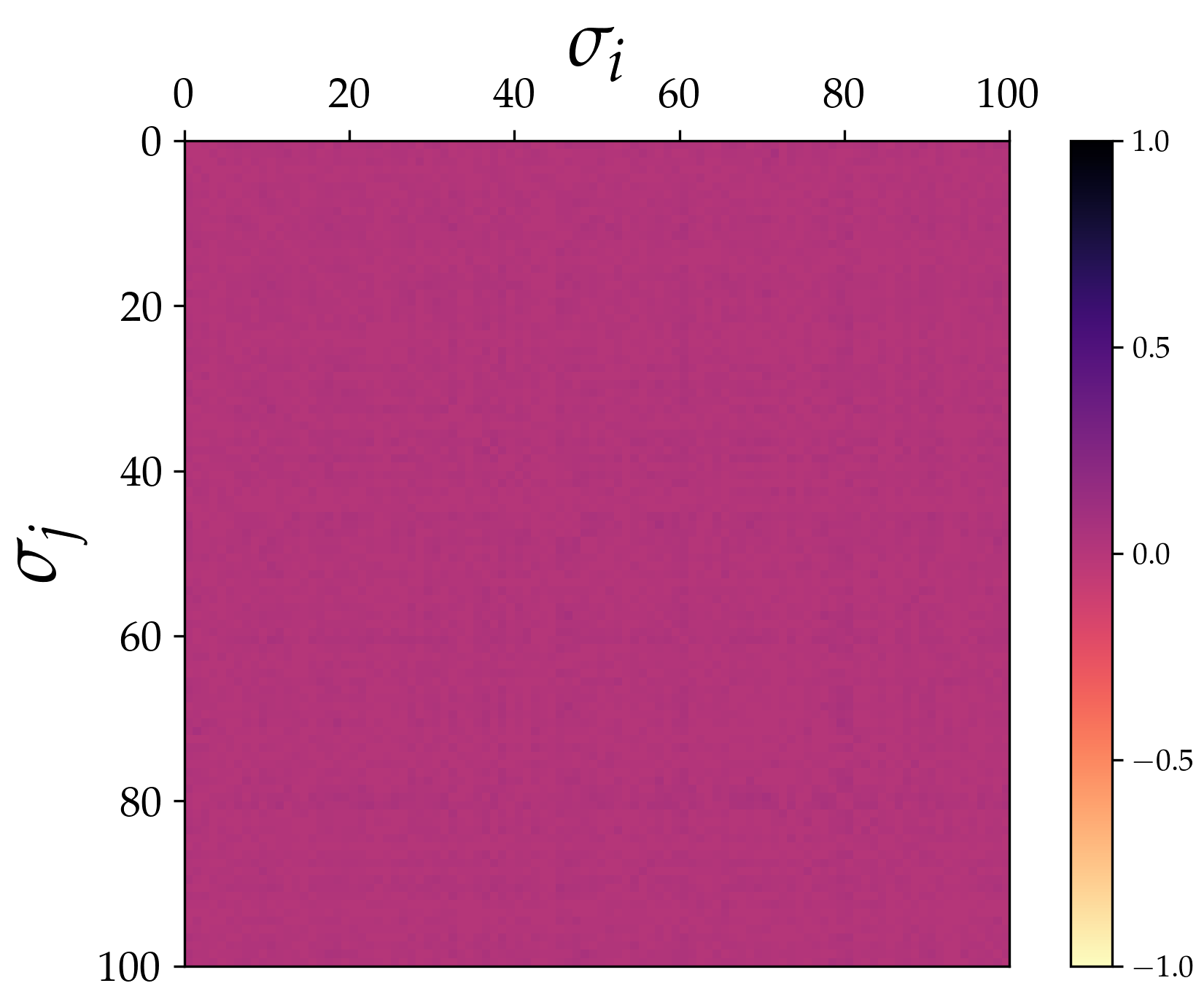}\\
         $\Delta_{ij}$
    \end{minipage}
    \begin{minipage}{0.26\textwidth}
         \includegraphics[width=\textwidth]{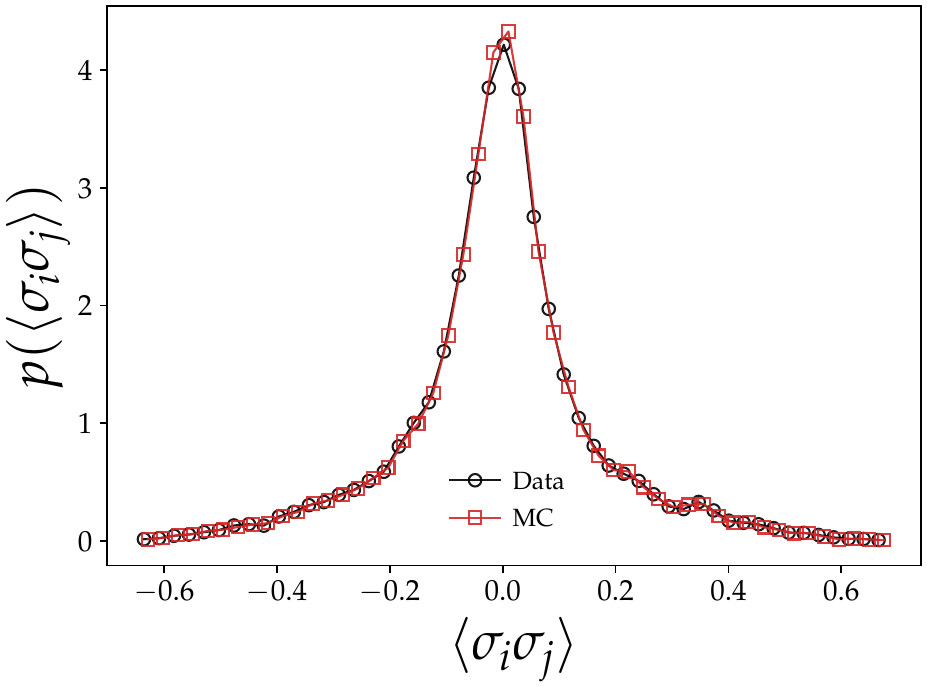}\\[10px]
         \includegraphics[width=\textwidth]{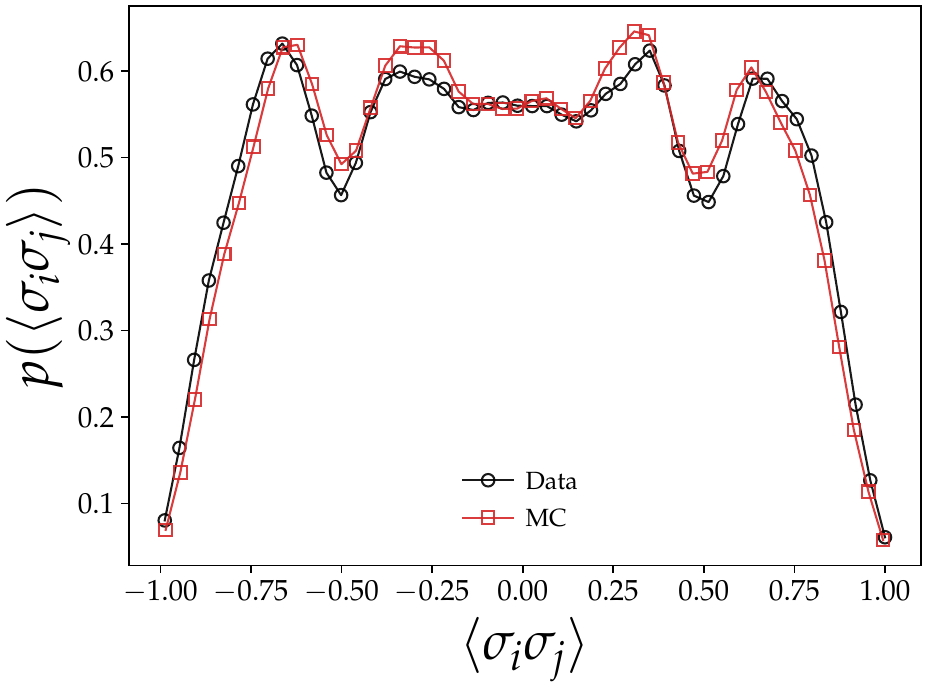}
    \end{minipage}
    \caption{\label{fig_matSij} Comparison of the correlations between the pore network simulation data and the MC results with BML for $S_n=0.3$ and two different values of ${\rm Ca}$. The first two columns show the matrix plots of $\langle \sigma_i\sigma_j \rangle$ for DPN and BML respectively. The third column shows the difference $\Delta_{ij}=|\langle \sigma_i\sigma_j \rangle_{\rm Data} - \langle\sigma_i\sigma_j\rangle_{\rm MC}|$ which shows the values are around zero for all the pairs. The fourth column shows comparisons of the correlation distributions $p(\langle\sigma_i\sigma_j\rangle)$.}
\end{figure*}

\section{Boltzmann machine learning (BML)}
\label{sec_verif}
We adapt here the Inverse Ising modeling, also known as Boltzmann machine learning method \cite{meshulam2025statistical}, to determine the distribution of the local field couplings $h_i$ and coupling constants $J_{ij}$ related to the underlying Hamiltonian given in Eq. \ref{eq_H} from the DPN simulation. The method can be applied to a wide variety of problems ranging from biological networks \cite{erw20,fah24} and trade networks \cite{fpc26} to bird flocking \cite{b12, b14b}.

The idea is to obtain the values of $h_i$ and $J_{ij}$ for the probability distribution function $P(\{\sigma\})$, that best reproduce the configurations obtained from the simulation data. This is done by minimizing the difference between the target probability distribution $P^{\rm Data}$ related to the data (here the DPN simulations), and the probability distribution ($P^{\rm MC}$) that follow function $P(\{\sigma\})$ given in Eq. \ref{eq_P}. To generate configurations with $P(\{\sigma\})$ we use Monte Carlo (MC) simulations \cite{binder2025guide}. The distance between two statistical distributions are measured using Kullback-Leibler (KL) distance ($D_{\rm KL}$) given by,
\begin{equation}
    \label{eq_KL}
    \displaystyle
    D_{\rm KL}\left(P^{\rm Data}\parallel P^{\rm MC}\right) = \sum_k P_k^{\rm Data} \ln\left(\frac{P_k^{\rm Data}}{P_k^{\rm MC}}\right)\;,
\end{equation}
where the sum is over all configurations. To minimize $D_{\rm KL}$ with respect to $h_i$ and $J_{ij}$, we use the iterative gradient descent method,
\begin{equation}
    \begin{aligned}
        \label{eq_gd}
        \displaystyle
        h_i (\tau+1)    & = h_i(\tau) - \eta\left.\frac{\partial D_{\rm KL}}{\partial h_i}\right|_\tau\quad\text{and}\\
        J_{ij} (\tau+1) & = J_{ij}(\tau) - \eta\left. \frac{\partial D_{\rm KL}}{\partial J_{ij}}\right|_\tau\quad,
    \end{aligned}
\end{equation}
where $\tau$ is the iteration step, and $\eta$ is the learning rate. By using derivatives of $D_{\rm KL}$, and the definition of $P^{\rm MC}\equiv P(\{\sigma\})$ given in Eq. \ref{eq_P}, we find from Eqs. \ref{eq_KL} and \ref{eq_gd} that
\begin{equation}
    \begin{aligned}
        \label{eq_bml}
        \displaystyle
        J_{ij} (\tau+1) & = J_{ij}(\tau) + \eta\left[\langle \sigma_i\sigma_j \rangle^{\rm Data} - \langle \sigma_i\sigma_j \rangle_\tau^{\rm MC}\right]\; {\rm and}\\
        h_i (\tau+1)    & = h_i(\tau) + \eta\left[\langle \sigma_i \rangle^{\rm Data} - \langle \sigma_i \rangle_\tau^{\rm MC}\right]\;,
    \end{aligned}
\end{equation}
where the quantities $\langle\ldots\rangle^{\rm Data}$ are measured from the DPN simulation data at the beginning of BML iterations, and $\langle\ldots\rangle^{\rm MC}$ are measured from the MC simulations at every iteration $\tau$.

\begin{figure*}
    \centerline{\hfill
        \includegraphics[width=0.25\textwidth]{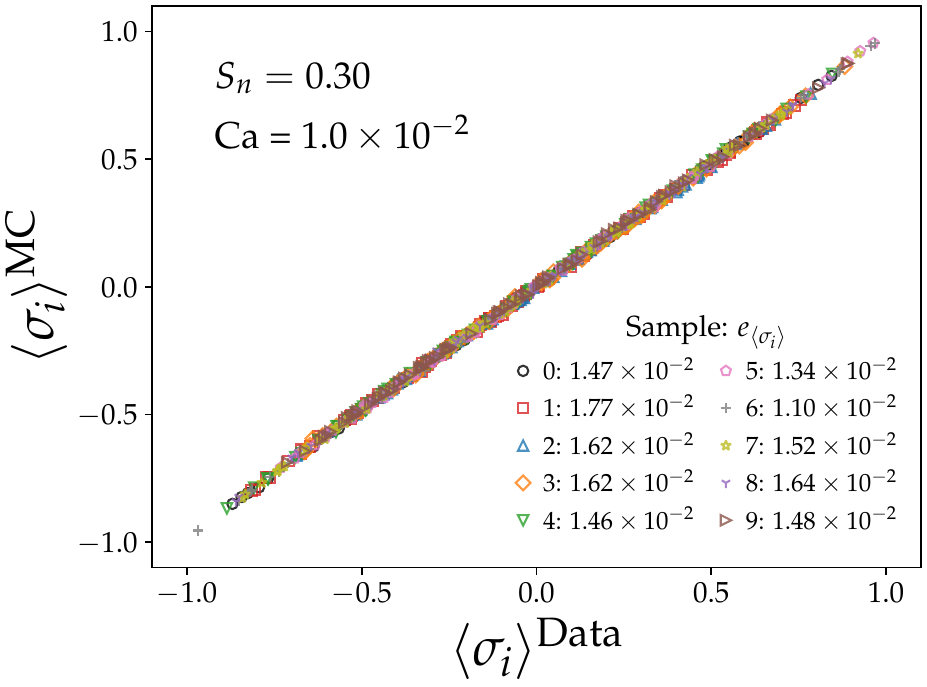}\hfill
        \includegraphics[width=0.25\textwidth]{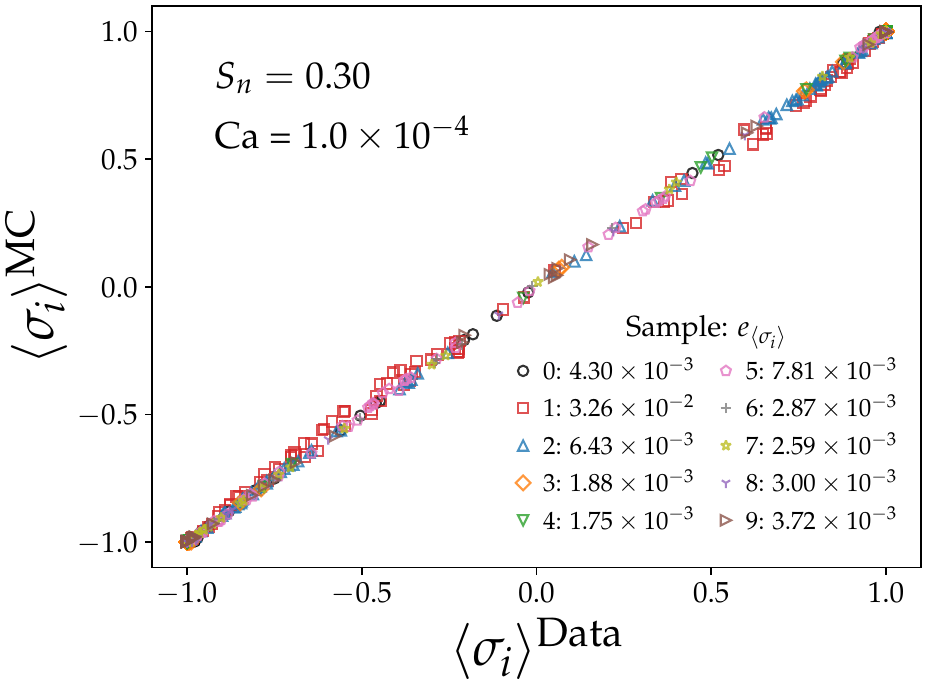}\hfill
        \includegraphics[width=0.25\textwidth]{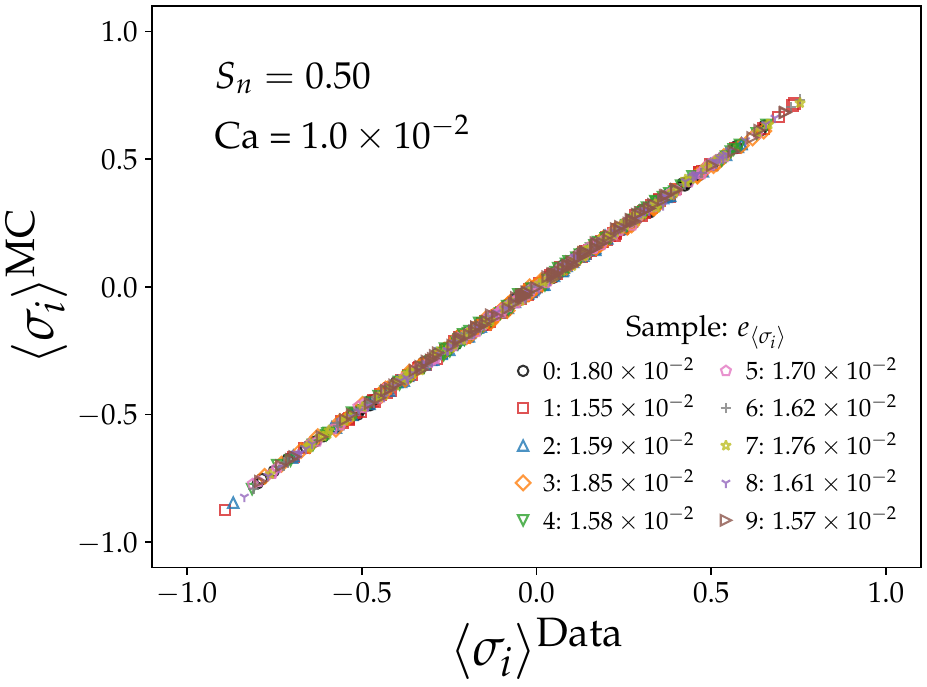}\hfill
        \includegraphics[width=0.25\textwidth]{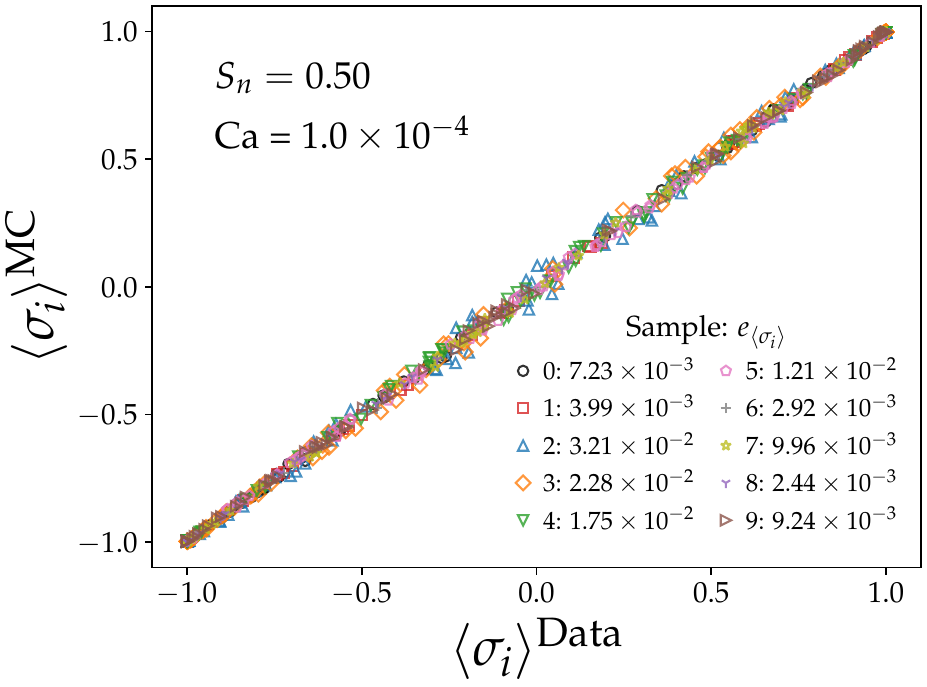}\hfill
    }
    \centerline{\hfill
        \includegraphics[width=0.25\textwidth]{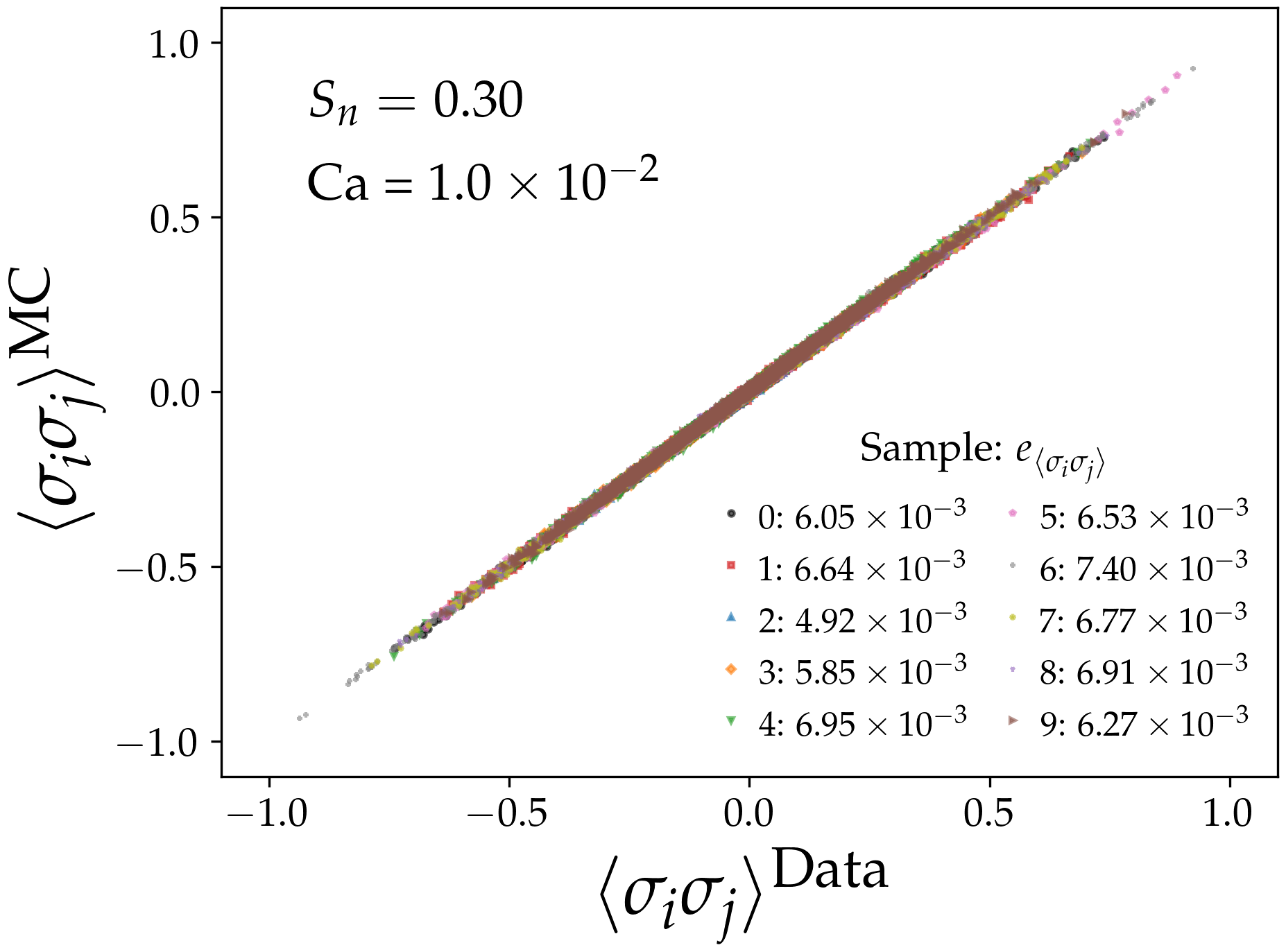}\hfill
        \includegraphics[width=0.25\textwidth]{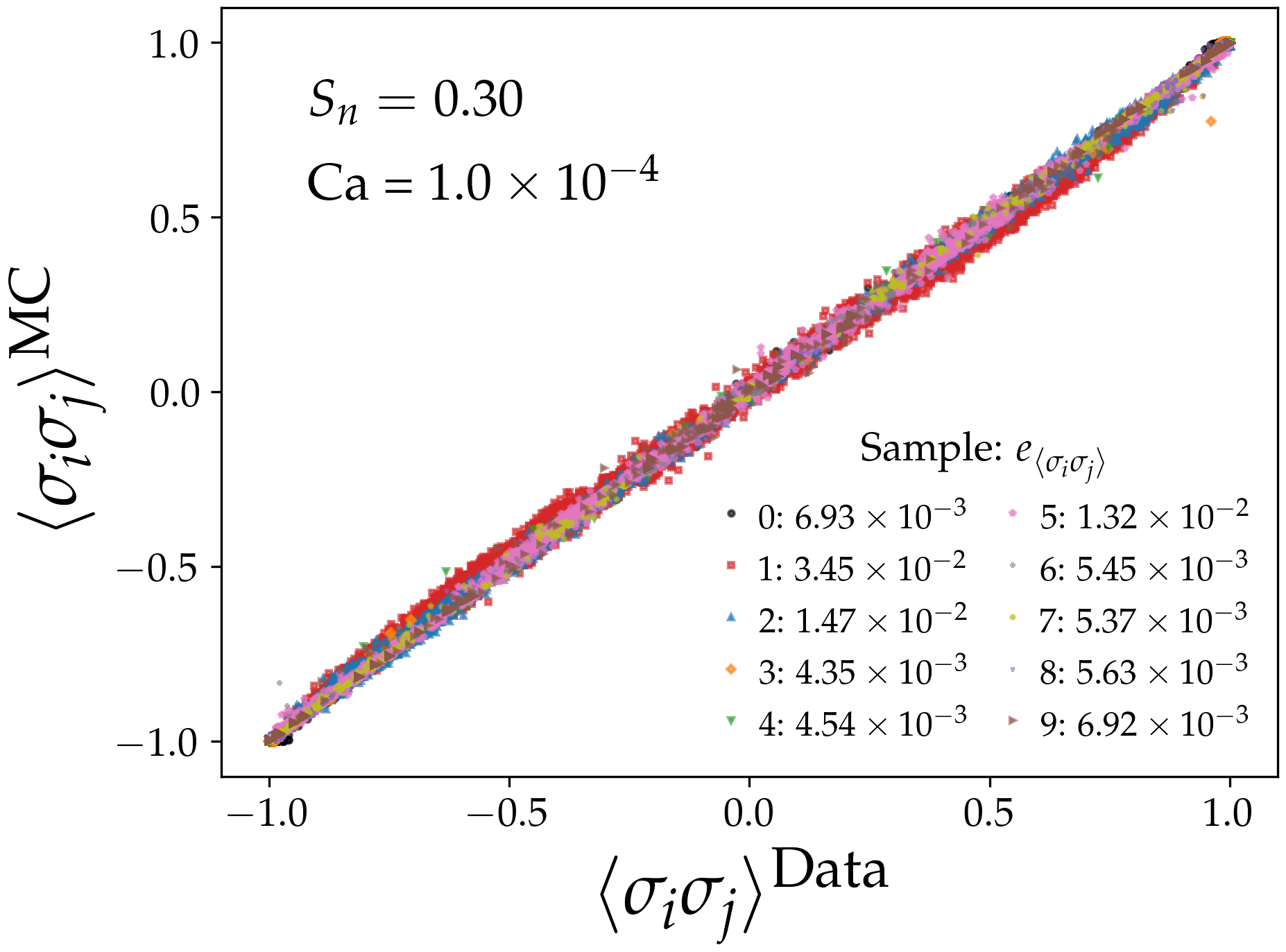}\hfill
        \includegraphics[width=0.25\textwidth]{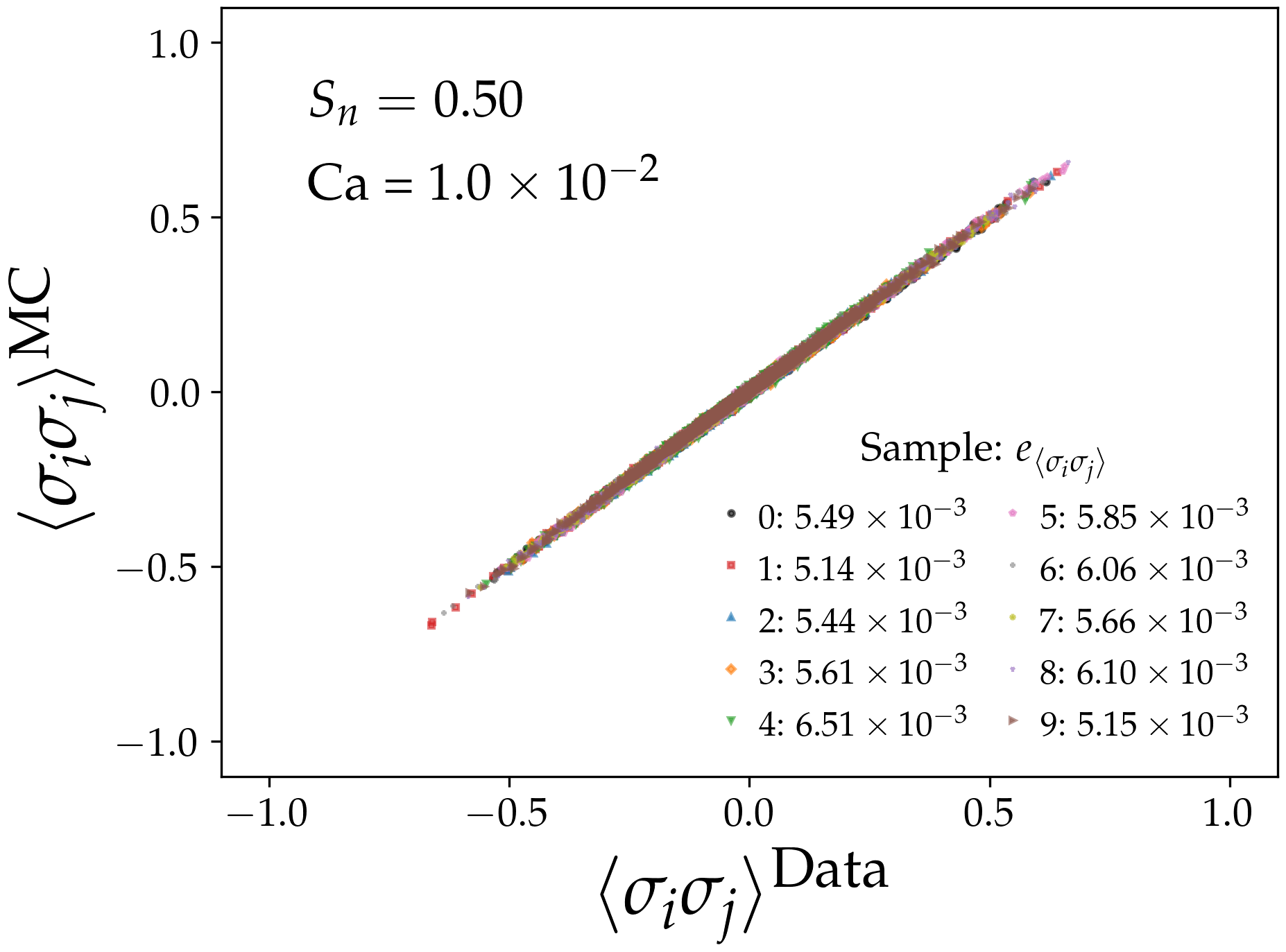}\hfill
        \includegraphics[width=0.25\textwidth]{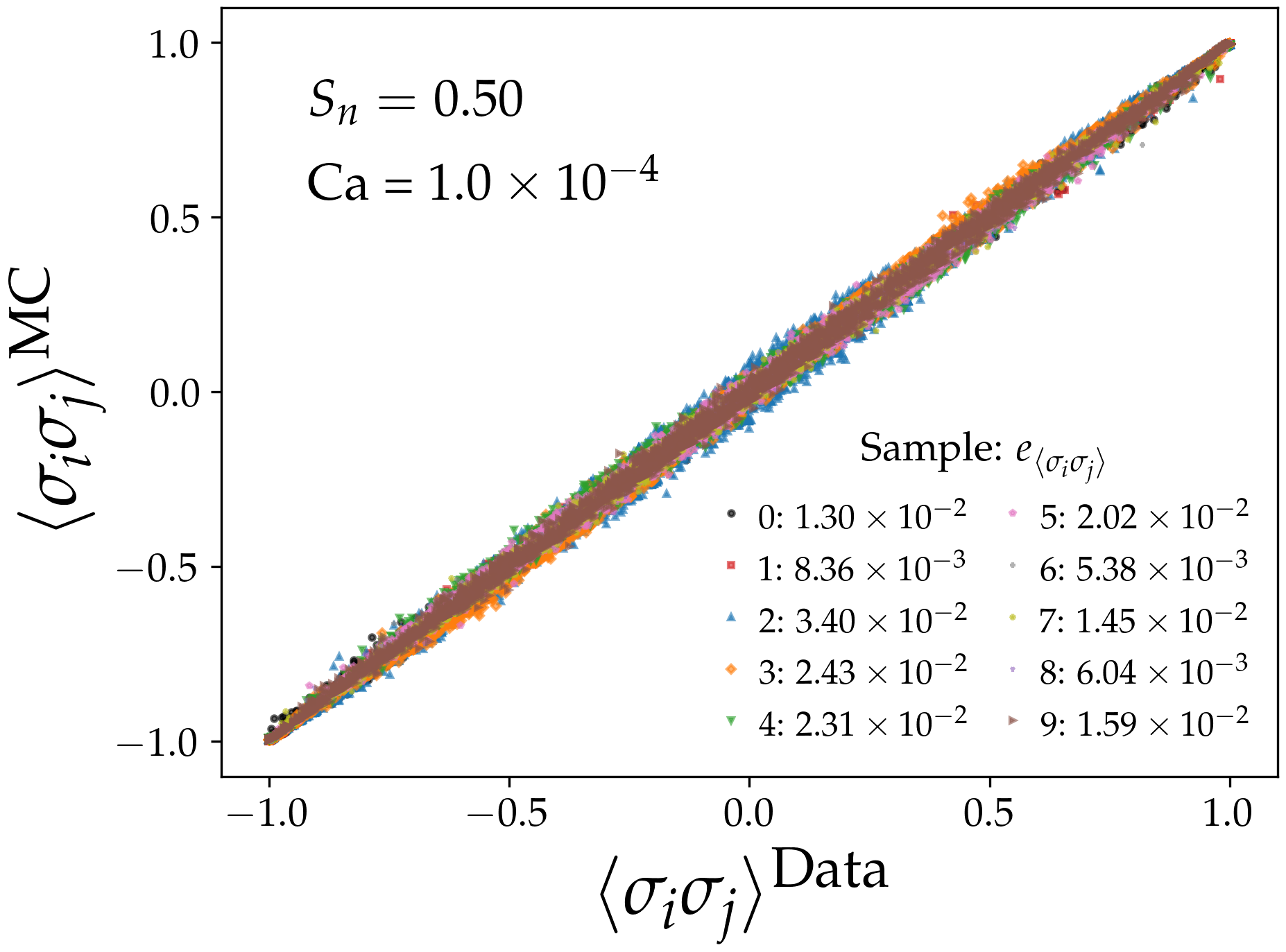}\hfill
    }
    \caption{\label{fig_Sij} Plots of $\langle\sigma_i\rangle^{\rm Data}$ vs $\langle\sigma_i\rangle^{\rm MC}$ (top row), and $\langle\sigma_i\sigma_j\rangle^{\rm Data}$ vs $\langle\sigma_i\sigma_j\rangle^{\rm MC}$ (bottom row), showing the comparison between DPN simulation data and BML computations. We show data for two saturations $S_n=0.3$ and $0.5$, with two capillary numbers, ${\rm Ca} = 1.0\times 10^{-2}$ and $1.0\times 10^{-4}$ for each $S_n$. Each plot contains results for $10$ different network samples indicated by different symbols. The legends show the total error ($e$) between the DPN data and BML computations for different samples.}
\end{figure*}
\begin{figure*}
    \centerline{\hfill
         \includegraphics[width=0.25\textwidth]{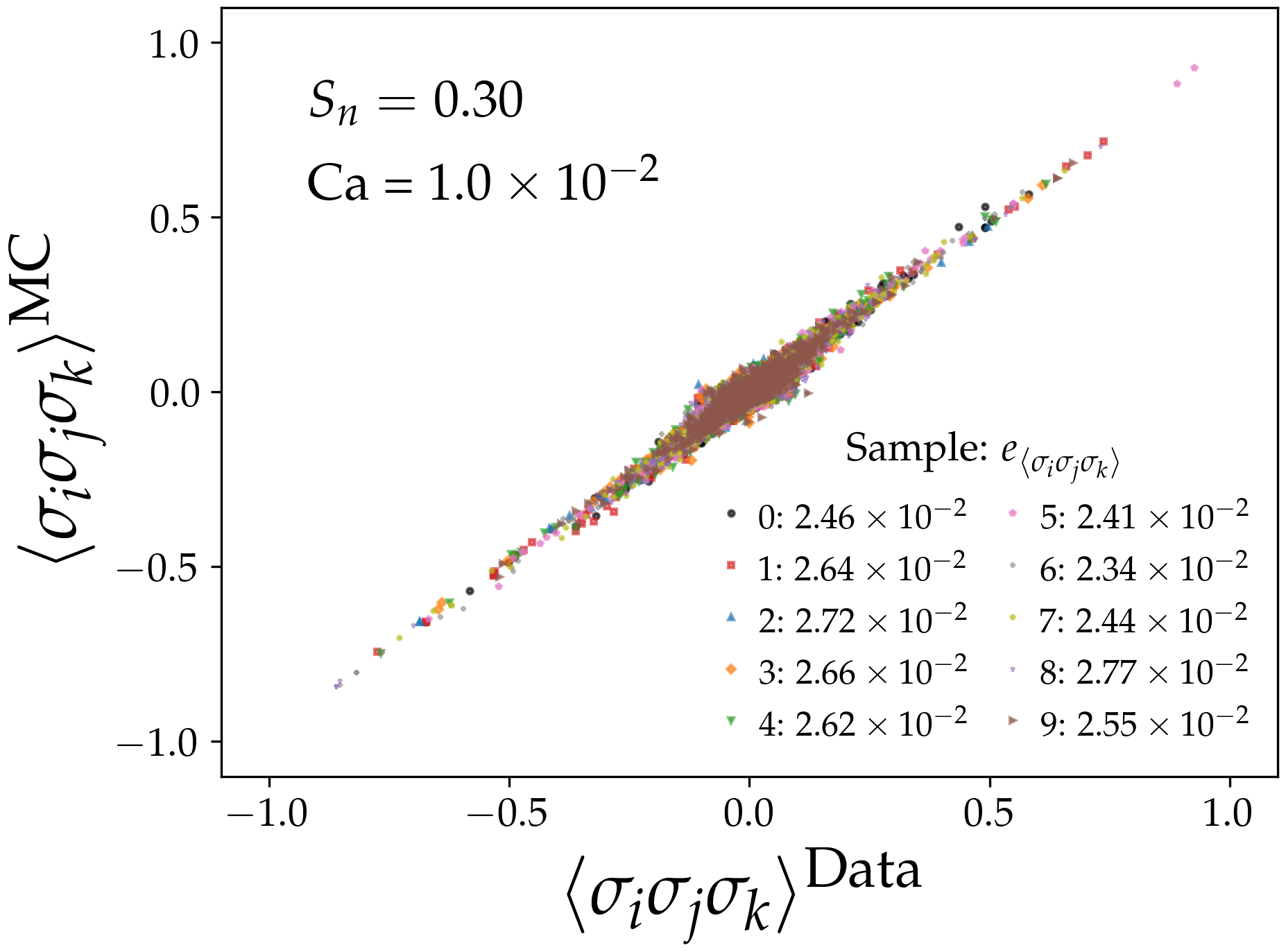}\hfill
         \includegraphics[width=0.25\textwidth]{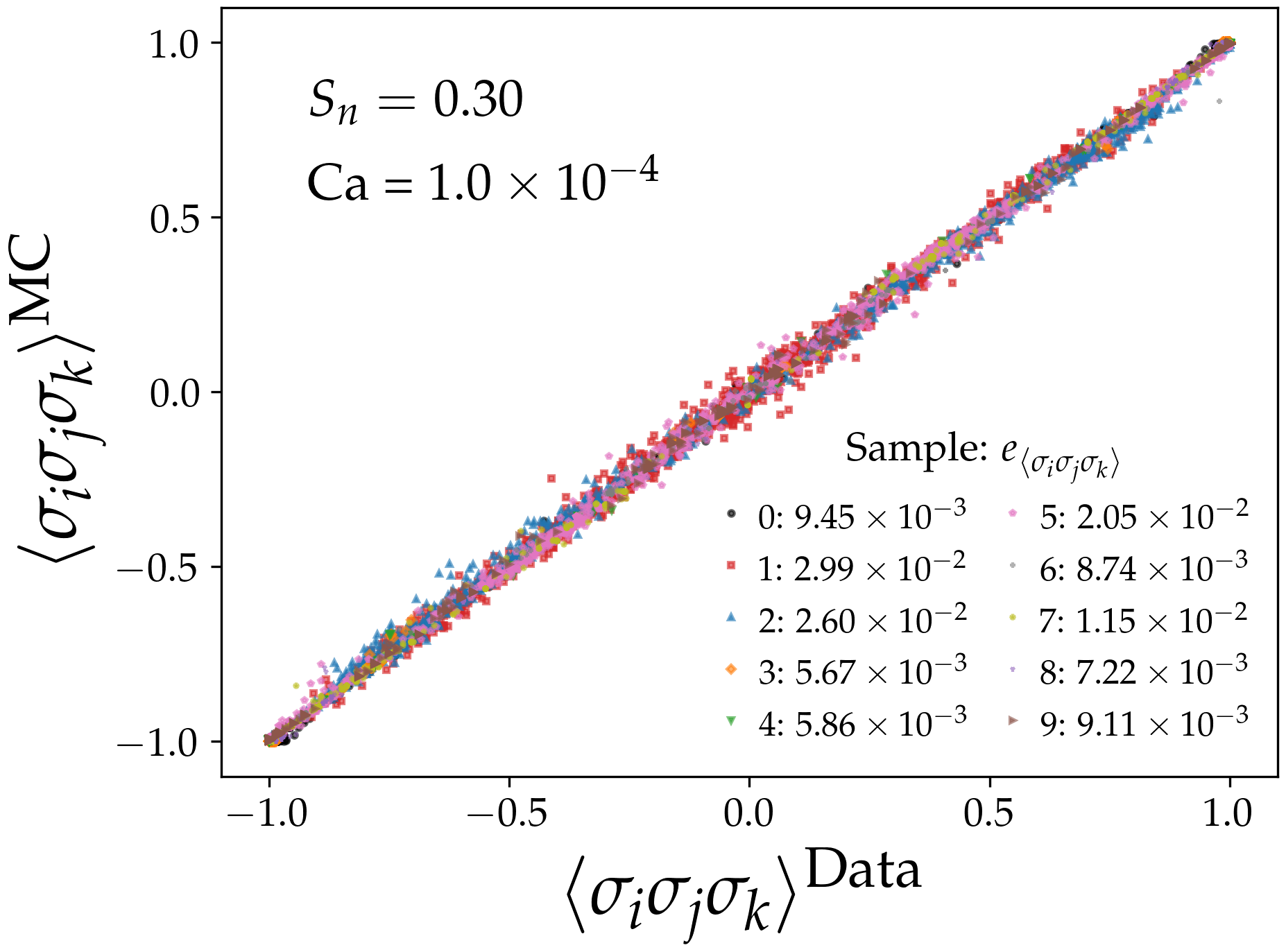}\hfill
         \includegraphics[width=0.25\textwidth]{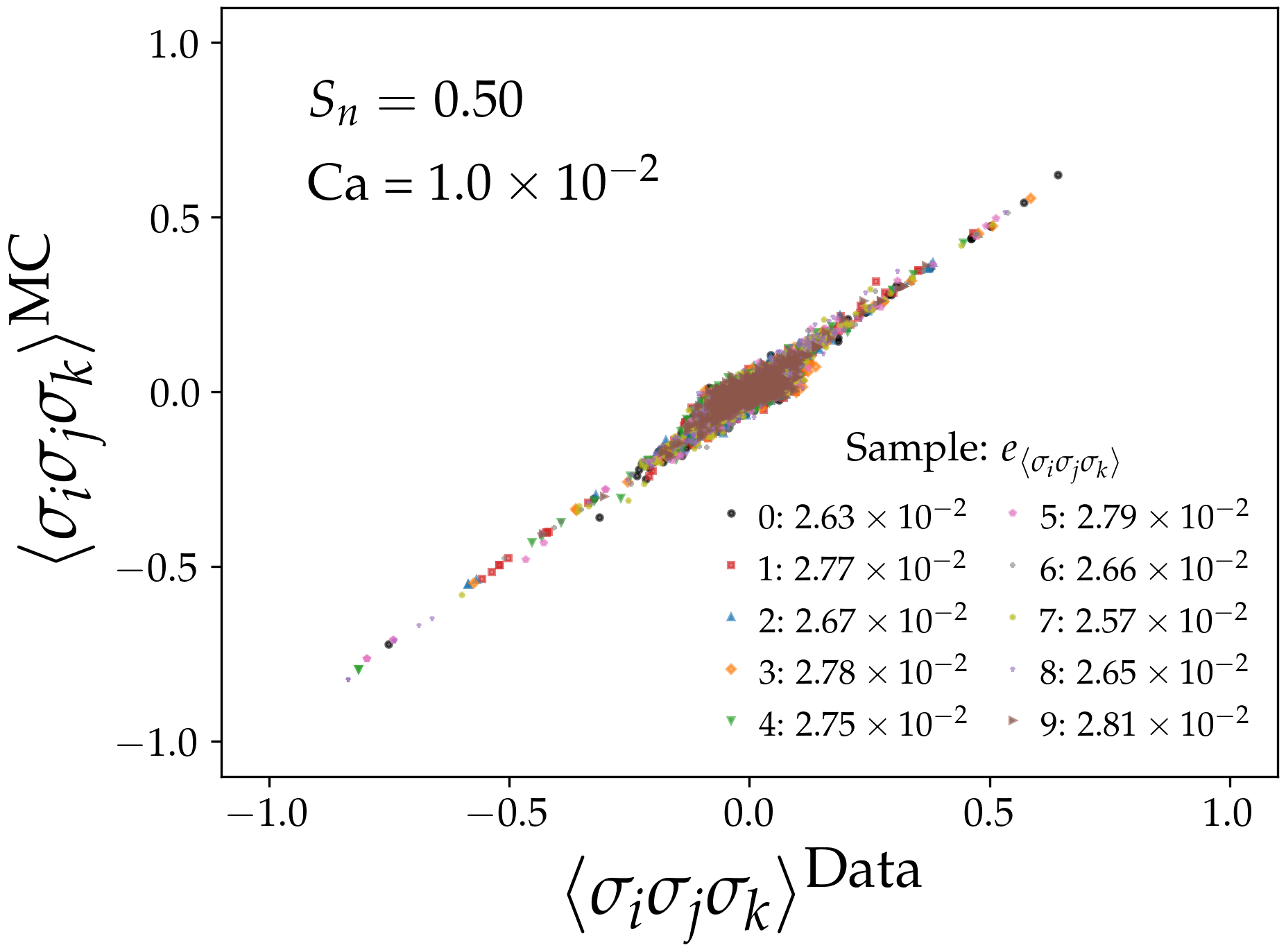}\hfill
         \includegraphics[width=0.25\textwidth]{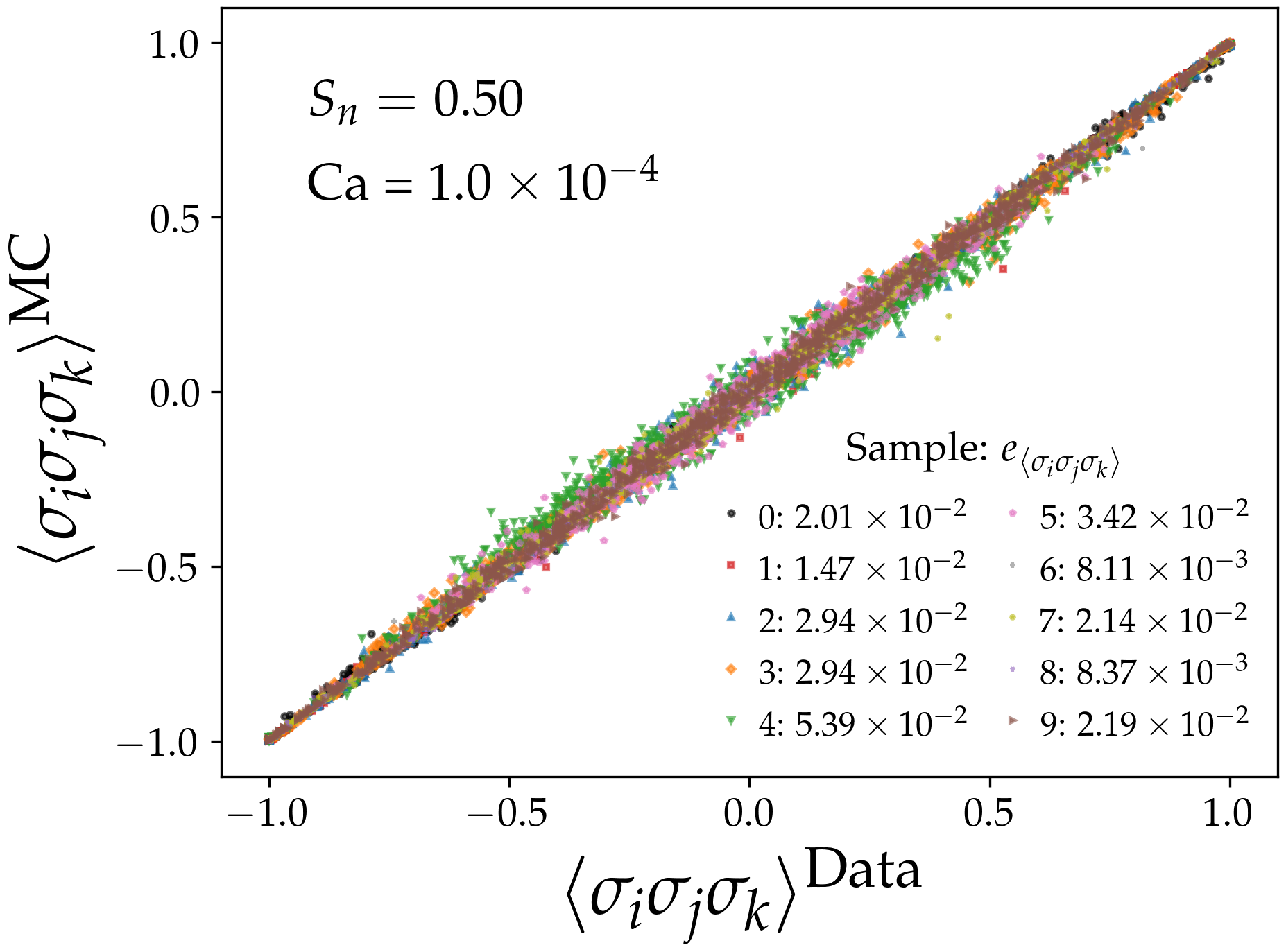}\hfill
    }
    \caption{\label{fig_Sijk}Plots of 3-point correlations $\langle \sigma_i\sigma_j\sigma_k \rangle$ compared between the DPN data and MC samples. We show results for two saturations $S_n=0.3$ and $0.5$, and two capillary numbers ${\rm Ca}=1.0\times 10^{-2}$ and $1.0\times 10^{-4}$ for each $S_n$. Each plot contains results for $10$ different network samples where we plotted data only for $1000$ randomly chosen unique triples \{$ijk$\} for each sample. The legends show the total error ($e$) between the DPN data and BML computations for different samples. We have also repeated the analysis with $160$ independent random sets of $1000$ triplets and found the median error across all sets is $<4\%$ for most of the network samples, ensuring the error distribution is stable under independent random resampling of the triplets.}
\end{figure*}

The learning rate $\eta$ determines the rate of update during the gradient descent method, which is a crucial parameter for the optimal performance of the algorithm. This affects the speed and stability of the convergence. A too high or too low learning rate can result in overshoot or very slow convergence, making the simulations impossible to reach solutions with good accuracy. A general practice is to start with a relatively large value of $\eta$, and then reduce it gradually using a learning rate scheduler. There are different types of schedulers which work best for different types of problems. By trying a few different schedulers, we found that the scheduler {\it ReduceLROnPlateau\/} \cite{p12} works best for this study. This method monitors the loss ($E$) during the iteration, and reduces $\eta$ by a factor when $E$ stops improving for a specific number of iterations. The loss $E$ is defined as, 
\begin{equation}
    \begin{aligned}
        \label{eq_loss}
        \displaystyle
        E = & \frac{1}{N}\sum_i \left(\langle\sigma_i\rangle^{\rm Data} - \langle\sigma_i\rangle_\tau^{\rm MC} \right)^2 \\
            & + \frac{1}{N(N-1)}\sum_i\sum_{j<i} \left(\langle \sigma_i\sigma_j\rangle^{\rm Data} - \langle\sigma_i\sigma_j\rangle_\tau^{\rm MC} \right)^2\;,
    \end{aligned}
\end{equation} 
which calculates the mean squared difference of the correlations and the averages between the data and MC.   

\begin{figure*}
    \centerline{\hfill
         \includegraphics[width=0.2\textwidth]{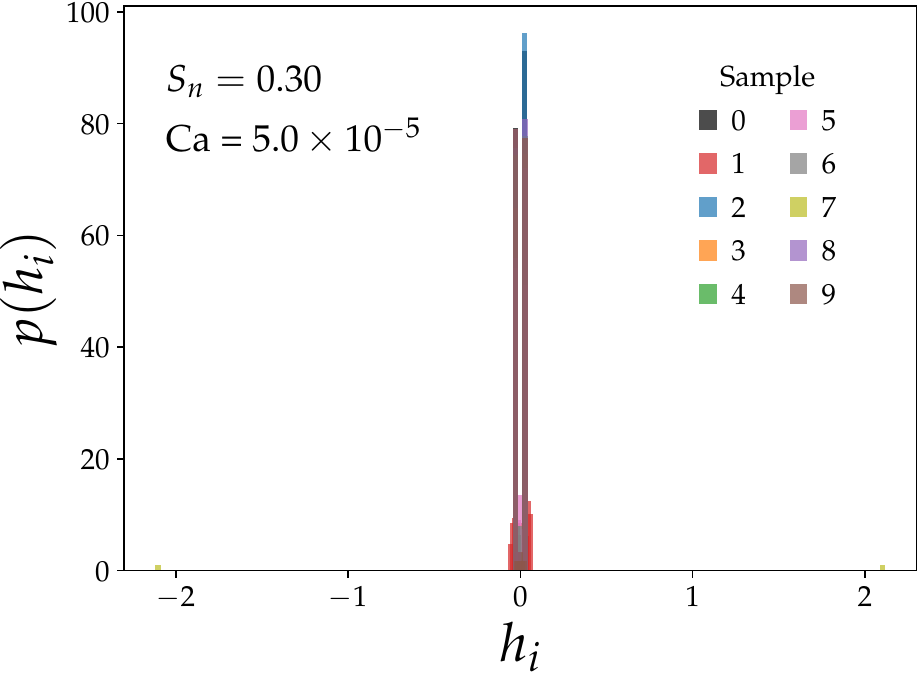}\hfill
         \includegraphics[width=0.2\textwidth]{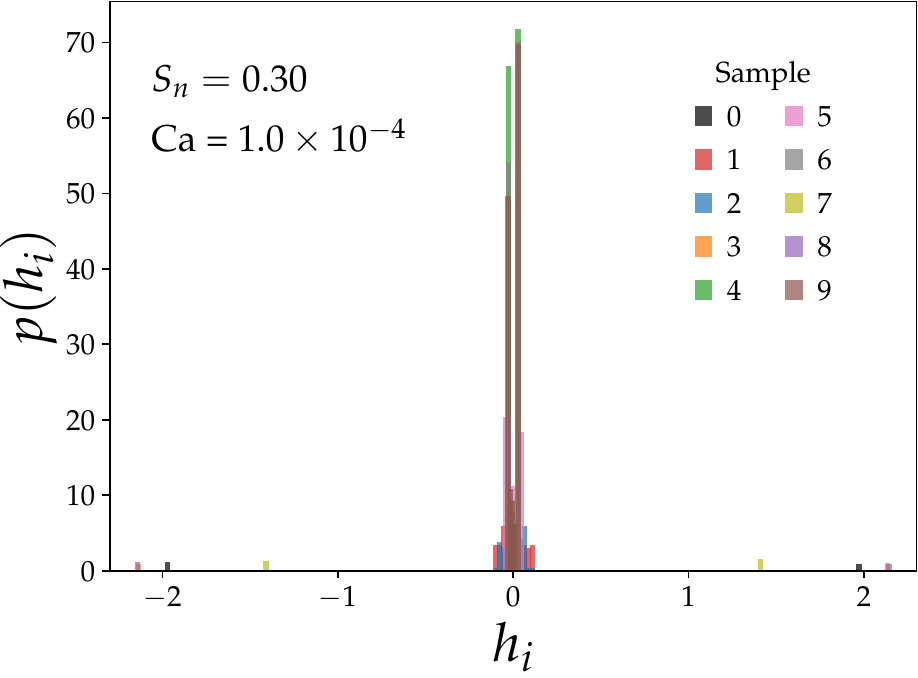}\hfill
         \includegraphics[width=0.2\textwidth]{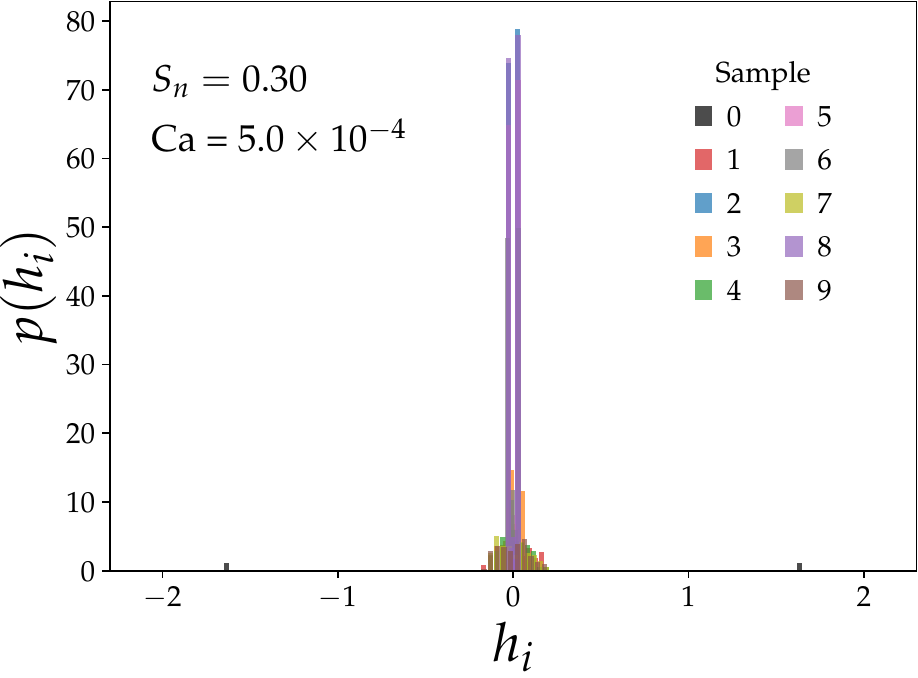}\hfill
         \includegraphics[width=0.2\textwidth]{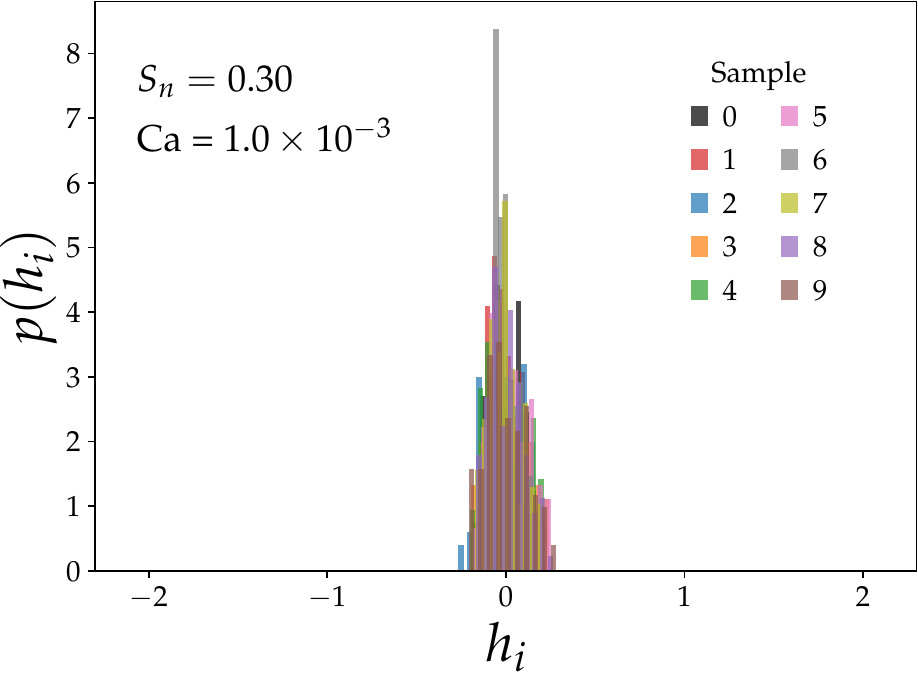}\hfill
         \includegraphics[width=0.2\textwidth]{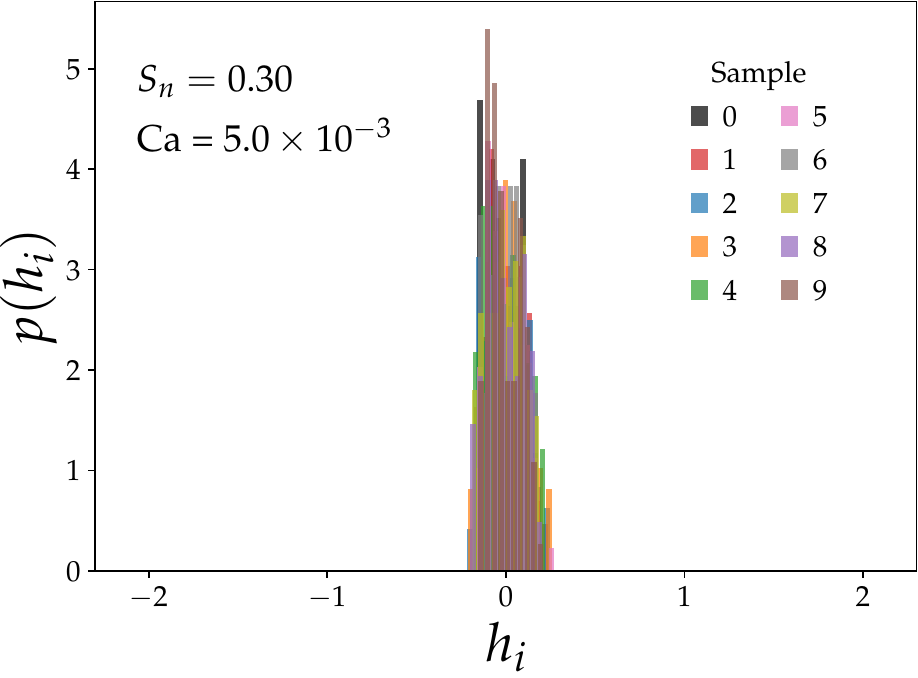}\hfill
    }
    \caption{\label{fig_hi}Histograms of local field constants $h_i$ for $S_n=0.3$ at different capillary numbers. Each plot shows data for $10$ different network samples.}
\end{figure*}

\begin{figure*}
    \centerline{\hfill
         \includegraphics[width=0.2\textwidth]{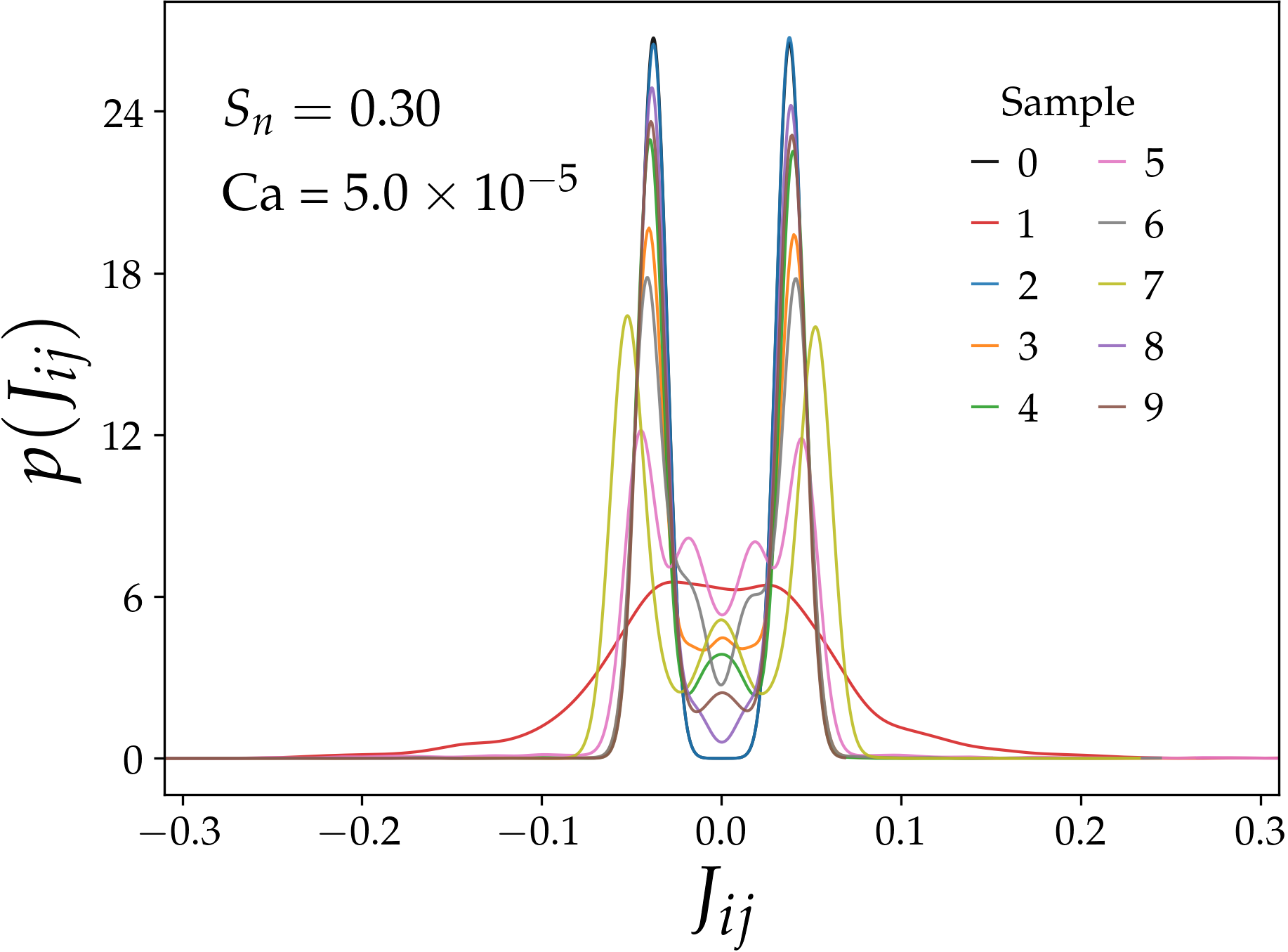}\hfill
         \includegraphics[width=0.2\textwidth]{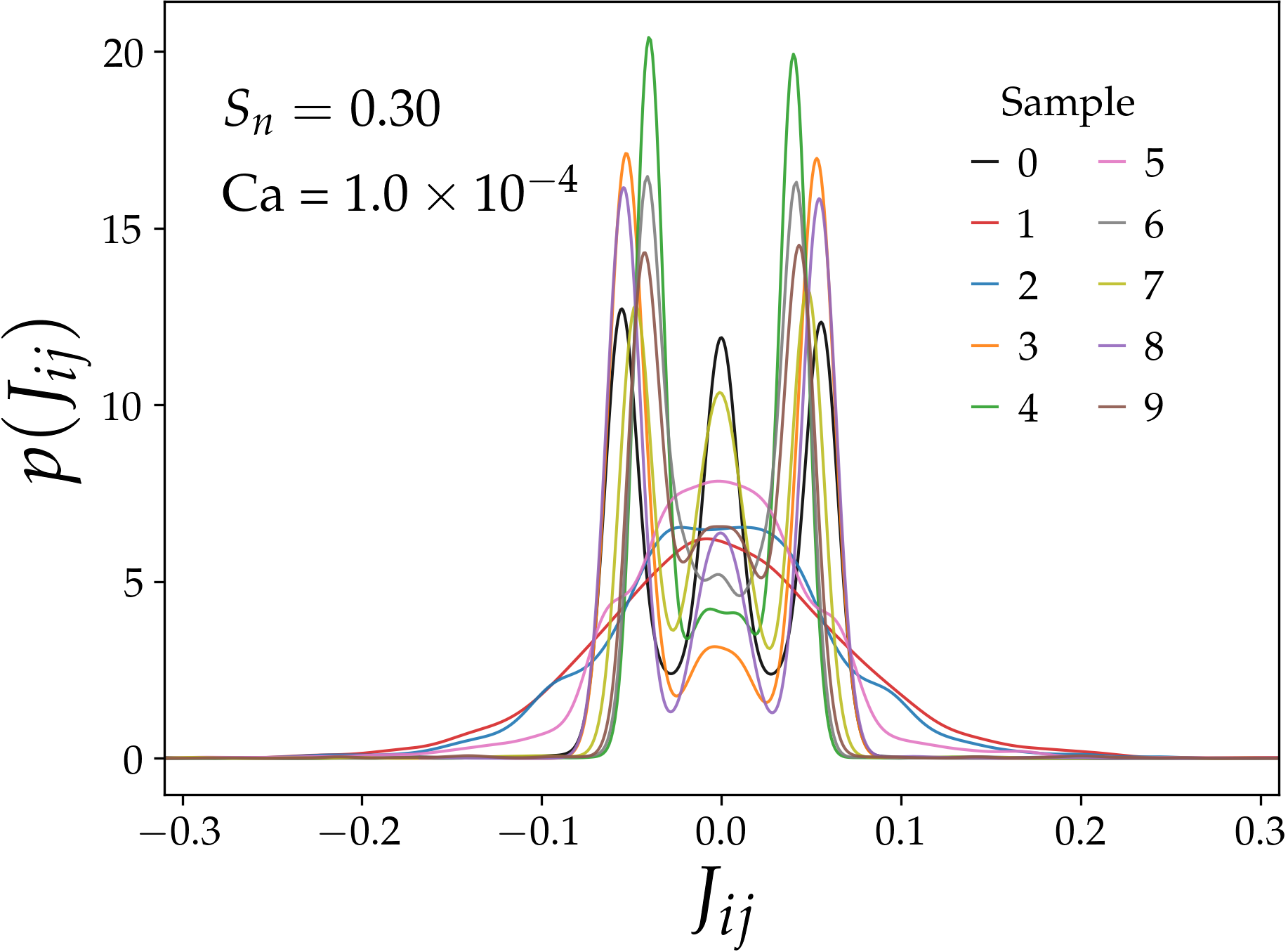}\hfill
         \includegraphics[width=0.2\textwidth]{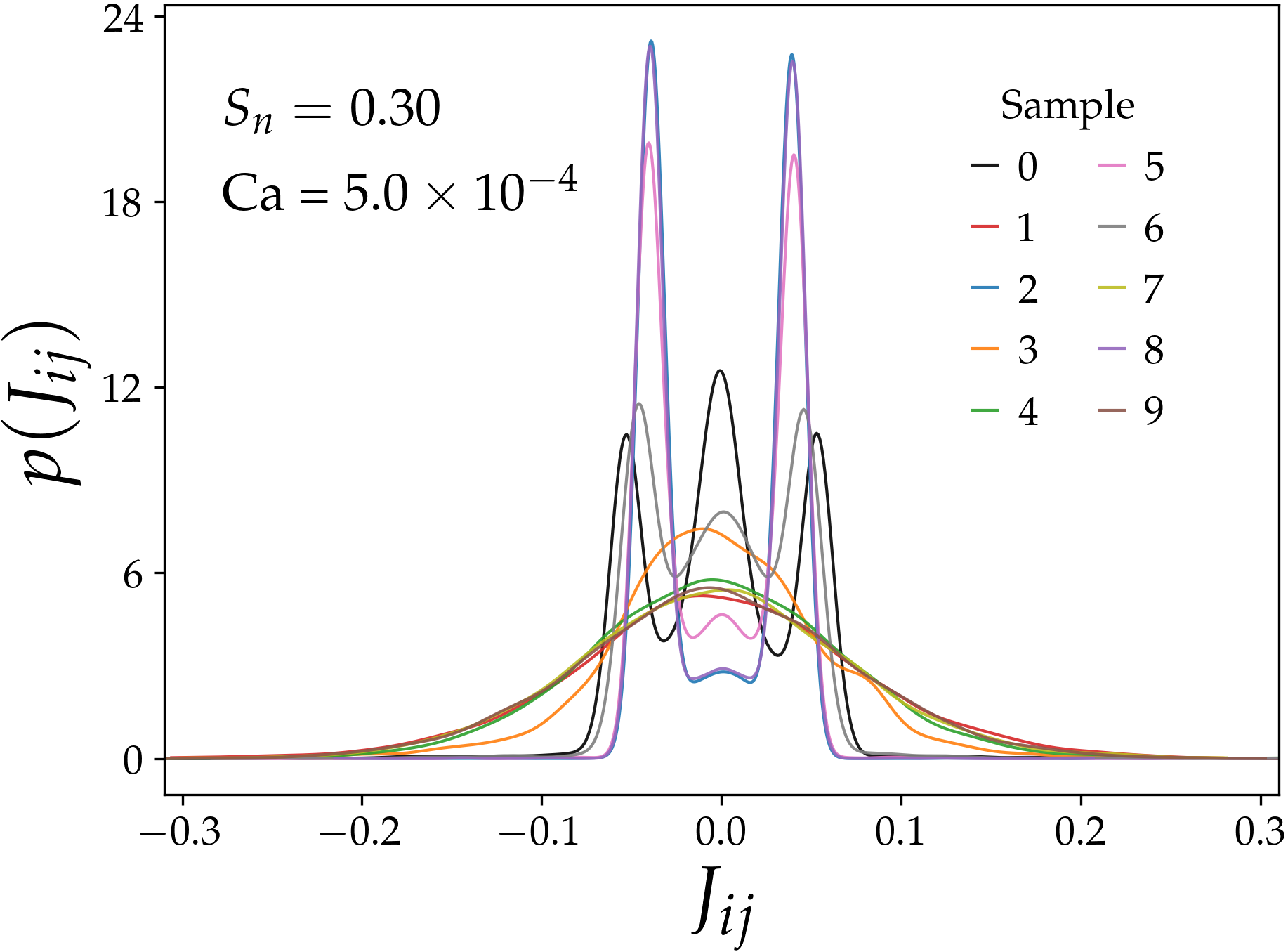}\hfill
         \includegraphics[width=0.2\textwidth]{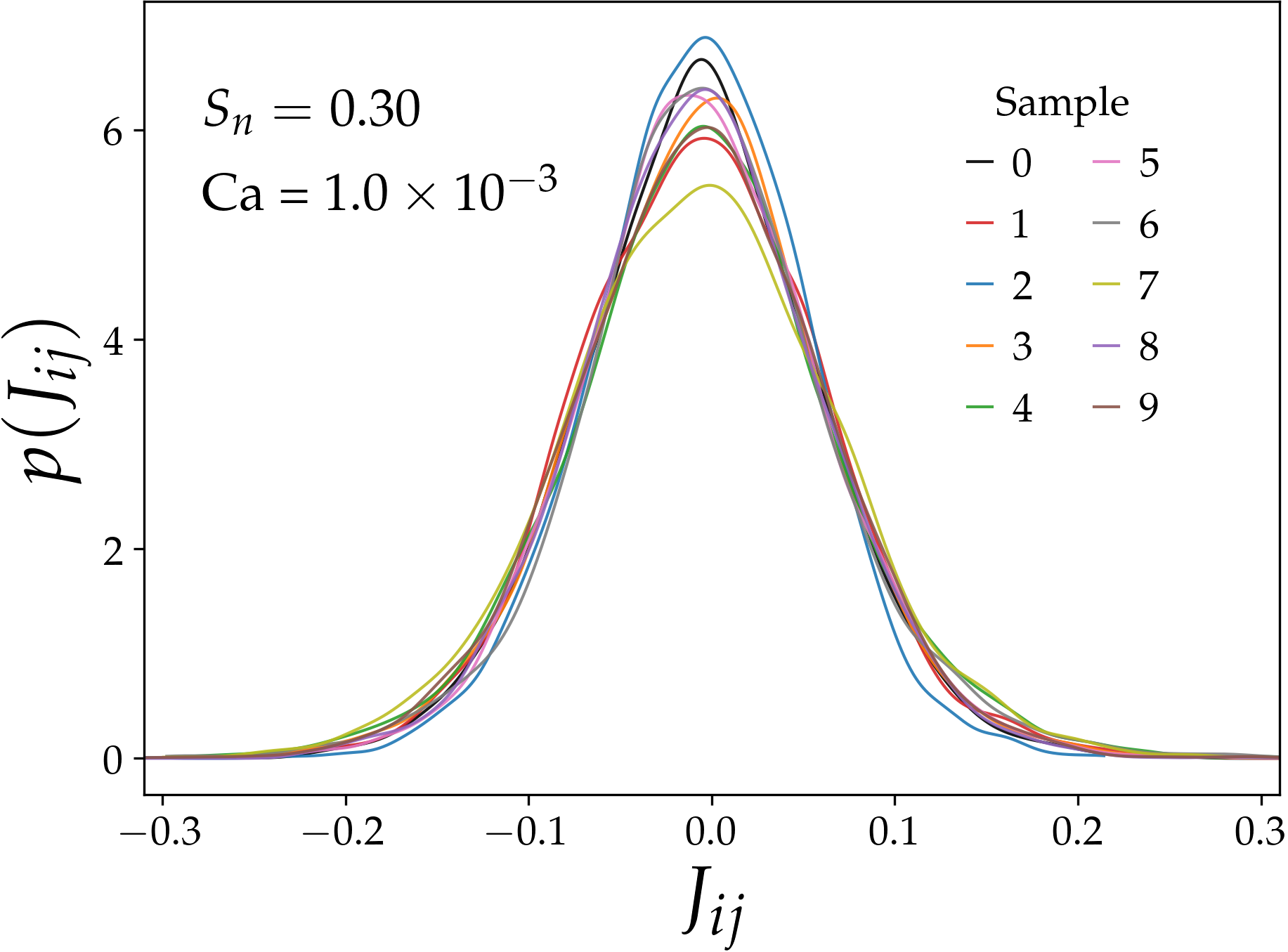}\hfill
         \includegraphics[width=0.2\textwidth]{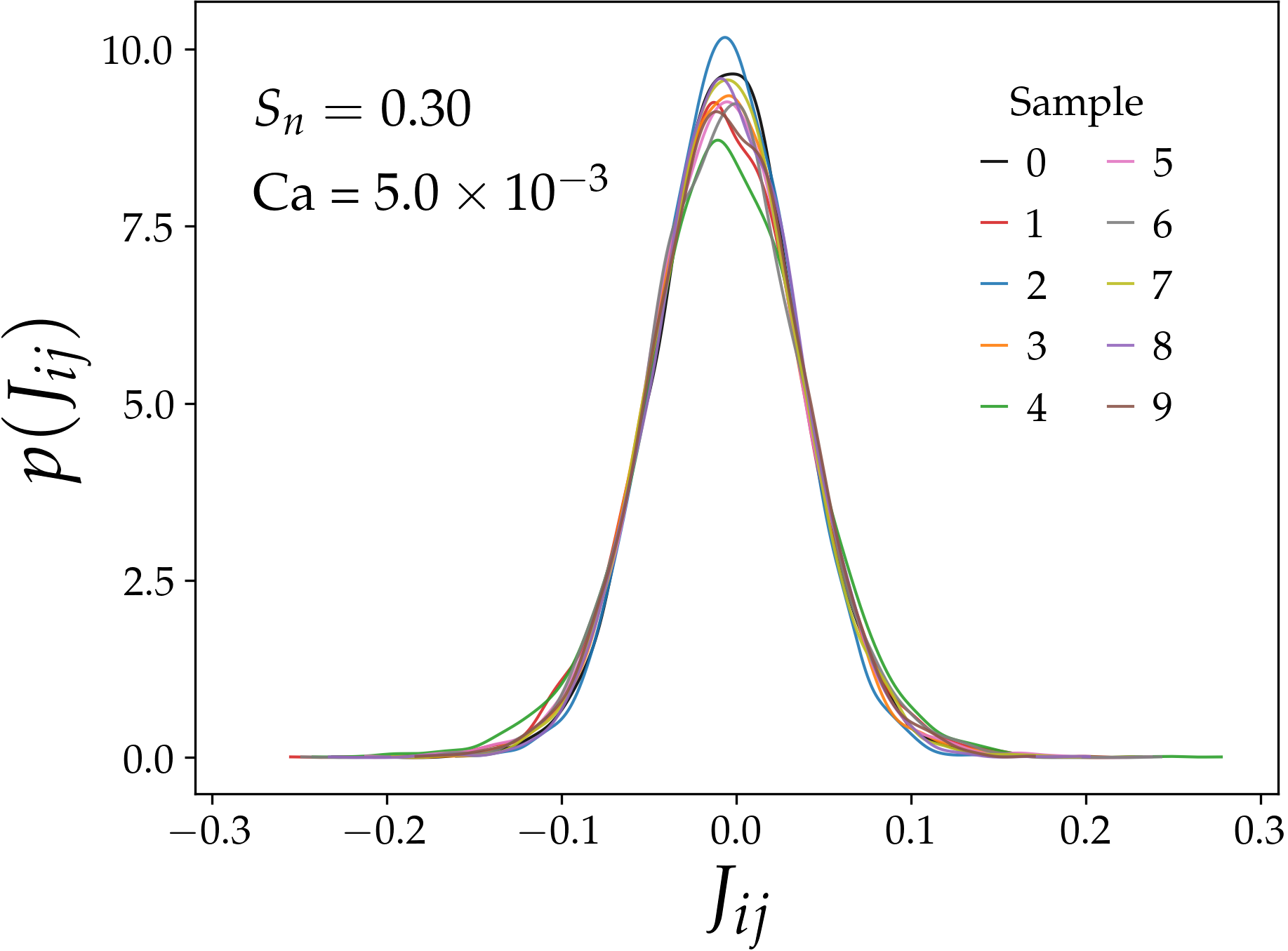}\hfill
    }
    \caption{\label{fig_Jij} Plots of the distributions of the coupling constant $J_{ij}$ for $S_n = 0.30$. Here we show results for $5$ capillary numbers. Each plot contains results for $10$ different network samples. Apparently, all the distributions have a mean around zero, however at low $\rm Ca$ most samples show multiple peaks, indicating groups of highly correlated or anti-correlated couplings. There is also dominant sample-to-sample variations at low Ca. At high Ca, the coupling constants are spread around zero with a Gaussian-like distribution for all the samples.}
\end{figure*}

We performed BML computations using data sets obtained from DPN simulations for different values of $S_n$ and Ca. The values of $\langle\sigma_i\rangle^{\rm MC}$ and $\langle\sigma_i\sigma_j\rangle^{\rm MC}$ are calculated with Metropolis Monte Carlo simulation with the probability distribution function given in Eq. \ref{eq_P}. This requires measuring of $N(N-1)/2$ values for the correlations for each MC sample at every BML step which is computationally expensive. Our DPN simulations were performed with a $40\times 40$ network which would lead to $1279200$ values of couplings. Furthermore, a large number of MC samplings is necessary in the spin-glass phase, as the configurations can get stuck in a local minima for a long time. We also have a large parameter space (the complete detail of the parameter sets and the size of the data sets are provided in the next section). To reduce the computational time that is feasible to available resources, we coarse grained the system to a bi-periodic $10\times 10$ network for the BML computations. This leads to $N=100$ spins and $4950$ values of couplings. We used $2\times 10^5$ MC sweeps for MC thermalization and $10^5$ MC configurations for averaging in each BML iteration. We performed a minimum of $10^4$ BML iterations for each case and then terminated the iterations when the loss $E$ in Eq. \ref{eq_loss} reached a small enough value. The final values of $h_i$ and $J_{ij}$ are then considered at the end of the BML iterations.

To verify how the Hamiltonian in Eq. \ref{eq_H} and the corresponding configurational probability $P(\{\sigma\})$ in Eq. \ref{eq_P} reproduces two-phase flow data obtained from the DPN model, we show in Fig. \ref{fig_matSij} a visual comparison between the DPN data and the BML computations with the matrix plots of $\langle\sigma_i\sigma_j\rangle$ for two cases, together with their distributions. We show for two values of Ca, and for both cases the results from BML match with the DPN data. We furthermore note that the correlations are very different in high and low Ca. At high Ca, the correlations indicate a continuous variation around zero, whereas at the low $\rm Ca$ wee see clustering of high positive or negative values. This indicates that the flow in different pores are highly correlated or anti-correlated at low Ca.

A more quantitative comparison between the DPN data and BML results are provided in Fig. \ref{fig_Sij}, where we plot $\langle\sigma_i\rangle$ and $\langle \sigma_i\sigma_j \rangle$ between data and MC for all values of $i$ and $j$. We show results for two different saturations $S_n=0.3$ and $0.5$, each for two values of $\rm Ca$. Each plot contains data points for $10$ different samples of the network. The data points follow the diagonal line with minimal deviations. This indicates that configurational probability $P(\{\sigma\})$, Eq. \ref{eq_P}, agrees well with DPN data.

The probability distribution $P(\{\sigma\})$ in Eq. \ref{eq_P} was obtained with the constraints up to two point correlations. In order to therefore verify whether the model can predict higher order correlations, we plotted in Fig. \ref{fig_Sijk} the three point correlations $\langle\sigma_i\sigma_j\sigma_k\rangle$ obtained from the DPN data and from MC, and the plots show that the model demonstrate the data with minimum deviations. The agreement may possible to improve further by longer BML simulations with higher number of MC sweeps.

The values of local fields $h_i$ and the coupling constants $J_{ij}$ obtained from the BML computations with the DPN data are plotted in Figs. \ref{fig_hi} and \ref{fig_Jij}. Each plot contains data from $10$ different network samples. The histograms of $h_i$ indicate that the values are mostly spread around zero, though at low $\rm Ca$ there are a few spins with very high positive or negative values of $h_i$. Such spins do not appear at high Ca, however the spread of $h_i$ widens.

\begin{figure*}
    \hrule
    \begin{minipage}{0.49\textwidth}
        \centering
        \vspace{0.4em}
        (a) ${\rm Ca} = 2\times 10^{-1}$\\
        \vspace{0.4em}    
        \hrule
        \vspace{0.2em}
        \begin{minipage}{0.33\textwidth}
            \centering
            \includegraphics[height=0.86\textwidth]{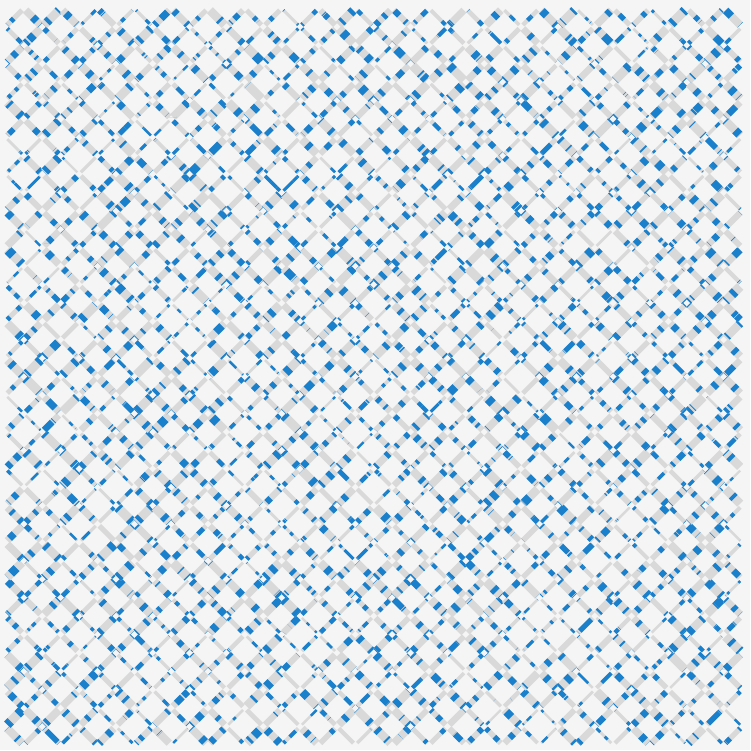}\\i
        \end{minipage}%
        \begin{minipage}{0.33\textwidth}
            \centering
            \includegraphics[height=0.86\textwidth]{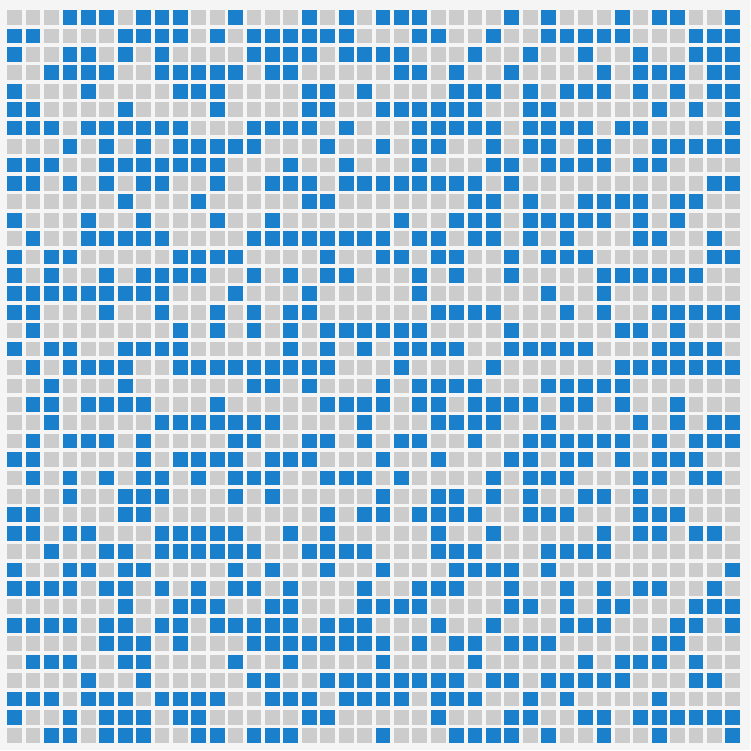}\\ii
        \end{minipage}%
        \begin{minipage}{0.33\textwidth}
            \centering
            \includegraphics[height=0.86\textwidth]{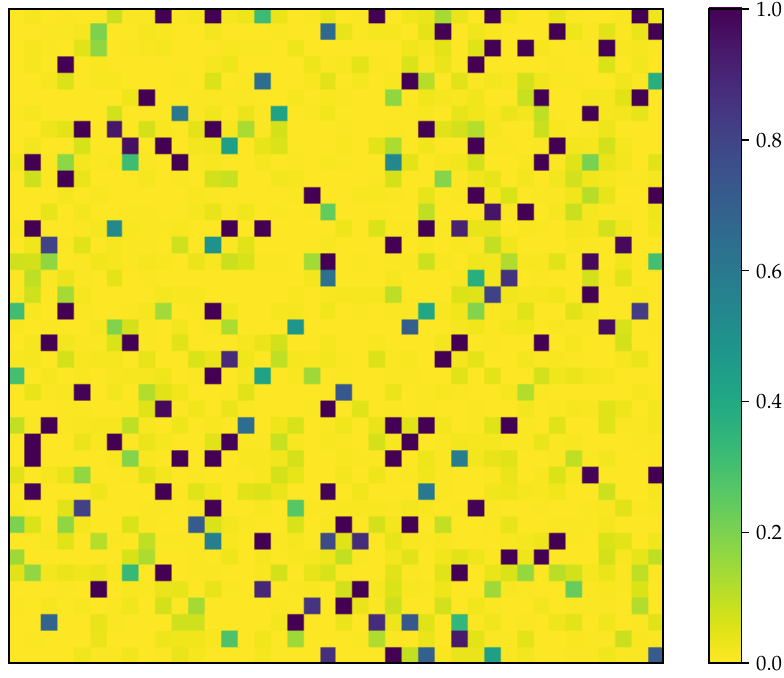}\\iii
        \end{minipage}\\
        \vspace{0.2em}
    \end{minipage}%
    \hfill\vline\hfill    
    \begin{minipage}{0.49\textwidth}
        \centering
        \vspace{0.4em}
        (b) ${\rm Ca} = 5\times 10^{-4}$\\
        \vspace{0.4em}
        \hrule
        \vspace{0.2em}
        \begin{minipage}{0.33\textwidth}
            \centering
            \includegraphics[height=0.86\textwidth]{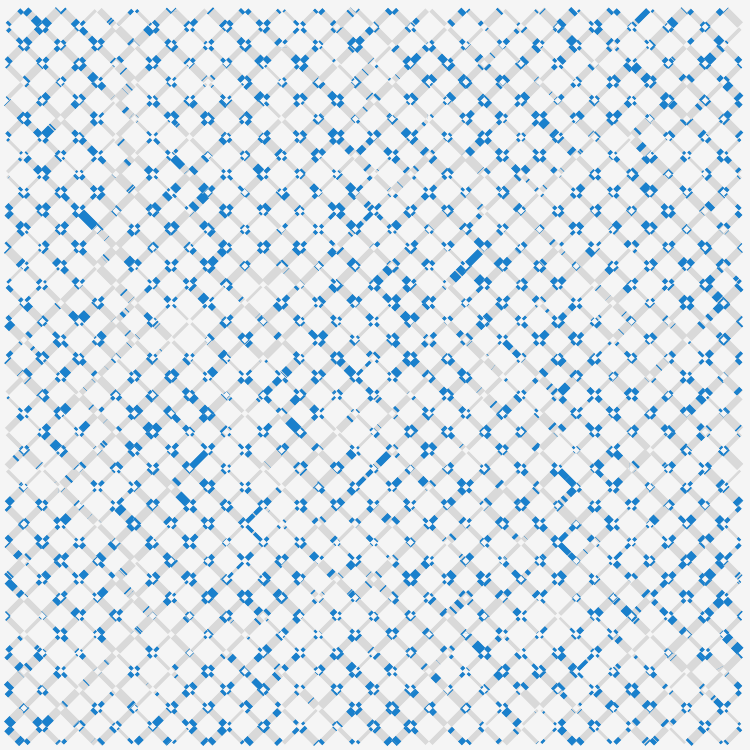}\\i
        \end{minipage}%
        \begin{minipage}{0.33\textwidth}
            \centering
            \includegraphics[height=0.86\textwidth]{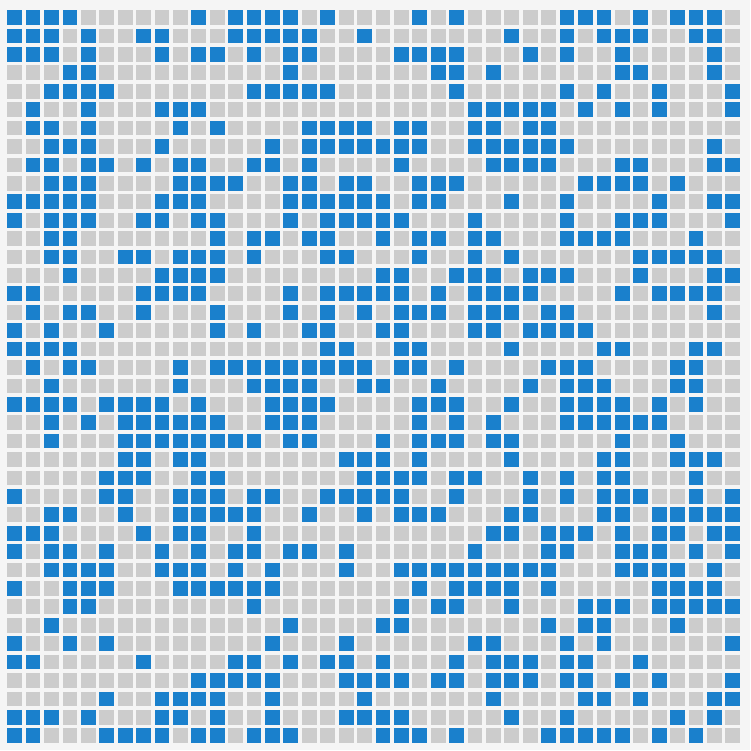}\\ii
        \end{minipage}%
        \begin{minipage}{0.33\textwidth}
            \centering
            \includegraphics[height=0.86\textwidth]{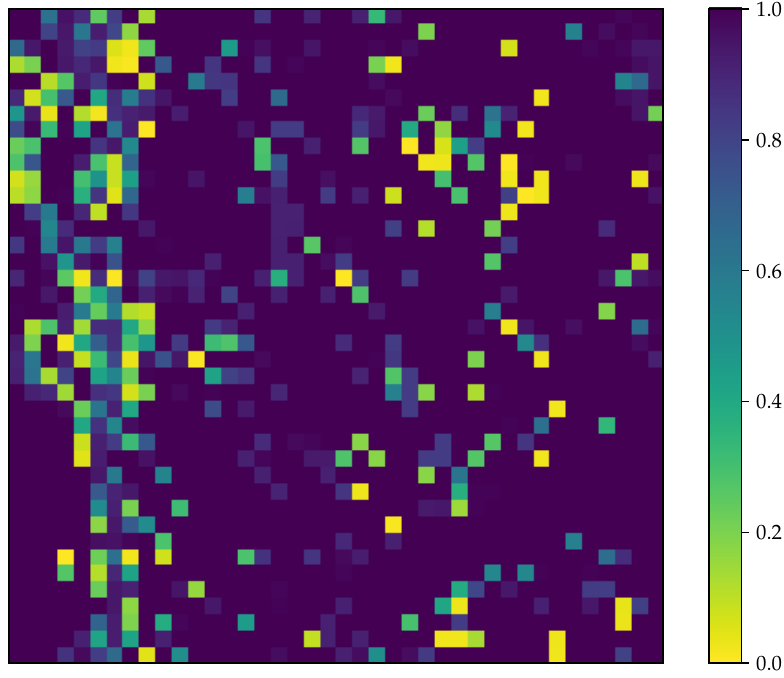}\\iii
        \end{minipage}\\
        \vspace{0.2em}  
    \end{minipage}
    \rule{\textwidth}{0.5pt}
    \caption{\label{fig_snap} Snapshots from dynamic pore network simulations illustrating the differences between steady-state flow that give rise to paramagnetic and spin-glass phases in spin model. The simulations were performed with a network of $40\times 40=1600$ links, and a global non-wetting saturation of $S_n = 0.3$. We show two cases corresponding to the capillary numbers (a) ${\rm Ca} = 2\times 10^{-1}$ and (b) ${\rm Ca} = 5\times 10^{-4}$ for the same network. There are $3$ figures for each case. Snapshots in (i) show the positions of wetting (gray) and non-wetting (blue) fluids in the network at a certain time step in steady state. The corresponding spin configurations, obtained using Eq. \ref{eq_spindef} are shown in (ii) where the blue and gray dots represent $\sigma_i = +1$ and $-1$ respectively. The average $\sum_i \sigma_i/N$ over different time steps in the steady state returns low values for both capillary numbers, resulting the single-sample magnetization $\approx 0.026$ and $\approx 0.057$ for (a) and (b) respectively. However, a noticeable difference is observed in the snapshots in (iii) which show $\langle\sigma_i\rangle^2$, highlighting local frozen spins at the low capillary number. This leads to a high value of single-sample Edwards-Anderson order parameter $\approx 0.884$ for (b), whereas a low $\approx 0.088$ for (a).}
\end{figure*}

The distributions $p(J_{ij})$ shown in Fig. \ref{fig_Jij} reveal a nontrivial form. The distributions for all the sets show the coupling constants are spread around zero with both positive and negative values, indicating a spin-glass type system. However, at low Ca, the $p(J_{ij})$ shows bimodal type distributions with two peaks for most of the samples, indicating there are two sets of correlated and anti-correlated pores. While increasing Ca, peaks at $J_{ij}=0$ start appearing, and at high $\rm Ca$ all the distributions lead to a unimodal distribution.

\section{The glass transition and glassy flow state}
\label{sec_glass}

The paramagnetic (P) and ferromagnetic (F) phases in a spin system are characterized by zero and non-zero values of the total average magnetization ($m$) respectively, implying that the spin alignments to be random or in a specific direction. The characterization of a spin-glass (SG) phase on the other hand is subtler. There is no global alignment of spins is a specific direction similar to a paramagnetic phase, but the spins are now locally oriented. This is caused by the presence of both ferromagnetic ($J_{ij}>0$) and antiferromagnetic ($J_{ij}<0$) couplings (though it is not a sufficient condition) in the system, resulting in an inability to satisfy the coupling conditions for every spin pairs. This leads to {\it frustration\/} and rugged energy landscapes \cite{binder1986spin}, with the appearance of disorderly frozen spin patterns making the spins locally magnetized.

Using data obtained from DPN simulations, we illustrate in Fig. \ref{fig_snap} the existence of a spin-glass phase at low capillary number (${\rm Ca} = 0.0005$), whereas such a phase is not present at high capillary number (${\rm Ca} = 0.2$). The global saturation of the non-wetting fluid is $S_n=0.3$ here. Fig. \ref{fig_snap} (i) and (ii) show the bubble distributions and the respective spin mappings (Eq. \ref{eq_spindef}), whereas Fig. \ref{fig_snap} (iii) shows the degree of local magnetization $\langle \sigma_i \rangle^2$. The applied global pressure drop is in the vertical directions in these figures. The spin representations for these two low and high capillary numbers show similar random spin orientations giving low values of total magnetization for this sample. However the degree of local magnetization $\langle\sigma_i\rangle^2$ is significantly different. At high Ca, $\langle\sigma_i\rangle^2$ shows  uniformly zero values across the system for majority of spins, whereas at low $\rm Ca$ the majority of the spins have very high degrees of local magnetization. Furthermore, patches of small value of $\langle\sigma_i\rangle^2$ can be observed at low $\rm Ca$ (b), which indicates the dominance of continuous pathway flow. These maps therefore seem to capture the different pore-scale flow patterns similar to what was described by Avraam and Payatakes \cite{avraam1995flow}. We will discuss this further later on.

\begin{figure*}
    \centerline{\hfill
        \includegraphics[width=0.25\textwidth]{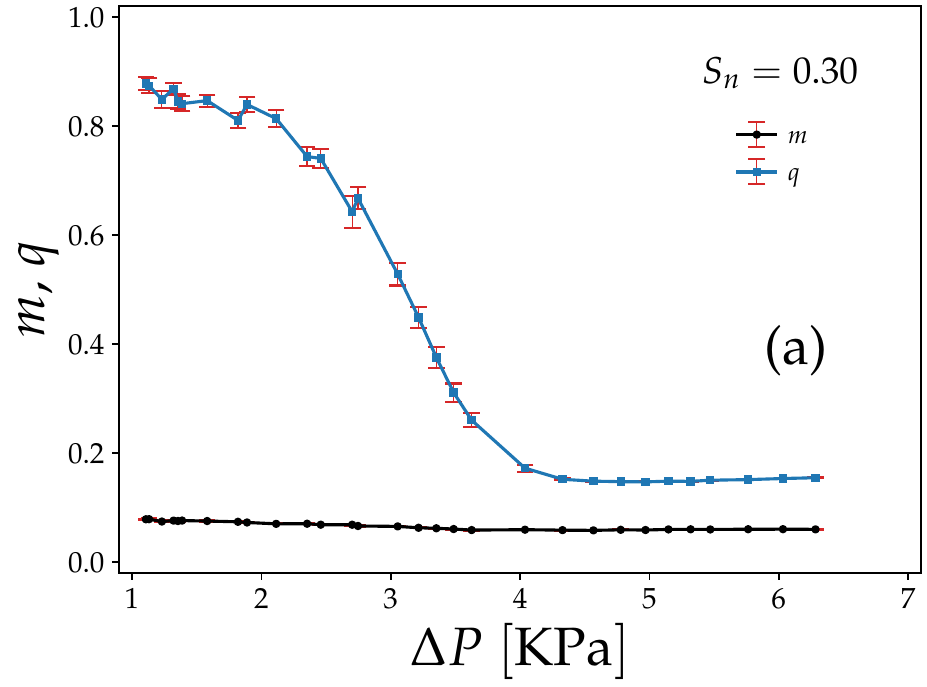}\hfill
        \includegraphics[width=0.25\textwidth]{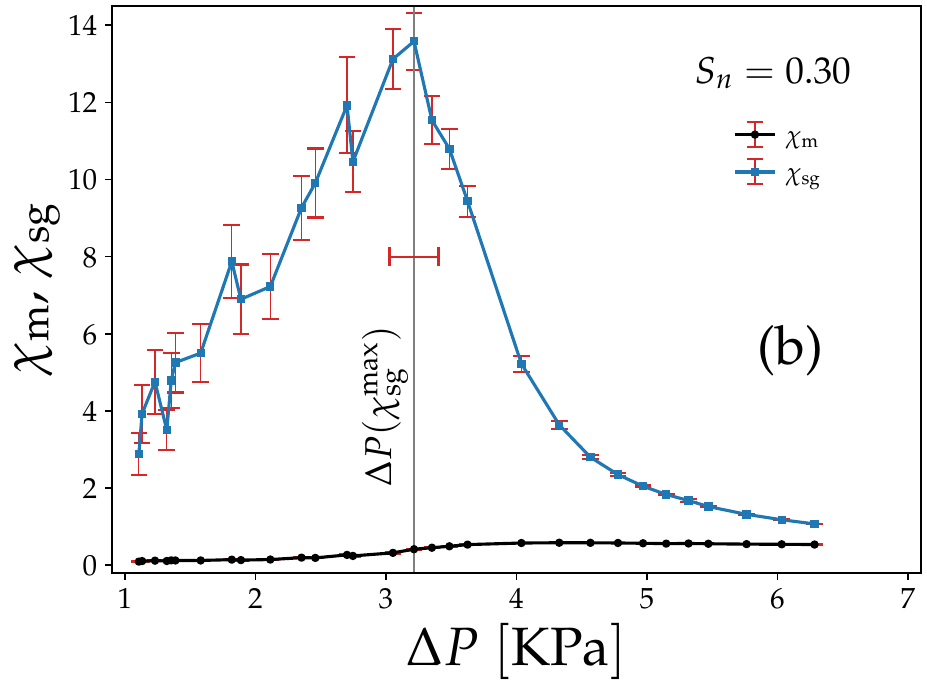}\hfill
        \includegraphics[width=0.24\textwidth]{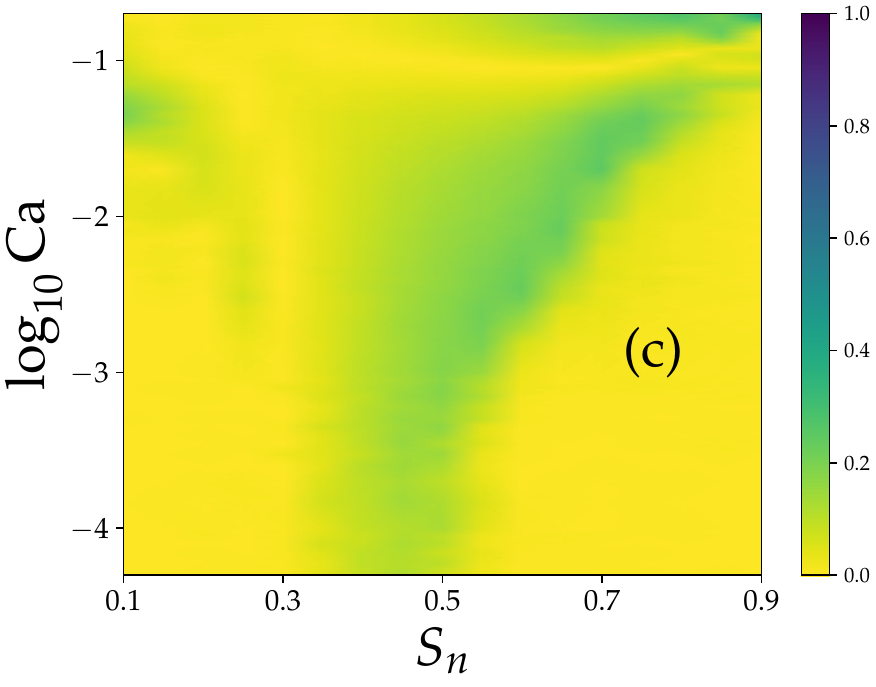}\hfill
        \includegraphics[width=0.24\textwidth]{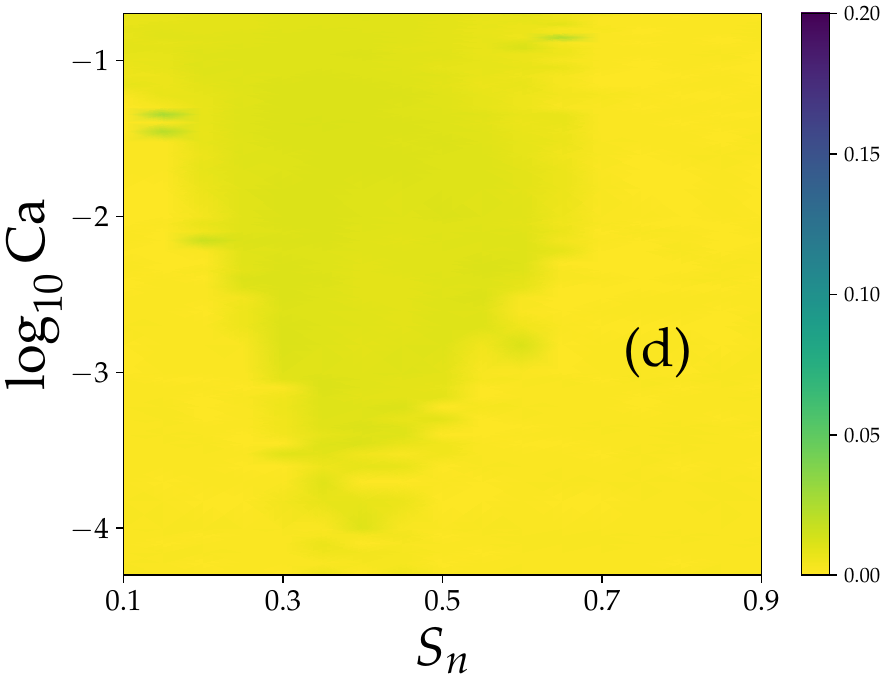}\hfill
    }
    \caption{\label{fig_mag} (a) The ferromagnetic and spin-glass order parameters $m$ and $q$, and (b) the uniform and spin-glass susceptibilities $\chi_{\rm m}$ and $\chi_{\rm sg}$, measured from the DPN data of $40\times 40$ system, plotted as a function of the average steady state pressure drop $\Delta P$ for global non-wetting saturation $S_n=0.3$. The magnetization $m$ remains low for the whole range of $\Delta P$ whereas $q$ rises sharply with the decrease of $\Delta P$, indicating a spin-glass transition. This is supported by the plots in (b), where the uniform susceptibility $\chi_{\rm m}$ remains low for the whole range of $\Delta P$ whereas $\chi_{\rm sg}$ shows a distinct peak at the pressure drop $\Delta P(\chi_{\rm sg}^{\rm max})$ around where $q$ rises. The vertical line indicates the position of $\Delta P(\chi_{\rm sg}^{\rm max})$ and the corresponding error is estimated by bootstrap resampling. The color map in (c) shows the difference in $q$ between the $40\times 40$ DPN data and the coarse-grained $10\times 10$ BML data, providing up to $20\%$ difference for certain data points. However, if we measure the difference between $10\times 10$ BML data and the coarse-grained $10\times 10$ DPN data, the difference is negligible as shown in (d), with $< 2\%$ difference, indicating the main source of difference in (c) is the coarse graining, and not BML. The observables to identify the phase transitions are measured from the $40\times 40$ DPN data, and therefore the critical points and the corresponding phase boundaries do not depend on coarse graining.}
\end{figure*}

Two main observables to characterize these phases are total ferromagnetic magnetization $m$, and the Edwards-Anderson order parameter $q$. These are two order parameters which characterize the transitions between paramagnetic and ferromagnetic phases, and between paramagnetic and spin-glass phases respectively. The spin-glass order parameter $q$ represents the degree of local magnetization. They are respectively, 
\begin{equation}
    \label{eq_op}
    \displaystyle
        m = \frac{1}{N}\left[\sum_i\langle \sigma_i \rangle\right]_a \quad \text{and} \quad 
        q =  \frac{1}{N}\left[\sum_i\langle \sigma_i \rangle^2\right]_a \;,
\end{equation}
where the sum is over all the $N$ spins. The average $\langle \ldots \rangle$ represents the average over different configurations for a specific sample, whereas $\left[ \ldots \right]_a$ represents the average over different network samples. The paramagnetic phase is characterized by $m=0$ and $q=0$, the ferromagnetic phase is characterized by $m\neq 0$ and $q>0$, and the spin-glass phase is characterized by $m=0$ and $q>0$. The other two relevant observables are the uniform susceptibility ($\chi_{\rm m}$) and spin-glass susceptibility ($\chi_{\rm sg}$), which are defined as \cite{fh91},
\begin{equation}
    \begin{aligned}
        \label{eq_sus}
        \displaystyle
        \chi_{\rm m}  & = \frac{1}{N^2}\left[\langle M^2\rangle - \langle M \rangle^2 \right]_{a}  \quad \text{and}\\
        \chi_{\rm sg} & = \frac{1}{N^2}\left[\sum_{\langle i,j\rangle}\left(\langle \sigma_i\sigma_j\rangle - \langle\sigma_i\rangle\langle\sigma_j\rangle\right)^2 \right]_{a}\;,
    \end{aligned}
\end{equation}
where $M = \sum_i \sigma_i$. The uniform susceptibility $\chi_{\rm m}$ diverges at the boundary between the P and F phases whereas the spin-glass susceptibility $\chi_{\rm sg}$ diverges at the boundary between the P and SG phases. 

To measure these observables we performed DPN simulations with constant capillary number (Ca) using networks containing $40\times 40=1600$ links. Simulations were performed for $17$ values of total non-wetting saturations from $S_n=0.1$ to $0.9$ with interval of $0.05$. For each saturation, we performed simulations with $55$ different values of $\rm Ca$ in the range of $5\times 10^{-5}$ to $2\times 10^{-1}$. This leads to a total of $935$ independent simulation parameters. For each parameter set, we considered $100$ different samples (realizations) of the network corresponding to the average $\left[ \ldots \right]_a$ in the Eqs. \ref{eq_op} and \ref{eq_sus}. For each sample, $1000$ different fluid configurations were considered in steady state (the average $\langle\ldots\rangle$). This leads to a total of $10^5$ steady-state configurations for each parameter set. To reach the steady state, we typically skipped $10$ pore-volumes of flow, which we found sufficient for a random initial distribution of the two fluids. In the steady state, we only measured at time intervals sufficient to avoid correlations between them (see Section  \ref{sec_model}).  

The steady-state fluid configurations were mapped onto spin configurations, from which we measure the observables defined in Eqs. \ref{eq_op} and \ref{eq_sus}. We show the results in Fig. \ref{fig_mag} for $S_n=0.3$, where we plot $m$, $q$, $\chi_{\rm m}$ and $\chi_{\rm sg}$ as a function of the average steady-state pressure drop $\Delta P$. We see that the magnetization $m$ remains essentially unchanged at a low value, wheres the spin-glass order parameter $q$ rises sharply while decreasing the pressure drop. This indicates high degree of local magnetization at low Ca, while the total magnetization remains unchanged at a low value.  We also see in Fig. \ref{fig_mag} (b), where the spin-glass susceptibility $\chi_{\rm sg}$ has a peak at a pressure value $\Delta P(\chi_{\rm sg}^{\rm max})$ similar to where the sharp increase in $q$ is observed. The uniform susceptibility $\chi_{\rm m}$ do not show such a peak.  These observations are exactly those that signal a critical point separating a spin-glass phase from a paramagnetic phase.  These results point towards the dynamic pore network model experiences a critical transition to a glassy state when $\rm Ca$ is lowered through the critical value. 

The data in Fig. \ref{fig_mag} (a) and (b) are directly from the DPN model. The reason for using these data rather than running equilibrium Monte Carlo simulations on the BML spin model is one of cost of computation. Using the DPN model data directly allows us to consider much larger systems than would be feasible using the BML model. We show in Fig. \ref{fig_mag} (c) and (d) the difference between the Edwards-Anderson order parameter $q$ calculated directly from the DPN data and from BML data. Fig. \ref{fig_mag} (c) shows the difference between $40\times 40$ DPN data and the BML data on a coarse-grained $10\times 10$ lattice. Fig. \ref{fig_mag} (d) shows the difference between $10\times 10$ DPN data and $10\times 10$ BML data. In the first case (c) where we have invoked coarse graining, there is a difference of up to 20\% between the data. In the second case (d) where we compare $10\times 10$ DPN data with $10\times 10$ BML data, the difference is less that 2\%. We conclude that the BML model works very well, whereas coarse graining introduces some errors.     

Another important question to raise is the following: The Edwards-Anderson order parameter $q$ picks up spins that do not move. Our interpretation is that these spins are stuck due to frustration. However, they could also be stuck due to geometric constraints built into the pore network. For example, there might be links that are so narrow that the non-wetting fluid cannot enter them.  Such links would be permanently assigned a spin equal to -1. Or we could have links that are part of a dead end, locking their spin values.  Yet another possibility is that there might be links that are so wide that the flow through them is so slow that their saturation do not change.  How can we ensure that it is frustration and not the geometry that causes spins to lock?

In order to answer this question, we note the following: We introduce the disorder in the DPN model through a probability distribution for the link radii. We have used a flat distribution on the interval $0.1l$ to $0.4l$ where $l$ is the length of the links. This is a moderate disorder that do not produce dead ends. If a link is too small to allow for the non-wetting fluid to enter, this will be due to the pressure drop across it being too small. The pressure drop is not determined by the link alone but by the fluid configuration surrounding it. Hence, the question of geometry vs.\ frustration becomes blurred. In the case of wide links, their slow flow would be caused by the links' high conductivity; they would essentially be short circuited. However, the disorder we are dealing with is moderate, so such links would be rare if they are at all present.  

We end this discussion by noting that the spin-glass susceptibility $\chi_{sg}$ will only register correlated spins. If a spin $\sigma_i$ is frozen, the contribution to the susceptibility is zero, as for any uncorrelated spin pair. Hence, the peak in Fig. \ref{fig_mag} (b) is not affected by geometrically frozen spins.

\begin{figure}
    \centerline{\hfill\includegraphics[width=\columnwidth]{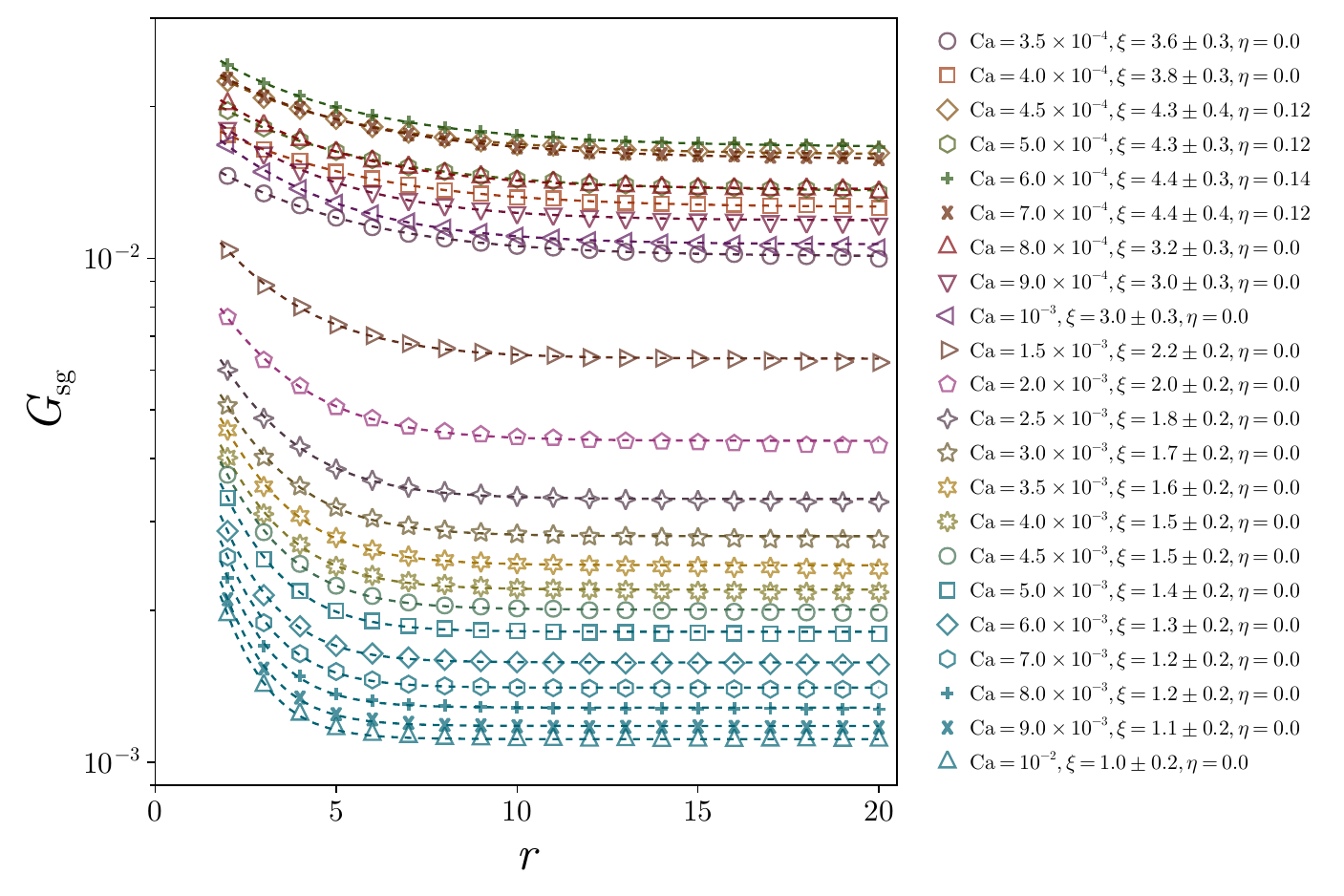}\hfill}
    \caption{\label{fig_Gr} Spin-glass spatial correlation function at saturation $S_{\mathrm{n}} = 0.3$. The disorder-averaged correlation function $G_{\mathrm{sg}}(r)$ is plotted on a logarithmic scale versus the DPN link separation $r$ for capillary numbers spanning $\mathrm{Ca}=3.5\times10^{-4}$ to $1\times10^{-2}$. The dashed curves are fits to Eq. \ref{eq_fit}, from which the correlation length $\xi$, the anomalous exponent $\Tilde{\eta}$, and the baseline $G_0$ are obtained.}
\end{figure}

We have so far established that $m\approx 0$, $q > 0$, $\chi_m\approx 0$, and there is a peak in $\chi_{sg}$. These are necessary but not sufficient criteria for establishing the spin-glass phase transition. A medium with spatially heterogeneous pore sizes, in which each pore independently freezes when Ca drops below its local mobilization threshold, would also satisfy the three first criteria. A concrete example where this would be true is the capillary fiber bundle model \cite{scheidegger1974physics}. In order to further establish that we are dealing with a collective phenomenon, we measured the correlation function \cite{McmillanSAn1983},
\begin{equation}
    \label{eqAH_67}
    G_{\rm sg}(r) = \left[\frac{1}{N_r}\sum_{j>i} (\langle\sigma_i\sigma_j\rangle -
                    \langle\sigma_i\rangle\langle\sigma_j\rangle)^2 \delta(r_{i,j} - r) \right]_a \;, 
\end{equation}
where $r_{ij}$ is the distance between spins $i$ and $j$, and $N_r$ is the number of spin pairs at a distance $r$. It is convenient to write the connected correlator as $\hat{C}_{ij} \equiv \langle\sigma_i \sigma_j\rangle - \langle\sigma_i\rangle\langle\sigma_j\rangle$, so that $G_{\rm sg}(r)=\bigl[(N_r)^{-1}\sum_{j>i}\hat{C}_{ij}^{2}\, \delta(r_{ij}-r)\bigr]_a$. The spin-glass susceptibility $\chi_{\rm sg}$ follows from
\begin{equation}
    \label{eq_chisg}
    \chi_{\mathrm{sg}} = \frac{1}{N}\sum_{r} N_r\, G_{\mathrm{sg}}(r) \;.
\end{equation}
We evaluated $G_{\mathrm{sg}}(r)$ from the DPN data for $40\times 40$ spins and averaged over $100$ independent network samples. Because the DPN model imposes periodic boundary conditions in both directions, we restrict the analysis to separations $0 \le r \le L/2 = 20$. We measured $G_{\mathrm{sg}}(r)$ at saturation $S_{\mathrm{n}} = 0.3$ over a range of capillary numbers $\mathrm{Ca}$ spanning both sides of the transition, near which the spin-glass susceptibility $\chi_{\mathrm{sg}}$ exhibits a peak at ($\mathrm{Ca}\approx 6\times10^{-4}$). The results are shown in Fig. \ref{fig_Gr}.

For all capillary numbers we find that the correlation function is well described by the standard scaling \cite{McmillanSAn1983, Lundow2018}:
\begin{equation}
    \label{eq_fit}
    G_{\rm{sg}}(r) = A \frac{ \mathrm{e}^{-r/\xi}}{r^{\Tilde{\eta}}} + G_{0},
\end{equation}
where $A$ is an amplitude, $\xi$ is the spin-glass correlation length, $\Tilde{\eta}$ is the anomalous-dimension exponent, and $G_0$ is a noise floor. At large $r$, the true connected correlations vanish, yet the {\it squared} sample correlator retains a positive bias $\langle\hat{C}_{ij}^2\rangle \simeq (1-m_i^2)(1-m_j^2)/K$ from the finite number $K$ of samples, so $G_{\rm sg}(r)$ saturates to a small positive constant $G_0$ rather than decaying to zero. The dashed lines in Fig. \ref{fig_Gr} correspond to fits using Eq. \ref{eq_fit}, with $R^2>0.9995$ in all cases.

\begin{figure}
    \centering
    \includegraphics[width = \columnwidth]{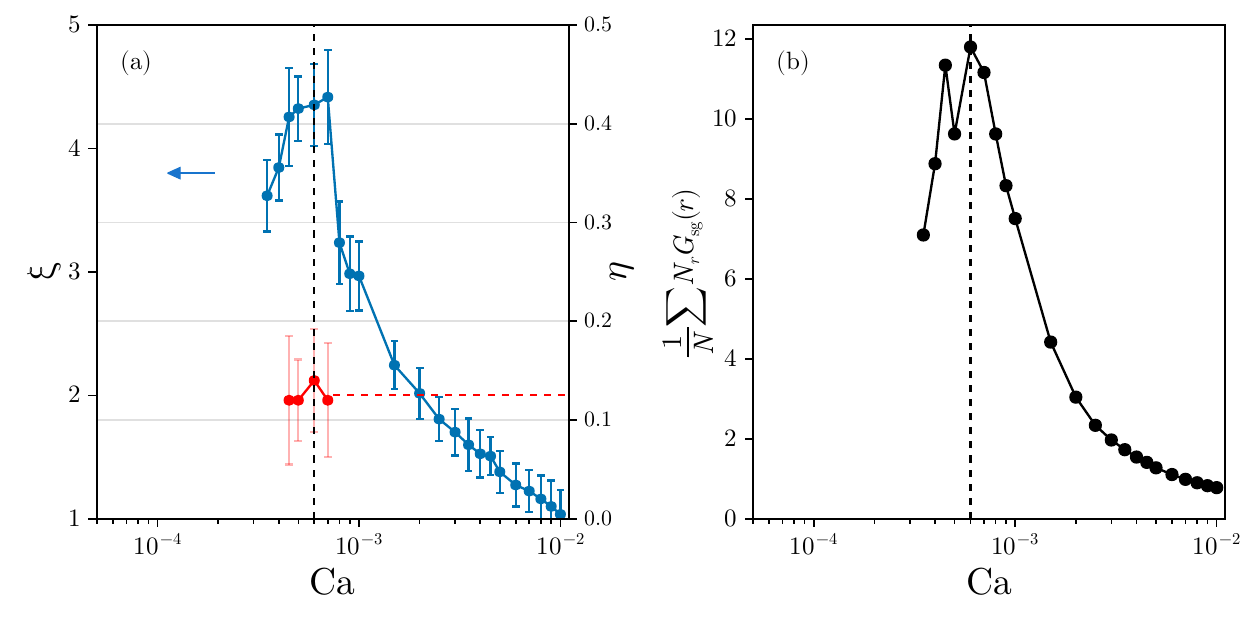}
    \caption{\label{fig_xi} (a) The blue markers show the fitted correlation length $\xi$ (left scale) as a function of ${\rm Ca}$ on a logarithmic scale. $\xi$ is non-monotonic: it grows through the capillary-dominated regime, reaches a maximum near ${\rm Ca}\approx 7\times10^{-4}$ (dashed line), and then decays in the higher Ca regime. The red markers show the fitted anomalous-dimension exponent $\Tilde{\eta}$ (right scale). For ${\rm Ca}<3\times10^{-4}$ most of the DPN realizations are frozen due to capillary arrest. For ${\rm Ca} > 7\times 10^{-4}$, the distributions of couplings $J_{ij}$ are Gaussian, and we fixed $\Tilde{\eta} = 0$ \cite{Fernandez2016}. (b) The spin-glass susceptibility $\chi_{\rm sg}$ computed from Eq. \ref{eq_chisg}, is shown as a function of $\rm Ca$ on a logarithmic scale. Even with the sum truncated at $r=20$, the result closely matches Fig. \ref{fig_mag} (b).}
\end{figure}

In Fig. \ref{fig_xi}(a), we show $\xi$ as a function of the capillary number $\rm Ca$ (blue markers, left axis). There is a clear peak at ${\rm Ca}\approx 6\times10^{-4}$. In the low-$\rm Ca$ (Ib) regime, the dynamics are nearly frozen: few links flip, so cooperative fluctuations cannot develop and $\xi$ is small despite strong capillary coupling. As $\rm Ca$ increases, the system approaches marginal stability, and capillary-mediated triggering organizes flips into system-spanning avalanches, causing $\xi$ to grow and peak near the Ib--II boundary. In the high-$\rm Ca$ (II) regime, the viscous forcing dominates: each link's occupancy is set by the local pressure field and fast advection rather than by its neighbors. The now-abundant flips become effectively independent, the fluctuations about the mean field decay over a microscopic scale, and $\xi$ collapses to a value comparable to the lattice spacing. The resulting peak of $\xi({\rm Ca})$ suggests that collective behavior dominates the spin freezing near the Ib--II boundary, since independent geometric pinning would produce a $\rm Ca$-independent characteristic length. The anomalous-dimension exponent $\Tilde{\eta}$ is also shown as a function of the capillary number in Fig. \ref{fig_xi}(a) (red markers, right axis). Near the transition, we find $\Tilde{\eta} \approx 0.125$. This value is consistent with numerical estimates of $\Tilde{\eta}$ in the range $0.14$--$0.22$ \cite{Katzgraber2005, Poulter2005, Hartmann2008} for two-dimensional Ising spin glasses with bimodal distribution of couplings, similar to those observed in Fig. \ref{fig_Jij}. Above the transition, for ${\rm Ca}>7\times10^{-4}$, we use the observation that the distribution of couplings $J_{ij}$ are Gaussian like and centered at zero (see Fig. \ref{fig_Jij}), and therefore fix anomalous-dimension exponent to its exact value $\Tilde{\eta} = 0$ \cite{Hartmann2008, Fernandez2016}.

\begin{figure*}
    \centerline{\hfill
        \includegraphics[width=0.25\textwidth]{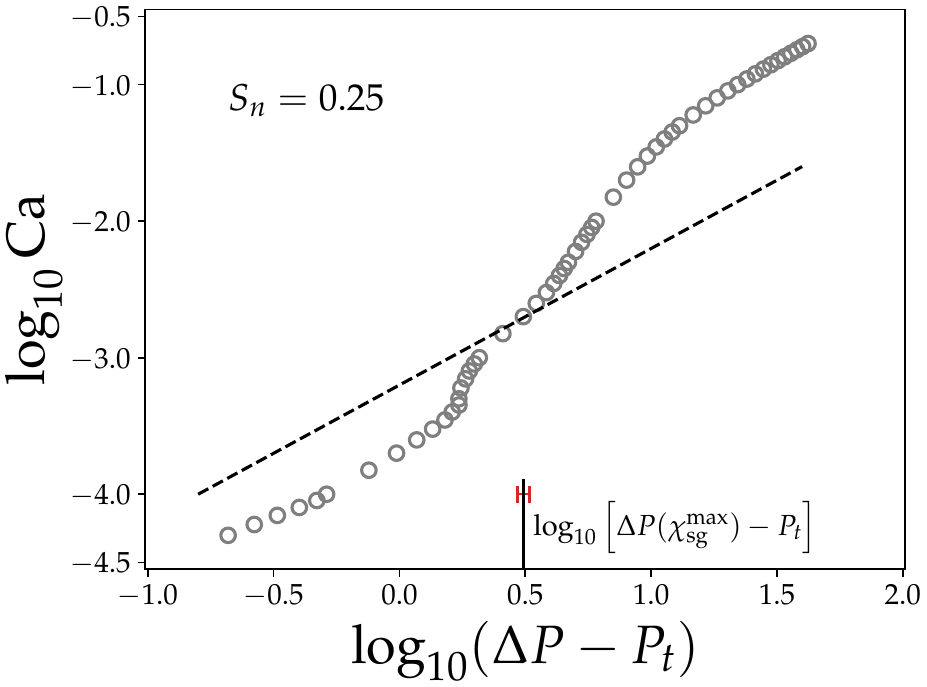}\hfill
        \includegraphics[width=0.25\textwidth]{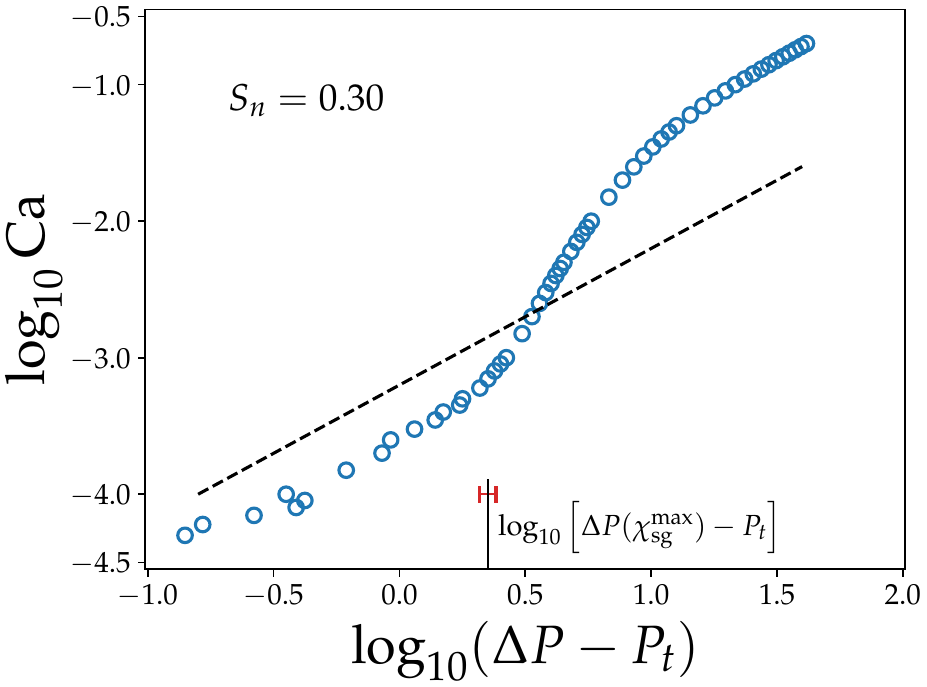}\hfill
        \includegraphics[width=0.25\textwidth]{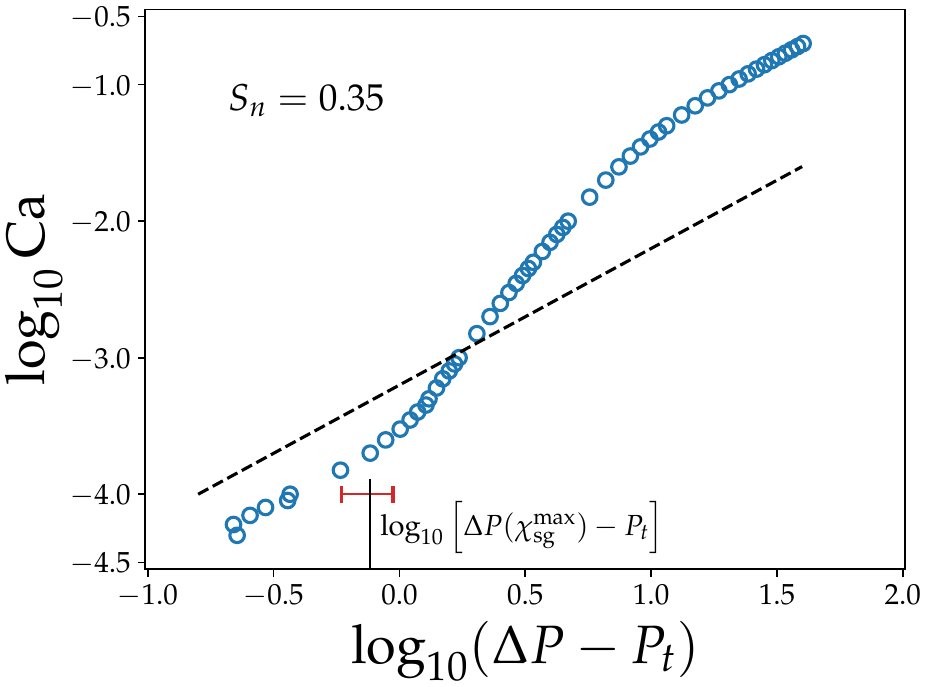}\hfill
        \includegraphics[width=0.25\textwidth]{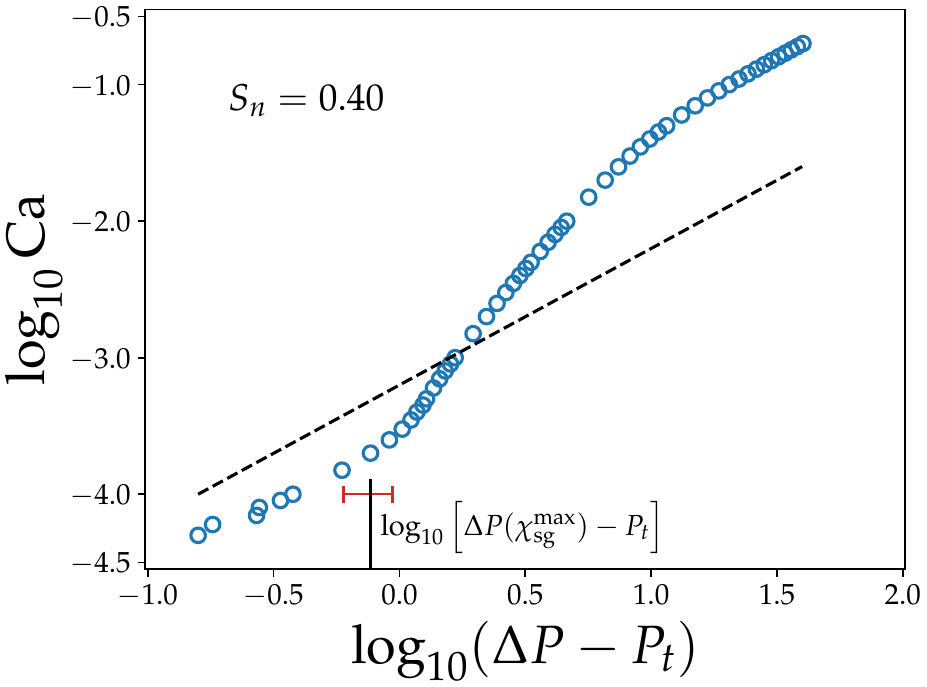}\hfill
    }
    \centerline{\hfill
        \includegraphics[width=0.25\textwidth]{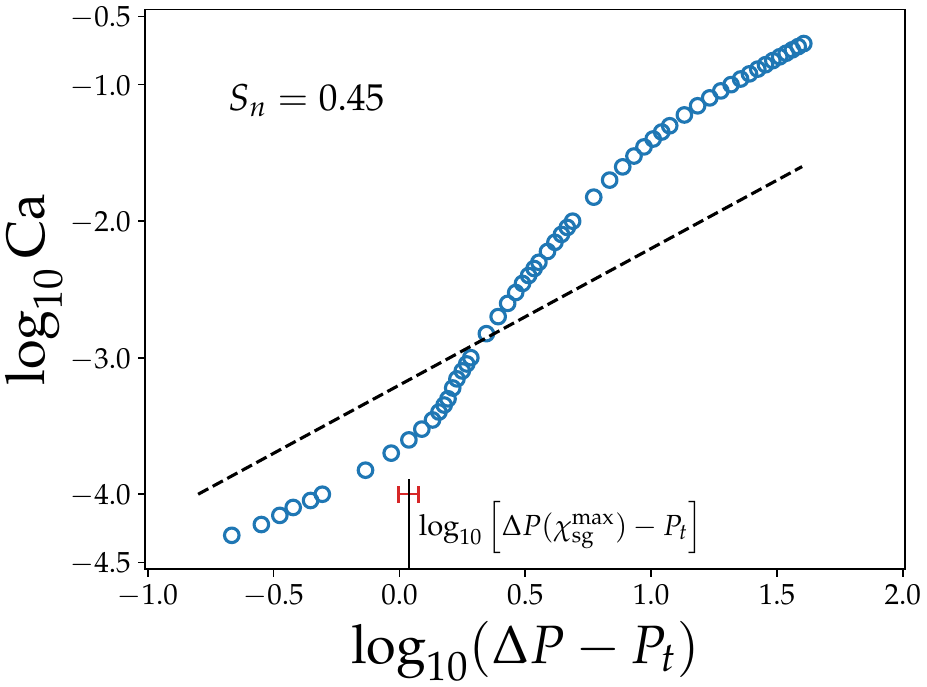}\hfill
        \includegraphics[width=0.25\textwidth]{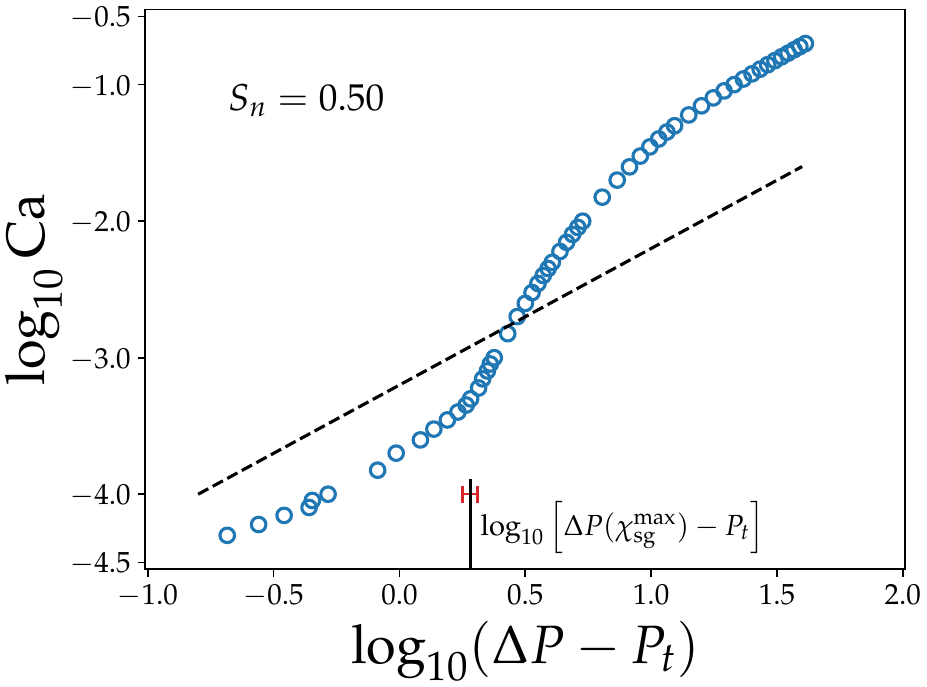}\hfill
        \includegraphics[width=0.25\textwidth]{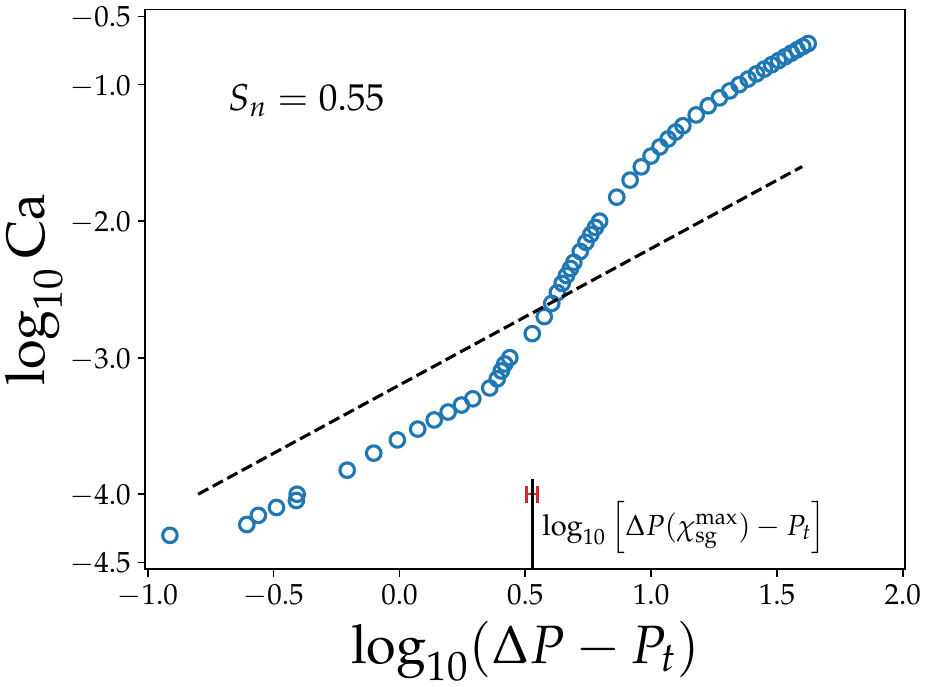}\hfill
        \includegraphics[width=0.25\textwidth]{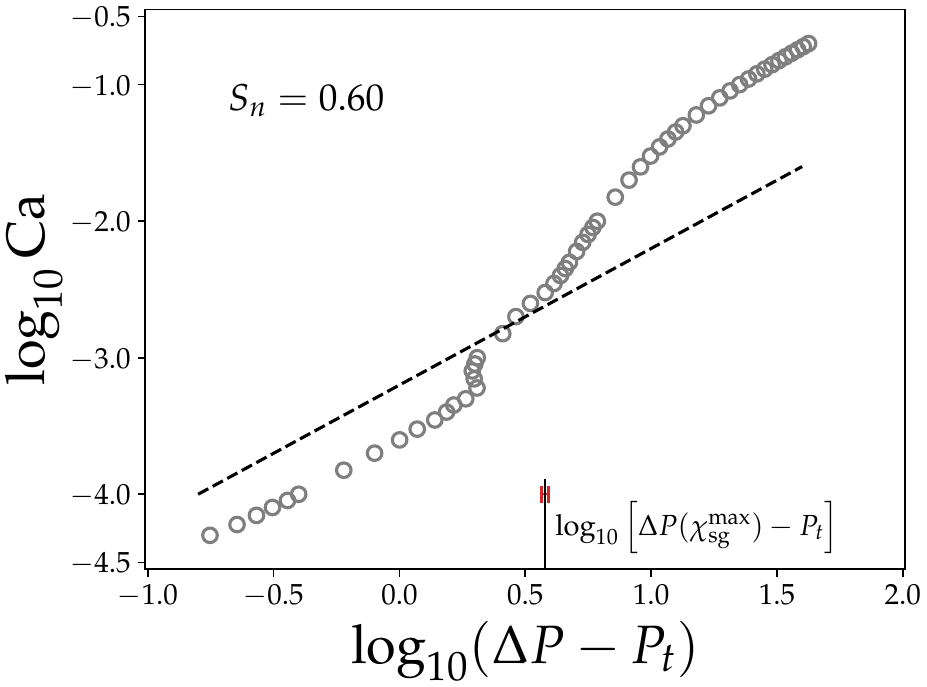}\hfill
    }
    \caption{\label{fig_pq} Plots of $\log_{10}\rm Ca$ as a function of $\log(\Delta P-P_t)$ in the steady state for different non-wetting saturations ($S_n$). The black dashed lines are drawn with slope $1$, indicating a linear relationship between the steady state flow rate $Q$ and the pressure drop $\Delta P$. The data show that there are three regimes, two linear regimes at low and high $\rm Ca$ values, and a non-linear regime in between, named regimes I, II and III by Berg et al.\ \cite{berg2026from}. The small vertical black lines above $x$-axis indicate the positions of pressure drops $\Delta P(\chi_{\rm sg}^{\max})$, where the spin-glass susceptibilities $\chi_{\rm sg}$ have the peak for respective saturations (see Fig. \ref{fig_mag} (b)). The corresponding error bars for $\Delta P(\chi_{\rm sg}^{\max})$ are indicated along the lines. We find that the positions $\Delta P(\chi_{\rm sg}^{\max})$ are comparable with the crossover points from the lower linear to the intermediate non-linear regime for intermediate saturations in the range $S_n = 0.30$ to $0.55$. This range is indicated by blue color in the plots. Outside this saturation range, the three regimes I, II and III are not distinct, and the values of $\Delta P(\chi_{\rm sg}^{\max})$ do not match with the crossover points.}
\end{figure*}

Is the transition from the paramagnetic to the glassy phase in the spin model, which is entirely characterized by the pore-scale fluid configurations, related to any transition or crossover observed at the Darcy scale? As described in the Introduction, Berg et al.\ \cite{berg2026from} based their classification of Darcy scale phases on the constitutive relation relating flow velocity to pressure gradient. In the laboratory, velocities and pressure gradients are not observed directly.  Rather, one observes the volumetric flow rate $Q$ as a function of the pressure drop $\Delta P$ across the porous medium sample under steady-state conditions. Experiments and computations have established that while increasing the Ca from a low value, the relationship between $Q$ and $\Delta P$ changes from a linear (regime I) to a non-linear power-law relationship, $Q\sim\Delta P^\alpha$, see Eq. \ref{eqAH1}, with $\alpha>1$ (regime II) \cite{rassi2011nuclear, sinha2012effective, yiotis2013blob,aursjo2014film, sinha2017effective, gao2017x, yiotis2019nonlinear, roy2019effective, gao2020pore, zbg21,fyhn2021rheology,zhang2022nonlinear, sales2022bubble, fsh23}. Certain systems have also shown the existence of a threshold pressure $P_t$ in this regime, such that $Q\sim(\Delta P-P_t)^\alpha$ \cite{sinha2012effective}. The origin of $P_t$ is the absence of continuous pathways \cite{shb13}, creating a net pressure drop for the medium which the viscous pressure drop must overcome to start the flow.  We note that $P_t\to 0$ as the system size increases \cite{feder2022physics,roy2024effective}. While increasing Ca further, the power law regime crosses over to another linear regime (regime III) displaying again a linear relationship between $Q$ and $\Delta P-P_t$. The origin of these three regimes may be found in the dynamics of the fluid-fluid interfaces as described in the Introduction. However, the behavior is generic for bi-viscous fluids \cite{rh87,talon2020effective}. As stated in the Introduction,  neither the transition from regime Ia to Ib nor from regime II to III are phase transitions.

\begin{figure*}
    \begin{minipage}{0.245\textwidth}
        \centering
        $m$\\
        \includegraphics[width=\textwidth]{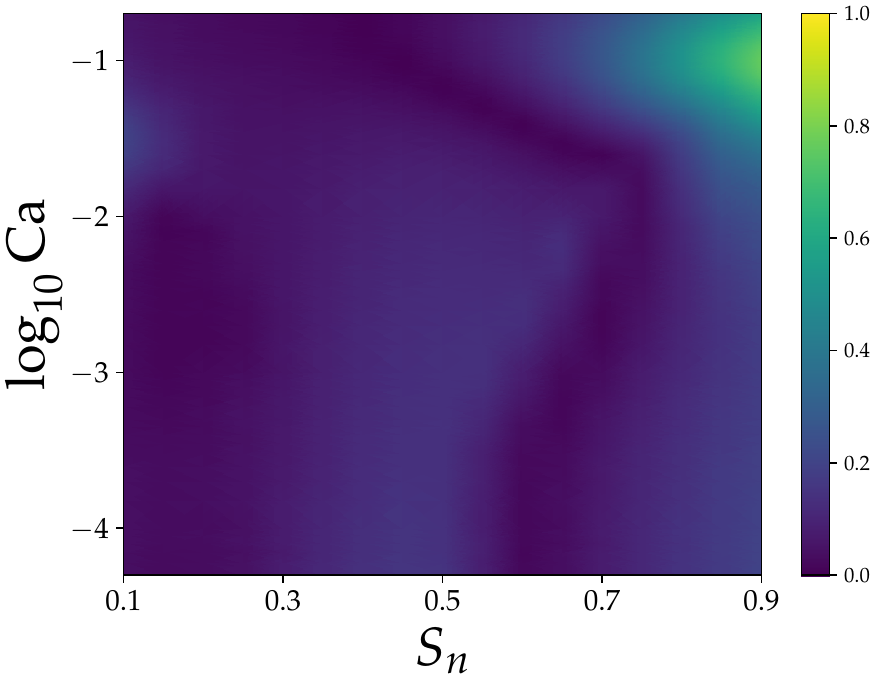}
    \end{minipage}
    \begin{minipage}{0.245\textwidth}
        \centering
        $q$\\
        \includegraphics[width=\textwidth]{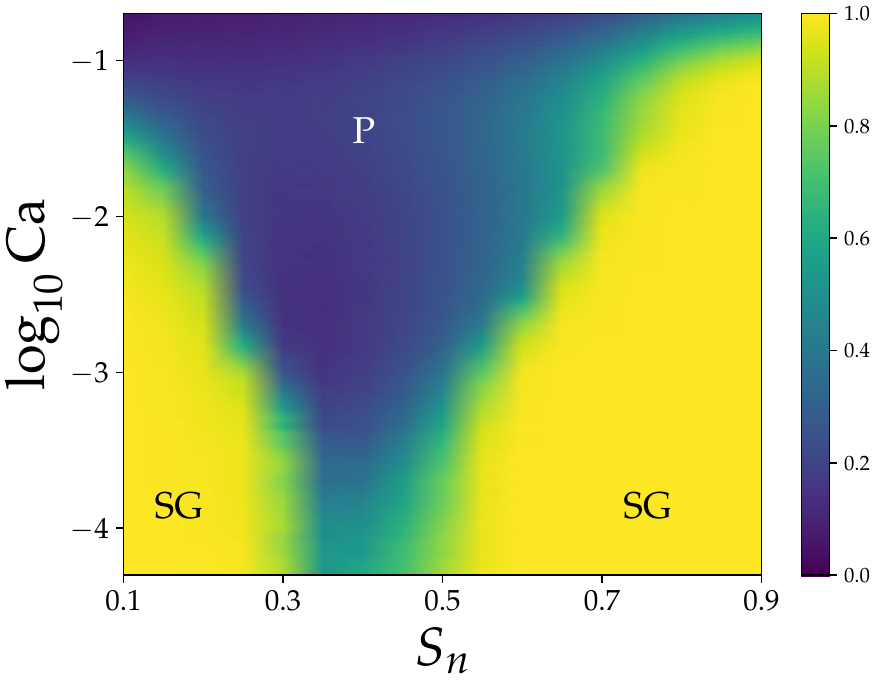}
    \end{minipage}
    \begin{minipage}{0.245\textwidth}
        \centering
        $\chi_{\rm m}$\\
        \includegraphics[width=\textwidth]{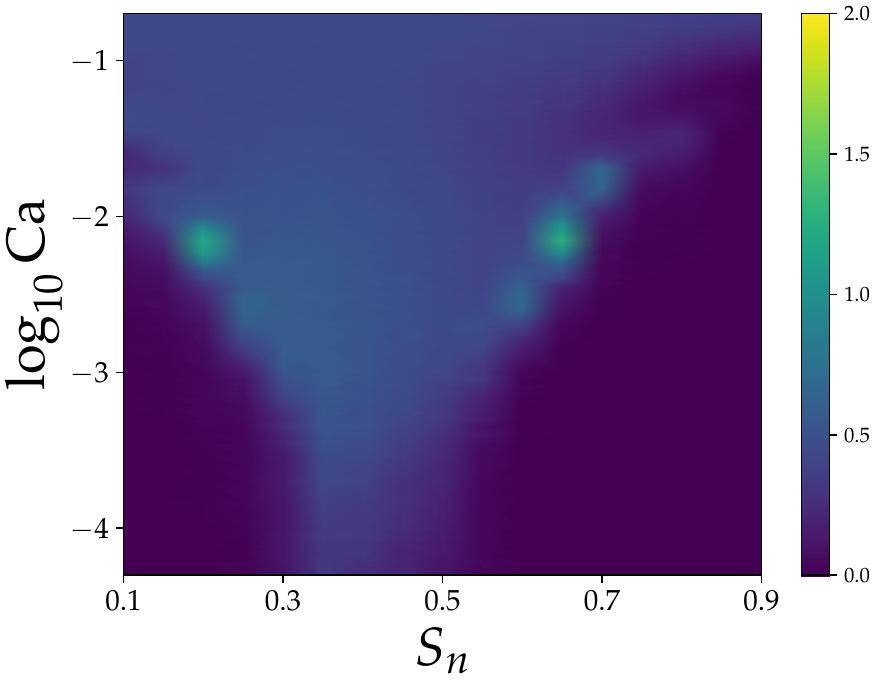}
    \end{minipage}
    \begin{minipage}{0.245\textwidth}
        \centering
        $\chi_{\rm sg}$\\
        \includegraphics[width=\textwidth]{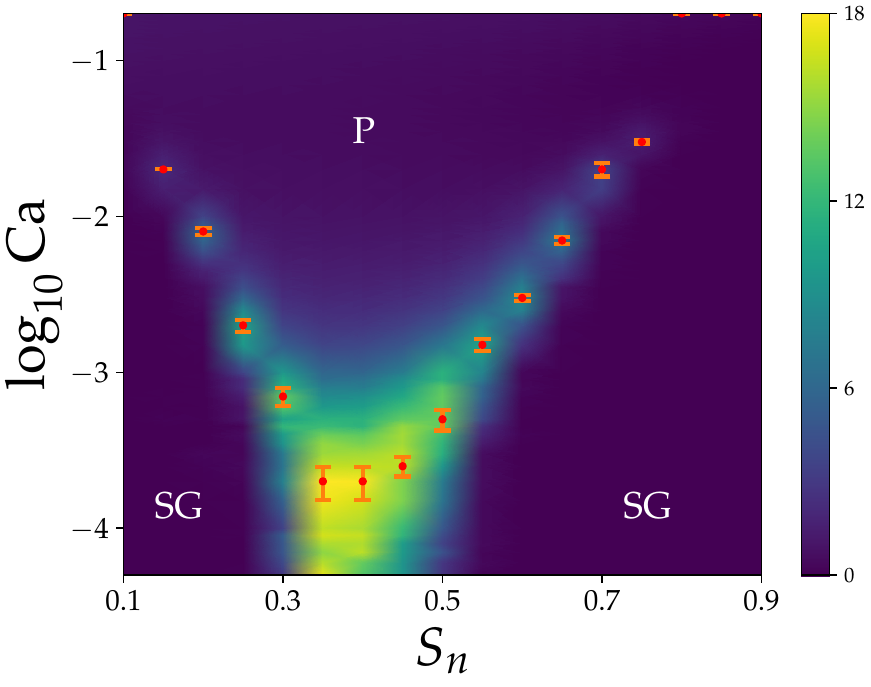}
    \end{minipage}
    \caption{\label{fig_map} Color maps showing the values of $m$, $q$, $\chi_{\rm m}$ and $\chi_{\rm sg}$ (top row), and the corresponding error bars (bottom row), plotted against $(S_n,\log_{10} {\rm Ca})$. Each plot contains 935 data points. The maps for $q$ and $\chi_{\rm sg}$ show the P and SG phases and the phase boundaries between these phases. We observe that the $\rm Ca$ for transition from P to SG phase decreases as $S_n$ approaches $\approx 0.4$. The maps for $m$ and $\chi_{\rm m}$ show some color gradient, however they are relatively very weak compared to $q$ and $\chi_{\rm sg}$, and do not indicate any transition between P and F phases. The error bars along the critical line are shown in the plot for $\chi_{\rm sg}$ which indicate the errors associated with the measurement of $\chi_{\rm sg}^{\rm max}$.}
\end{figure*}

In order to identify the connection between the Darcy scale flow regimes with the glass transition that we observed here for the pore network flow, we measured average steady-state pressure drops $\Delta P$ for all the simulations. The simulations were performed for constant capillary numbers Ca, which set the total flow rate $Q$ (Eq. \ref{eq_Ca}). We show these results in Fig. \ref{fig_pq}, where we plot $\log_{10}\rm Ca$ as a function of $\log_{10}(\Delta P-P_t)$ for different values of non-wetting saturation $S_n$. We see that for a wide range of intermediate saturations from $S_n=0.30$ to $0.55$, the plots show three distinct regimes, I, II and III, showing a non-linear regime in between two linear regimes. The dashed line is drawn with a slope of $1$, indicating a linear relationship between $Q$ and $\Delta P$. The threshold pressures $P_t$ are measured by fitting the data points of regime I to a straight line. We then added the small vertical red lines in the plots showing the values of $\Delta P(\chi_{\rm sg}^{\rm max})$ for each $S_n$, which indicate the tradition points to the glassy phase. The values of $\Delta P(\chi_{\rm sg}^{\rm max})$ are obtained from the peaks of the spin-glass susceptibilities $\chi_{\rm sg}$ measured for different $S_n$, as indicated in Fig. \ref{fig_mag} (b). Surprisingly, the positions related to the glass transition fall  at the values of $\Delta P$ where the crossover from regime I to regime II is observed. This explains the hysteretic behavior, the large fluctuations and the wide range of timescales seen in the intermittent regime: The transition Ib to II is a critical glass transition. 

We note that outside the intermediate range of $S_n=0.30$ to $0.55$, our DPN data do not reproduce the transition between the regimes I and II well, as shown by the black plots in Fig. \ref{fig_pq} for $S_n=0.25$ and $0.60$. There have, however, been no systematic studies of what happens to regime II for saturations for values of $S_n$ close to 0 or 1. It is therefore not clear what to expect at these values of $S_n$.

The complete phase diagram with all the $935$ parameters is shown in Fig. \ref{fig_map} where we plotted $m$, $q$, $\chi_{\rm m}$ and $\chi_{\rm sg}$ on $\log_{10}{\rm Ca}$ vs $S_n$ space . The color maps for $q$ and $\chi_{\rm sg}$ show distinct separation between regions indicating the boundaries between the SG and P phases. The respective phases are marked on these plots. The plots for $m$ and $\chi_{\rm m}$ on the other hand do not show any significant separation, therefore the presence of an F phase is not observed on this space. 

\begin{figure}
    \centerline{\hfill\includegraphics[width=0.9\columnwidth]{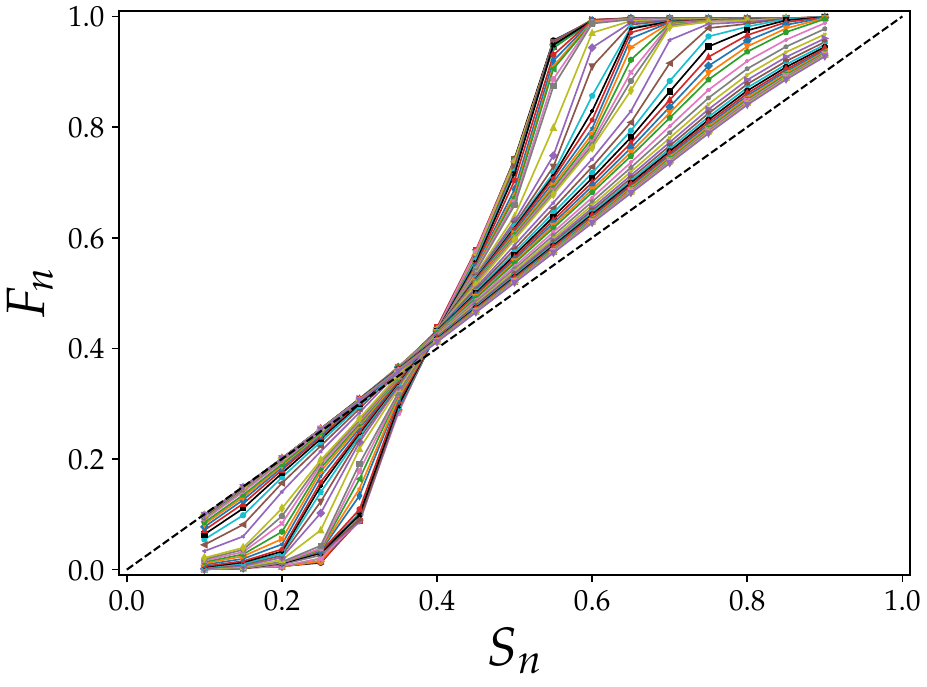}\hfill}
    \caption{\label{fig_FS} Non-wetting fractional flow ($F_n$) versus global non-wetting saturation ($S_n$) for different capillary numbers (Ca). The diagonal dashed line indicates $F_n=S_n$, along which the fluids will move with the same velocity. The curve closest to the diagonal corresponds to the highest $\rm Ca$, while the farthest with the most step-like form corresponds to the lowest Ca.}
\end{figure}

An observation from Fig. \ref{fig_map} is that the capillary number necessary for the transition to a glassy phase for any specific $S_n$ decreases as $S_n$ approaches to $\approx 0.4$ from both sides. This value of $S_n$ falls around the same value where $S_n$ crosses the $F_n=S_n$ line, as shown in Fig. \ref{fig_FS}. Here $F_n$ is the non-wetting fractional flow in the steady state defined as $F_n = Q_n/Q$, where $Q_n$ is the volumetric non-wetting flow rate. Different curves in Fig. \ref{fig_FS} correspond to different values $\rm Ca$, where the curves approaching towards the dashed $F_n=S_n$ line correspond to the increase in $\rm Ca$. Along the $F_n=S_n$ line, both the fluids flow with same velocity, similar to two miscible fluids as if there were no surface tension between the two fluids. We see in this figure that all the curves cross this line at around $S_n\approx 0.4$ for our DPN model. It therefor appears that as $S_n$ approaches this value, the capillary forces become less effective to hold any fluid ganglion in place, thus pushing the glass transition towards lower $\rm Ca$ numbers. 

\begin{figure*}
    \centerline{\hfill
        \includegraphics[width=0.25\textwidth]{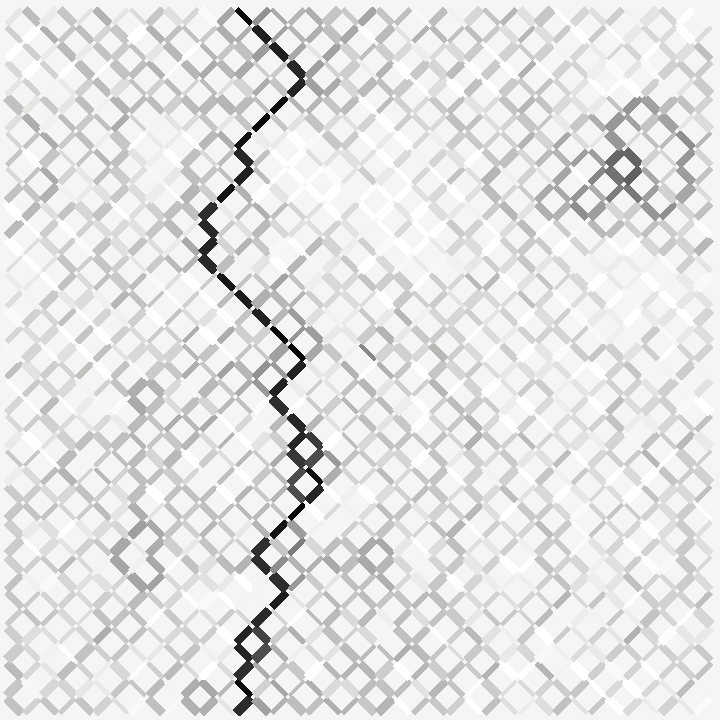}\hfill
        \includegraphics[width=0.25\textwidth]{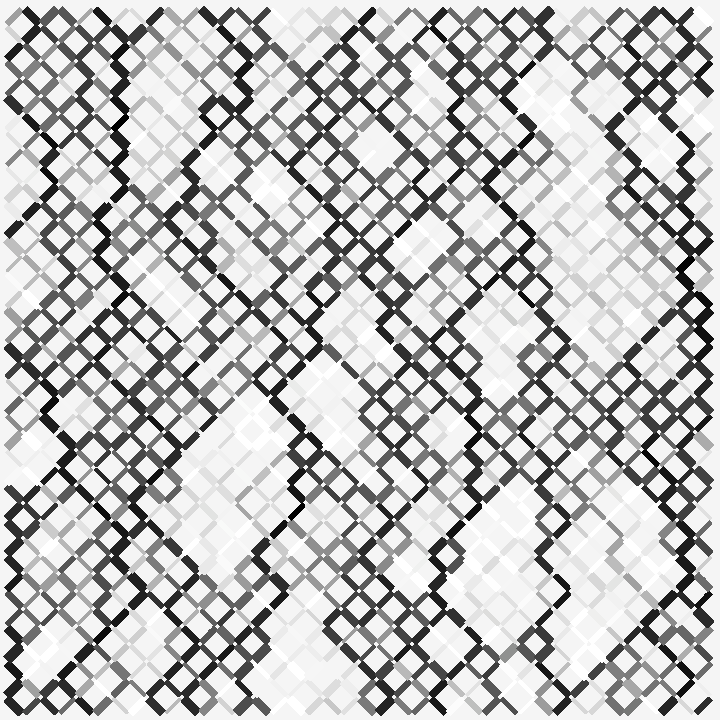}\hfill
        \includegraphics[width=0.25\textwidth]{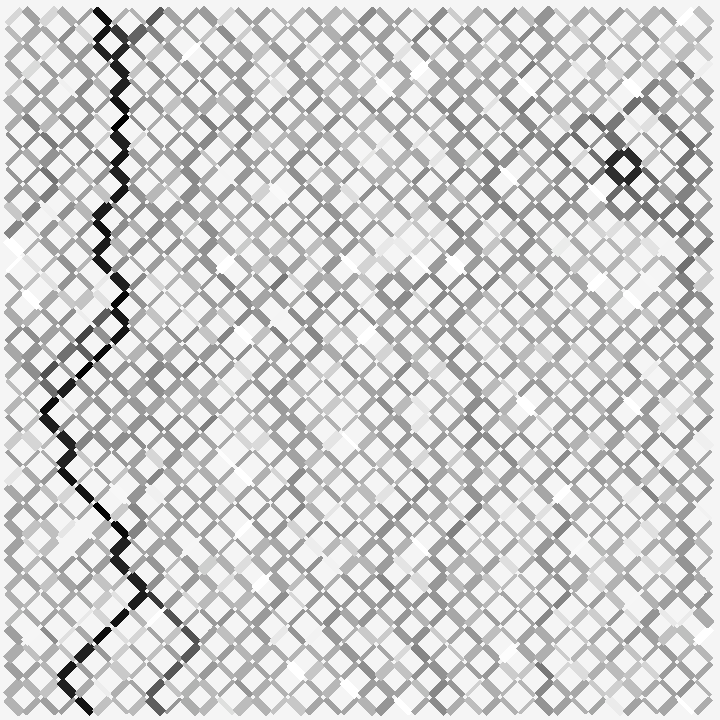}\hfill
    }
    \caption{\label{fig_frustration} Three snapshots showing the velocity profile of the DPN model, first being in steady state at Ca = $10^{-4}$ (left), then the capillary number is raised to Ca = $10^{-2}$ allowing the model to reach steady state again (middle), and lastly the capillary number is reduced back to Ca = $10^{-4}$ reaching steady state again (right). Darker color means higher velocity. In the left and right panels, the model is in phase Ib, whereas in the middle panel it is in phase II. The saturation was fixed at $S_n=0.55$.}
\end{figure*}

\section{Summary and discussion}
\label{sec_sum}
We have constructed the pore-scale configurational probability for immiscible two-phase flow in porous media and, on that basis, established a phase diagram for the steady state. The saturation patterns of a dynamic pore-network model were mapped onto a spin model that preserves the average saturation and the saturation–saturation correlations, and the spin-configurational probability was obtained from the Jaynes maximum entropy principle with the entropy of the fluid configurations. Under steady-state conditions this entropy is constant, so the spin model may be treated with equilibrium statistical mechanics. Training the spin model with Boltzmann machine learning and matching correlations up to third order, we showed that it reproduces the fluid configurations of the pore-network model. We then computed the thermodynamic variables over a large range of flow parameters and assembled the phase diagram shown in Fig. \ref{fig_map}, with axes the non-wetting saturation $S_n$ and the capillary number $\log_{10} {\rm Ca}$, at unit viscosity ratio. The magnetization order parameter $m$ and susceptibility $\chi_{\rm m}$ show little structure, whereas the Edwards–Anderson order parameter $q$ and the spin-glass susceptibility $\chi_{\rm sg}$ exhibit pronounced signatures, a spin-glass phase and a paramagnetic phase separated by critical lines.   

Berg et al.\ \cite{berg2026from} identify the Darcy-scale flow regimes Ia, Ib, II, and and III. The transition between regimes Ia and Ib is a kinetic arrest, and therefore not a proper transition.  The transition between regimes II and III is also not a proper transition, but rather a saturation phenomenon: all fluid-fluid interfaces that can move, are moving in phase III, whereas the non-linearity of the constitutive equation in phase II is due to increasing numbers of mobilized interfaces with increasing pressure gradient \cite{roy2024effective}. Fig. \ref{fig_pq} shows the constitutive relations between volumetric flow rate and excess pressure drop $\Delta P-P_t$ for the dynamic pore-network model at different saturations, with the spin-glass transition marked. Except at the lowest and highest saturations considered, the Ib-II transition coincides with the spin-glass transition. By contrast with regime II, which shows no hysteresis \cite{erpelding2013history}, regime Ib exhibits hysteresis as well as strong fluctuations and long relaxation times \cite{armstrong2016role, gao2020pore,zbg21}. We therefore identify regime Ib as a glassy flow phase.

A further observation supports this picture. In regime Ib the flow organizes into high-velocity pathways ("highways") that effectively short-circuit the network. If the flow rate is increased and then returned to its original value, the system does not recover the previous highways but settles into new ones, the difference depending on how far the flow rate was raised. This is illustrated in Fig. \ref{fig_frustration} showing the velocity distribution in the DPN model with the darker gray representing higher velocity. We start from a steady state at Ca = $10^{-4}$ (left panel), then we increase the capillary number to $10^{-2}$ and let it reach steady state (middle panel) before reducing the capillary number back to $10^{-4}$ and let the model reach steady state again (right panel). We see that the system does {\it not} return to the same state after increasing the capillary number. For immiscible two-phase flow in porous media, this is hysteresis. In the spin-glass context it is frustration on a rugged energy landscape, indicating that neither the in first nor in the last panel the model is in a globally optimal state, rather, both are in locally optimal states. This is consistent with the ultrametric structure of spin glasses, in which the distance between two states is set by how far the flow rate must be raised and lowered to move between them \cite{rammal1986ultrametricity}. The mapped steady-state flow problem thus exhibits the hallmarks of a rugged, hierarchically organized energy landscape with many locally stable states, as found in long-range interacting spin glasses \cite{parisi2023nobel}.

Specifying pore-scale configurational data with Ising-like spins and using them as input to Boltzmann machine learning provides a route for experiments and simulations to characterize Darcy-scale flow regimes from pore-scale measurements. Several questions remain: what else can the spin model reveal about the flow physics, is it meaningful to introduce a temperature in the spin model, and would it be related to the agiture (``agitation temperature") of the flow model \cite{hansen2023statistical, hansen2026immiscible}? What is the form of the overlap distribution? Addressing these would test how far the correspondence between the two systems extends.

\begin{acknowledgments}
We thank Steffen Berg, Daan Frenkel and Sauro Succi for helpful discussions. This work was partly supported by the Research Council of Norway through its Centers of Excellence funding scheme, project number 262644, and the INTPART program, project number 309139. SS and AH furthermore acknowledge funding from the European Research Council (Grant Agreement 101141323 AGIPORE). HC and JSA thank the Brazilian agencies CAPES, FUNCAP and the National Institute of Science and Technology for Complex Systems (INCT-SC) for financial support. JAS acknowledges support from CNPq, Award 303765/2017-8, Bolsa PQ.
\end{acknowledgments}

\end{document}